\documentclass[a4paper,11pt]{article}
\pdfoutput=1 % if your are submitting a pdflatex (i.e. if you have
\usepackage{jheppub} % for details on the use of the package, please
\usepackage[T1]{fontenc} % if needed

\usepackage{amsmath,amssymb,amsfonts,slashed}
\usepackage{cancel}
\usepackage{mathtools}
\usepackage[normalem]{ulem}

\title{\boldmath  Anti-collinear resummation in JIMWLK evolution in the linear regime}

\author[a]{Alex Kovner,}
\author[b]{Michael Lublinsky,}
\author[b,1]{Maxim Nefedov\note{Corresponding author.}}
\author[c]{and Vladimir Skokov}

\affiliation[a]{Physics Department, University of Connecticut, \\
196A Auditorium road, Storrs, CT 06269-3046, U.S.A.}
\affiliation[b]{Physics Department, Ben-Gurion University of the Negev, \\
Beer Sheva 84105, Israel}
\affiliation[c]{Department of Physics, North Carolina State University, \\
Raleigh, NC 27695, U.S.A.}

\emailAdd{alexander.kovner@uconn.edu}
\emailAdd{lublinm@bgu.ac.il}
\emailAdd{nefedov@post.bgu.ac.il}
\emailAdd{vskokov@ncsu.edu}

\abstract{The recently-proposed  resummation procedure for anti-collinear logarithms in the JIMWLK kernel~\cite{Kovner:2023vsy} is studied in the linear (BFKL) regime in the fixed-coupling approximation. Simple closed form expressions for the resummed momentum space kernel and characteristic function $\chi(n,\gamma)$ are found. We find that the anti-collinear pole in the leading order characteristic function at $\gamma=1$ disappears, and instead $\chi(\gamma=1)=\frac{12}{11}\frac{\pi}{\alpha_sN_c}$  for $n_F=0$. Comparison with the known NLO BFKL eigenvalue, with the target-Bjorken limit ($Q_T\gg Q_P$) of the $\gamma^*(Q_P)+\gamma^*(Q_T)$-scattering amplitude and with the ``all-poles'' resummation prescription are presented. }

\newcommand{\T}[1]{\boldsymbol{#1}}  % Transverse vector
\newcommand{\Tb}[1]{\bar{\boldsymbol{#1}}} % Transverse vector with a bar

\newcommand{\tr}{\mathop{\mbox{tr}}\nolimits}

\newcommand{\ket}[1]{\left\vert #1 \right\rangle}
\newcommand{\bra}[1]{\left\langle #1 \right\vert}

\newcommand{\new}[1]{{\color{red}{#1}}}

\begin{document} 
\maketitle
\flushbottom

%%%%%%%%%%%%%%%%%%%%%%%%%%%%%%%%%%%%%%%%%%%%%%%%%%%%%%%%%%%%%%%%%%%%%%%%%%%%%%%%%%%%%%%%%%%

\section{Introduction}\label{sec:intro}

The theory of evolution of hadronic observables at high energy has been around since the seminal BFKL works~\cite{BFKL1,BFKL2,BFKL3}. It deals with the multi-Regge regime, where the transverse momenta of produced particles are much smaller than their energy or longitudinal momenta. The original BFKL works have been extended over the years in several directions. NLO corrections to the BFKL equation have been calculated~\cite{NLO-BFKL,NLOCiafaloni1,NLOCiafaloni2}, and saturation effects due to multiple scattering effects have been included in the evolution (BK-JIMWLK formalism)~\cite{Balitsky:1998kc,Kovchegov:1999ua,Kovchegov:1999yj,Jalilian-Marian:1997qno,Jalilian-Marian:1997ubg,Jalilian-Marian:1997jhx,Kovner:2000pt,Kovner:1999bj,Iancu:2000hn,Iancu:2001ad,Ferreiro:2001qy,Weigert:2000gi}.

Both NLO BFKL \cite{NLO-BFKL,NLOCiafaloni1,NLOCiafaloni2} and NLO BK equations \cite{Balitsky:2007feb} suffer from a known malaise: they contain large transverse logarithms which lead to instabilities of solutions at very high energy. These large logarithms come in two distinct varieties. One set is due to energy nonconservation which is the property of these approaches at leading order, whose perturbative restoration turns out to be a large effect. The second type is related to DGLAP physics, which is assumed to be unimportant in multi-Regge kinematics, but turns out to be very important at NLO (and beyond). In fact, understanding the DGLAP type corrections to BFKL dynamics is an important physical question in its own right as it  is relevant for high energy processes that contain a hard scale. 
A lot of work has been done on resummation of these large transverse logarithms \cite{Kutak:2004ym,Motyka:2009gi,Beuf:2014uia,Iancu:2015vea,Iancu:2015joa,SabioVera:05,Ducloue:2019ezk,
Boussarie:2025bpq,Boussarie:2025mzh}. Although much has been understood, it is fair to say that the story there is far from complete.

In a recent work ~\cite{Kovner:2023vsy}, some of us have examined the question of DGLAP resummation within the JIMWLK evolution. We have identified the terms within the NLO JIMWLK kernel \cite{Kovner:2014lca,Kovner:2014xia,Lublinsky:2016meo} responsible for these processes in the anti-collinear regime (see below {, Sec.~\ref{sec:resumm-general}}), and have derived an equation that resums these corrections to all orders in $\alpha_s$. The physics of this resummation is completely transparent and is based directly on a DGLAP cascade in the projectile. The equation in question is nonlinear, and is thus not easy to solve. In ~\cite{Kovner:2023vsy}  it was solved approximately in the dilute regime of weak target color fields and in the opposite limit of saturated ``black disk'' target. In the present work we study  resummation in the weak field limit in much greater detail. 

We work in the dilute limit, where the dipole scattering amplitude can be expanded in $\alpha_s$ and the relevant equation becomes linear. In this linearized approximation, the energy evolution is also given by a linear (BFKL) equation. We thus explore the effect of the anti-collinear resummation on the BFKL evolution.

Our main result is the calculation of the characteristic function $\chi(\gamma)$ of the anti-collinearly resummed BFKL equation. We find that the resummation does not affect $\chi(\gamma)$ close to the collinear pole $\gamma=0$ as expected, but has a rather dramatic effect close to $\gamma=1$. In particular the LO BFKL pole at $\gamma=1$ disappears, so that $\chi(1)=a/\alpha_s$. Such behaviour has been anticipated in previous works, but the exact value of the constant $a$ we find is different than previously thought. We also check the known perturbative properties of $\chi(\gamma)$ when expanded near $\gamma=1$, and find that our solution reproduces those.

The paper has the following structure. In Sec.~\ref{sec:AC-res} we explain the general setup of anti-collinear resummation~\cite{Kovner:2023vsy} in the JIMWLK kernel (Sec.~\ref{sec:resumm-general}); linearize the kernel to obtain the resummed BKFL kernel (Sec.~\ref{sec:res-BFKL}) ({with certain sum-rules for the resummation functions being proven in the Appendix~\ref{append:R-sum-rules}}) in coordinate (Sec.~\ref{sec:res-BFKL-coord}) and momentum (Sec.~\ref{sec:res-BFKL-mom}) spaces. Details of these derivations are presented in the Appendix~\ref{append:Fourier}. The linearized resummation equations are derived and solved in (Sec.~\ref{sec:lin-res-coord}) and (Sec.~\ref{sec:lin-res-mom}).
Part of the  discussion irrelevant for the main content of the paper is relegated to  Appendix~\ref{append:lin-res-mom-1}. 

Using these results, in Sec.~\ref{sec:DGLAP} we derive the resummed BFKL characteristic function (Sec.~\ref{sec:DGLAP-chi}). Its behaviour in the anti-collinear limit  is compared with the known NLO BFKL eigenvalue in Sec.~\ref{sec:comp-NLO-BFKL}.  In Sec.~\ref{sec:gamma-gamma} we derive the  $\gamma^* \gamma^*$ scattering amplitude in the target-Bjorken limit and compare it with the asymptotic behaviour known from the DGLAP approach. In Sec.~\ref{sec:subl} we focus on  subleading-logarithmic effects, which originate from our resummation, and discuss their sensitivity to the details of the choice of scale for the DGLAP resummation. Finally in Sec.~\ref{sec:conclusions} we summarise our conclusions and present an outlook for future work.

 A standard discussion about different rapidity-factorisation schemes in the BFKL is flashed in Appendix~\ref{append:rap-schemes}, while details about the off-shell photon impact factors are presented in Appendix~\ref{append:gamma-IF}.
%% END OF INTRODUCTION %%%%%%%%%%%%%%%%%%%%%%%%%%%%%%%%%%%%%%%%%%%%%%%%%%%%%%%%%%%%%%%%%%%%

\section{Anti-collinear resummation in JIMWLK}\label{sec:AC-res}

\subsection{The resummation setup}\label{sec:resumm-general}

The JIMWLK equation \cite{Jalilian-Marian:1997qno,Jalilian-Marian:1997ubg,Jalilian-Marian:1997jhx,Kovner:2000pt,Kovner:1999bj,Iancu:2000hn,Iancu:2001ad,Ferreiro:2001qy,Weigert:2000gi}  can be understood as evolution of the projectile operator ${\cal O}[S]$ which is expressed in terms of Wilson-line factors $S^{ab}(\T{x})$.
The latter describes eikonal scattering of a projectile gluon at transverse position $\T{x}$  on a hadronic target. With boosting of the projectile, its large light-cone component of momentum ($P^+$) increases and the expectation value of the operator ${\cal O}_Y$ evolves according to
\begin{equation}
    \frac{\partial}{\partial Y} {\cal O}_Y[S] = -\hat{H}_{\text{JIMWLK}}\, {\cal O}_Y[S], \label{eq:JIMWLK} 
\end{equation}
with ``rapidity'' $Y=\ln P^+$. The JIMWLK Hamiltonian in the LO in $\alpha_s$ is 
\begin{equation}
    \hat{H}_{\text{JIMWLK}}= \frac{\alpha_s(\mu)}{2\pi^2} \int\limits_{\T{x},\T{y},\T{z}} K(\T{x},\T{y},\T{z}) \left[ J_L^a(\T{x}) J_L^a(\T{y}) + J_R^a(\T{x}) J_R^a(\T{y}) - 2S^{ab}(\T{z}) J_L^a(\T{x}) J^b_R(\T{y})   \right] \label{eq:H-JIMWLK}
\end{equation}
with the Weiczacker-Williams kernel
\begin{eqnarray}
    && K(\T{x},\T{y},\T{z}) = \frac{(\T{x}-\T{z})\cdot (\T{y}-\T{z})}{(\T{x}-\T{z})^2 (\T{y}-\T{z})^2} = \frac{1}{2} \left( \frac{1}{(\T{x}-\T{z})^2} + \frac{1}{(\T{y}-\T{z})^2} - K_D(\T{x},\T{y},\T{z})  \right), \\
    && K_D(\T{x},\T{y},\T{z})=\frac{(\T{x}-\T{y})^2}{(\T{x}-\T{z})^2 (\T{y}-\T{z})^2}. \label{eq:KD-LO-def}
\end{eqnarray}
The dipole kernel $-K_D(\T{x},\T{y},\T{z})/2$ can be substituted for $K$ in Eq.~(\ref{eq:H-JIMWLK}) if the Hamiltonian acts on color singlet combinations of Wilson lines, i.e. if the projectile state is a color singlet. The left-right color rotation operators act on the adjoint Wilson line (the eikonal gluon scattering matrix) as \cite{Kovner:2005jc}:
\begin{eqnarray}
   && \hspace{-12mm}J_L^a(\T{y}) S^{bc}(\T{x}) = [T^a S (\T{x}) ]^{bc}  \delta^{(2)}(\T{x}-\T{y}), ~~~~ J_L^a(\T{y}) (S^\dagger(\T{x}))^{bc} = [- S^\dagger (\T{x}) T^a ]^{bc}  \delta^{(2)}(\T{x}-\T{y}), \label{eq:JL-def} \\
   && \hspace{-12mm} J_R^a(\T{y}) S^{bc}(\T{x}) = [S (\T{x}) T^a]^{bc}  \delta^{(2)}(\T{x}-\T{y}), ~~~~ J_R^a(\T{y}) (S^\dagger(\T{x}))^{bc} = [-T^a S^\dagger (\T{x})]^{bc}  \delta^{(2)}(\T{x}-\T{y}), \label{eq:JR-def} 
\end{eqnarray} 
where the $SU(N_c)$ generators in the adjoint representation are $T_{bc}^a=-if^{abc}$. 
%\old{Similar commutation relations hold between} 
 $J_{L(R)}$ act in a similar way on fundamental Wilson lines $V^{ij}(\T{x})$ which describe eikonal propagation of quarks.
Another important property of the color rotation operators is
\begin{eqnarray}
    S^{ab}(\T{x}) J_{R}^{b}(\T{x}) = J^a_{L}(\T{x}),~~~~~
 S^{ab}(\T{x}) J_{L}^{a}(\T{x})   = J^b_{R}(\T{x}), \label{eq:bare-SJl-eqn}
\end{eqnarray}
which in particular ensures the UV-finiteness of the Hamiltonian at $\T{z}\to \T{x}$ or $\T{z}\to\T{y}$. 
%\old{if one replaces $K\to -K_D/2$} in Eq. (\ref{eq:H-JIMWLK}).

 The next-to-leading order (NLO) correction to the JIMWLK Hamiltonian in QCD were first derived  in Refs.~\cite{Kovner:2013ona,Kovner:2014lca,Kovner:2014xia} using the results of Refs.~\cite{Balitsky:2007feb,Grabovsky:2013mba} and then re-derived independently from Light-Cone Perturbation Theory (LCPT) in Ref.~\cite{Lublinsky:2016meo}. Beyond the LO in $\alpha_s$, the Hamiltonian (\ref{eq:H-JIMWLK}) receives corrections, many of which are enhanced by ``transverse'' logarithms: $\alpha_s \ln((\T{x}-\T{z})^2)$, $\alpha_s \ln((\T{x}-\T{y})^2)$ etc. These higher order corrections give $O(1)$ contribution to the cross section if the inverse correlation lengths in the projectile ($Q^{-1}_P$) and target ($Q^{-1}_T$) become hierarchical\footnote{Here we do not deal with double logarithmic terms in the NLO JIMWLK, which have a different physical origin, see \cite{Boussarie:2025mzh,Boussarie:2025bpq} for the most recent discussion.}. 

  Customarily one thinks of the projectile to be a small object, e.g. highly virtual photon, so that the regime $Q_P\gg Q_T$ corresponds to the usual Bjorken limit, typically studied in the case of DIS. In this case the DGLAP (or {\it collinear}) resummation of $\ln(Q_P/Q_T)$-enhanced corrections is mandatory.   Large transverse logarithms also arise in the opposite, {\it anti-collinear} regime: $Q_T\gg Q_P$. DGLAP resummation in the anti-collinear regime within JIMWLK approach was formulated in  Ref.~\cite{Kovner:2023vsy}.   
 In the present paper we study in detail how the anti-collinear resummation %\old{ put forward in Ref.~\cite{Kovner:2023vsy}} 
 affects the weak field (a.k.a BFKL) dynamics\new{\footnote {Which term ``collinear'' or ``anti collinear'' to use is a matter of choice corresponding to which one of the colliding objects we choose to call ``target'' and which one ``projectile''. Eventually of course one should resum both types of logarithms in the JIMWLK framework, but this requires a significant generalization of the approach of \cite{Kovner:2023vsy} and is left for future work.}}. 

The anti-collinear resummation in the JIMWLK kernel has a particularly concise form, noted in Ref.~\cite{Kovner:2023vsy}. Large transverse logarithms are resummed by replacing the basic degrees of freedom of JIMWLK -- $S^{ab}(\T{x})$  by their {\it dressed} version $\Bbb{S}_Q^{ab}(\T{x}) \equiv \Bbb{S}^{ab}(\T{x},Q)$, which depends on the appropriately chosen transverse resolution scale in the projectile $Q$, and otherwise keeping the functional form of LO JIMWLK Hamiltonian:
\begin{equation}
\hat{H}^{\text{(res.)}}_{\text{JIMWLK}}= \frac{\alpha_s(\mu)}{2\pi^2} \int\limits_{\T{x},\T{y},\T{z}} K(\T{x},\T{y},\T{z}) \left[ J_L^a(\T{x}) J_L^a(\T{y}) + J_R^a(\T{x}) J_R^a(\T{y}) - 2\Bbb{S}_Q^{ab}(\T{z}) J_L^a(\T{x}) J^b_R(\T{y})   \right]. \label{eq:H-JIMWLK-res}
\end{equation}
The dressed adjoint Wilson line is found by  solving the following evolution equation  ~\cite{Kovner:2023vsy}:
\begin{eqnarray}
    && \frac{\partial}{\partial \ln Q^2} \Bbb{S}^{ab}(\T{x},Q) = -a_s\beta_0 \Bbb{S}^{ab}(\T{x},Q) \nonumber \\
    && - a_s \int\limits_0^1 d\xi\int\limits_0^{2\pi} \frac{d\phi}{2\pi} \bigg[  2N_c p_{gg}(\xi) \Bbb{D}^{ab}\left( \T{x} + (1-\xi) Q^{-1} \T{n}_\phi, \T{x} - \xi Q^{-1} \T{n}_\phi, Q \right) \nonumber \\
    && + 2T_F n_F p_{qg}(\xi) \Bbb{D}^{(F),ab}\left( \T{x} + (1-\xi) Q^{-1} \T{n}_\phi, \T{x} - \xi Q^{-1} \T{n}_\phi, Q \right) \bigg],\label{eq:eqn-SQ-quarks}
\end{eqnarray}
where $\T{n}_\phi$ is the radial unit vector relative to point $\T{x}$ in the transverse plane at azimuthal angle $\phi$.  Equation \eqref{eq:eqn-SQ-quarks} is written for the theory with $n_F$ massless quark flavours and fixed coupling, so that  $a_s=\alpha_s(\mu)/(4\pi)$, $\beta_0=\beta_0^{(g)}+\beta_0^{(q)}$ with $\beta_0^{(g)}=11C_A/3$, $C_A=N_c$, $\beta_0^{(q)}=-4n_FT_F/3$, $T_F=1/2$. The other functions appearing in 
 {Eq.} (\ref{eq:eqn-SQ-quarks}) are:
\begin{eqnarray}
        && \Bbb{D}^{ab}(\T{x},\T{y}, Q)= \frac{1}{N_c} \tr\left[ T^a \Bbb{S}(\T{x},Q) T^b \Bbb{S}^\dagger (\T{y},Q) \right], \\
        &&  \Bbb{D}^{(F),ab}(\T{x},\T{y}, Q)=2\tr\left[ t^a \Bbb{V}(\T{x},Q) t^b \Bbb{V}^\dagger (\T{y},Q) \right], \\
        && p_{gg}(z)=\frac{g(z)}{z_+ (1-z)_+}, \\
        && g(z)= z^2+(1-z)^2+z^2(1-z)^2, \\
        && p_{qg}(z)=z^2+(1-z)^2,
\end{eqnarray}
where $T^a_{bc}$ and $t^a_{ij}$ are $SU(N_c)$-generators in the adjoint and fundamental representations respectively. The functions $p_{gg}(z)$ and $p_{qg}(z)$ are the usual DGLAP splitting functions $P_{gg}(z)$ and $P_{qg}(z)$ respectively. The $P_{gg}(z)$ is taken without the $\delta(1-z)$-term and with the high-energy $1/z$-pole subtracted. This subtraction avoids double-counting with the JIMWLK evolution.  The two-sided plus distribution in $p_{gg}(z)$ is defined as: 
\begin{equation}
\int\limits_0^1 \frac{f(z) dz}{z_+(1-z)_+} = \int\limits_0^1 \bigg[\frac{f(z)}{z(1-z)} - \frac{f(0)}{z}-\frac{f(1)}{1-z}\bigg].
\end{equation}

 Equation~\eqref{eq:eqn-SQ-quarks} is supplemented by  the evolution equation for the dressed fundamental Wilson line ($\Bbb{V}_Q^{ij}$)
\begin{eqnarray}
    && \frac{\partial}{\partial \ln Q^2} \Bbb{V}_{ij}(\T{x},Q) = -3a_s C_F   \Bbb{V}_{ij}(\T{x},Q) \nonumber \\
    && - 2a_s \int\limits_0^1 d\xi\int\limits_0^{2\pi} \frac{d\phi}{2\pi} p_{gq}(\xi) \Bbb{D}^{(FA)}_{ij}\left( \T{x} + (1-\xi) Q^{-1} \T{n}_\phi, \T{x} - \xi Q^{-1} \T{n}_\phi, Q \right) , \label{eq:eqn-VQ}
\end{eqnarray}
where $C_F=(N_c^2-1)/(2N_c)$ and
\begin{eqnarray}
        &&  \Bbb{D}^{(FA)}_{ij}(\T{x},\T{y}, Q)=\big[ t^a  \Bbb{V} (\T{y},Q) t^b \big]_{ij} \Bbb{S}^{ab}(\T{x},Q) , \\
        && p_{gq}(z)=\frac{1+(1-z)^2}{z_+}.
\end{eqnarray}
 Here the $p_{gq}(z)$ is the DGLAP splitting function $P_{gq}(z)$,  with its leading $1/z$-pole regularised by the standard plus-prescription. 

The initial conditions for {Eqs.} (\ref{eq:eqn-SQ-quarks}) and (\ref{eq:eqn-VQ}) are set at $Q=Q_T$ such that $$\Bbb{S}^{ab}_{Q_T}(\T{x}) = S^{ab}(\T{x}) ~~~~ {\rm and} ~~~~  \Bbb{V}_{ij,Q_T}(\T{x}) = V_{ij}(\T{x}).$$  
These evolution equations have an {interesting} 
%\old{ counter-intuitive} 
property that allows one to set the initial conditions  at any scale $\Lambda>Q_T$. For any such $\Lambda$ the distance ($\sim \Lambda^{-1}$) between the two  Wilson lines in $\Bbb{D}(\T{x}_1,\T{x}_2)$ in Eqs.~(\ref{eq:eqn-SQ-quarks}) and (\ref{eq:eqn-VQ})  is smaller than the correlation length of the target $Q_T^{-1}$ \footnote{This qualitative argument relies on eventual averaging of the correlator of the Wilson lines over the target. We will see in the following that the conclusion holds quantitatively as well.}. Since e.g. $\Bbb{D}^{ab}(\T{x},\T{y})\to \Bbb{S}^{ab}(\T{x})$ when $\T{y}\to\T{x}$, the real and virtual terms in Eqs. (\ref{eq:eqn-SQ-quarks}) and (\ref{eq:eqn-VQ})   cancel each other so that effectively there  is no evolution between  the scales $\Lambda$ and  $Q_T$. As a consequence  one does not  need to know the actual value of $Q_T$ and can set the initial condition $\Bbb{S}^{ab}_{\Lambda}(\T{x}) = S^{ab}(\T{x})$ and  $\Bbb{V}_{ij,\Lambda}(\T{x}) = V_{ij}(\T{x})$ at $\Lambda\to\infty$, as we will do below. 

The issue of choosing the scale $Q$ in Eq.~(\ref{eq:H-JIMWLK-res}) is more interesting. 
 Physically it is clear that $Q$ should be of order $Q_P$, but this is an ``average'' quantity. In practice one notes that the relevant transverse logarithms are completely eliminated from the NLO JIMWLK kernel if one stops the evolution once the transverse size of the dressed gluon (or quark) becomes equal to the distance between the gluon and the nearest source. On the other hand it is clear that in order to preserve the UV finiteness of the Hamiltonian the scale $Q^2$ should blow-up when $X^2=(\T{x}-\T{z})^2\to 0$ or $Y^2=(\T{y}-\T{z})^2\to 0$ such that $\Bbb{S}^{ab}_Q(\T{x})\to S^{ab}(\T{x})$.  These considerations lead to the choice made in  Ref.~\cite{Kovner:2023vsy}:
\begin{equation}
    Q^2_\star(\T{x},\T{y},\T{z})=\max(X^{-2},Y^{-2}). \label{eq:scale choice}
\end{equation}
%with $X^2\equiv(\T{x}-\T{z})^2$ and $Y^2\equiv(\T{y}-\T{z})^2$.

\begin{figure}
    \centering
   \includegraphics[width=0.6\linewidth]{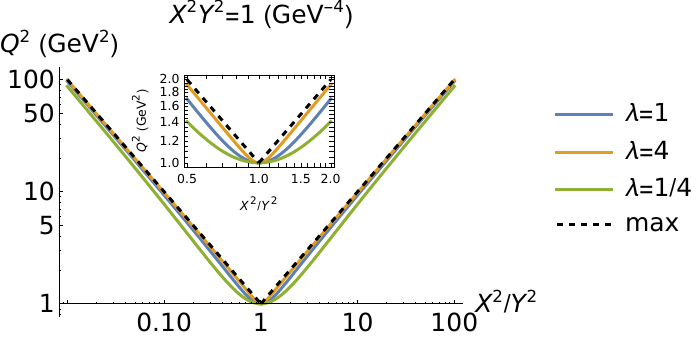}
    \caption{Solid lines -- the scale choice function (\ref{eq:scale choice-smooth}) as function of $X^2/Y^2$ at fixed $X^2Y^2$ for several values of $\lambda$. Dashed line (marked as "max")  -- the same plot but for the function (\ref{eq:scale choice}). {In the inset, the neighbourhood of the point $X^2/Y^2=1$ is zoomed-in.} }
    \label{fig:smooth-max}
\end{figure}

 The piecewise smooth expression for the scale (\ref{eq:scale choice}) is convenient also from the analytic point of view: it makes it possible to obtain the momentum space BFKL kernel in a closed form, see Sec.~\ref{sec:res-BFKL-mom} and Appendix~\ref{append:Fourier}. However, it turns out that the non-smooth behaviour of the scale (\ref{eq:scale choice}) at $X=Y$ leads to unphysical artifacts away from the anti-collinear limit and in particular to  appearance of an unphysical pole at $\gamma=1/2$ in the BFKL characteristic function (Sec.~\ref{sec:subl-mom-space}). Although in principle this is not an issue, as this regime is outside the validity of the resummation considered here, from the practical point of view it is more convenient to work with a choice of $Q$ which is free of such unphysical singularities. As it is demonstrated  in Sec.~\ref{sec:subl-coord-space},  to avoid such awkward behaviour it is sufficient to make the scale choice (\ref{eq:scale choice}) a smooth function e.g. as follows: 
\begin{equation}
    Q^2_{\star\lambda}(\T{x},\T{y},\T{z})=Y^{-2}\Theta_{\lambda}(X^{2},Y^{2}) + X^{-2}\Theta_{\lambda}(Y^{2},X^{2}), \label{eq:scale choice-smooth}
\end{equation}
where the function $\Theta_\lambda(x,y)$ is a smooth version of $\theta(x>y)$. It can be chosen for example as:
\begin{equation}
\Theta_\lambda(x,y)=\frac{1}{2}+\frac{1}{\pi}\arctan\big(\lambda \ln(x/y) \big),    
\end{equation}
where $\lambda$ is a smearing parameter and for $\lambda\to\infty$ the function $\Theta_\lambda(x,y)\to \theta(x>y)$. The $\lambda$-dependence of the scale choice (\ref{eq:scale choice-smooth}) is illustrated in the Fig.~\ref{fig:smooth-max}. It is shown in Sec.~\ref{sec:subl-mom-space} that the leading logarithmic approximation (LLA) for the anti-collinearly resummed BFKL kernel is free from any unphysical behaviour and does not depend on  exact choice of $Q_{\star}$.

In Ref.~\cite{Kovner:2023vsy} most of the discussion is carried out in a particular approximation, where the exact dependence of the ``dipole-functions'' $\Bbb{D}$ in Eqs.~(\ref{eq:eqn-SQ-quarks}) and (\ref{eq:eqn-VQ}) on the longitudinal momentum fraction $\xi$ is neglected: one replaced $\Bbb{D}_Q\left( \T{x} + (1-\xi) Q^{-1} \T{n}_\phi, \T{x} - \xi Q^{-1} \T{n}_\phi \right)\to \Bbb{D}_Q\left( \T{x} +  Q^{-1} \T{n}_\phi/2, \T{x} -  Q^{-1} \T{n}_\phi/2 \right)$. This {\it $\xi=1/2$-approximation}  is perfectly adequate for the discussion of the leading logarithmic corrections in  Ref.~\cite{Kovner:2023vsy}.  In the present paper, however,  we retain the full $\xi$-dependence of all quantities, as this poses no additional difficulties in our calculations. %\old{We will comment on the relation of the results with those using  the $\xi=1/2$-approximation whenever appropriate. }

 Finally we note that in the present paper we study the resummation of Ref.~\cite{Kovner:2023vsy} in the {\it fixed coupling approximation}. This  means that factors of $\alpha_s$ both in the resummed JIMWLK Hamiltonian~(\ref{eq:H-JIMWLK-res}) and in the resummation equations (\ref{eq:eqn-SQ-quarks}) and (\ref{eq:eqn-VQ})  are taken at a fixed scale $\mu^2$, unrelated to the coordinates $\T{x}$, $\T{y}$, $\T{z}$. While the running of the coupling in Eqs.~(\ref{eq:eqn-SQ-quarks}) and (\ref{eq:eqn-VQ}) is formally an NNLO effect and thus this approximation is fully justified, the running of $\alpha_s$ in front of the JIMWLK kernel is of the same order as the effects included in our resummation. Nevertheless such an approximation is often employed in studies of the relation between the DGLAP and BFKL equations~\cite{Jaroszewicz:1982gr,Altarelli:1999vw}. It is self-consistent as long as the scale invariance of the BFKL kernel is not violated, i.e. explicit terms dependent on $\ln\mu^2$ do not appear. This is the case for all the computations in the present paper. The running-coupling effects in the resummed JIMWLK kernel, which also had been touched upon in Ref.~\cite{Kovner:2023vsy}, and their mapping to the running-coupling effects in DGLAP evolution will be discussed in our forthcoming paper. 

This concludes our discussion of the general setup of the anti-collinear resummation in the JIMWLK kernel and now we are in a position to study how it affects the linearised energy evolution.

\subsection{Resummed BFKL kernel}\label{sec:res-BFKL}

In this section we expand the resummed JIMWLK Hamiltonian (\ref{eq:H-JIMWLK-res}) to quadratic order in terms of the Reggeized gluons. In terms of basic degrees of freedom of JIMWLK, the Reggeized gluon field $\alpha^a(\T{x})$ is defined as~\cite{Kovner:2005uw,Caron-Huot:2013fea} \footnote{ The problem of operator definitions of Reggeized gluon and its very existence beyond Next-to-Leading Logarithmic approximation attracts considerable attention nowadays, in particular in the context of ongoing computation of NNLO corrections to the kernel of BFKL equation~\cite{Byrne:2023nqx,Fadin:2023roz,Byrne:2022wzk,Caola:2021izf,Falcioni:2021dgr, DelDuca:2021vjq,Abreu:2024xoh,Buccioni:2024gzo}. Since in the present paper we are concerned with all-order in $\alpha_s$ effects in the BFKL kernel, it is important for us to stick to a particular definition, and we choose the Eq.~(\ref{eq:S-alpha-exp}). As it is noted in Ref.~\cite{Hentschinski:2018rrf}, this definition is closely related with the one given by Lipatov's gauge-invariant EFT for Multi-Regge processes in QCD~\cite{Lipatov95} which had been used in a number of theoretical and phenomenological applications, see e.g.~\cite{Chachamis:2013hma,Hentschinski:2021lsh,Nefedov:dijet,NS_PRA,He:di-Jpsi,Nefedov:2024swu}. Possible differences between different definitions are likely to appear only at three loop level (see the Appendix of Ref.~\cite{Nefedov:2019mrg}) and do not influence the leading-logarithmic collinear asymptotics of the kernel. We will come back to the interesting issue of comparing various definitions of Reggeized gluon elsewhere.}:
\begin{eqnarray}
S^{ab}(\T{x}) &=& \exp [ig T^c \alpha^c(\T{x})]^{ab} = \hat{1} + ig (T^{c_1})^{ab} \alpha^{c_1}(\T{x}) \nonumber \\
&& + \frac{(ig)^2}{2} (T^{c_1} T^{c_2})^{ab} \alpha^{c_1}(\T{x}) \alpha^{c_2}(\T{x})+O(g^3),    \label{eq:S-alpha-exp}
\end{eqnarray}
the color rotation operators ($J_{L,R}^a$) then can be defined by enforcing the properties (\ref{eq:JL-def}) and (\ref{eq:JR-def}) to all orders in $g$, leading to:
\begin{eqnarray}
  && (ig) J_{(L,R)}^a(\T{x}) = \frac{\delta}{\delta \alpha^a(\T{x})} \nonumber \\
  && + \sum\limits_{n=1}^\infty g^n c_n^{(L,R)} \left( f^{a b_1 c_1} f^{b_1 b_2 c_2} \ldots f^{b_{n-1} b c_n} \right) \left[ \alpha^{c_1} (\T{x}) \ldots \alpha^{c_n}(\T{x}) \right] \frac{\delta}{\delta \alpha^b(\T{x})}, \label{eq:JLR-exp}
\end{eqnarray}
where $c_1^{(L)}=-c_1^{(R)} = +1/2$, $c_2^{(L)}=c_2^{(R)} = 1/12$, $c_3^{(L)}=c_3^{(R)} = 0$,  $c_4^{(L)}=c_4^{(R)} = -1/720$, $c_5^{(L)}=c_5^{(R)} = 0$, $c_6^{(L)}=c_6^{(R)}= 1/30240$ and so on. The series can be resummed as \cite{Altinoluk:2013rua}:
\begin{eqnarray}
    (ig) J^a_{(L,R)} &=& \left[ F(\pm i T^c \alpha^c(\T{x})) \right]^{ab} \frac{\delta}{\delta \alpha^b(\T{x})},
\end{eqnarray}
with $F(x) = x/(1-e^{-x})$, where the $(+)$ sign is for $J_L$ and $(-)$ is for $J_R$.

For the dressed Wilson lines, the expansion (\ref{eq:S-alpha-exp}) is not valid since, as noted in Refs.~\cite{Kovner:2023vsy,Armesto:2025dwd}, the dressed Wilson line $\Bbb{S}^{ab}_Q$ for $Q<\Lambda$ is no longer a unitary matrix. Therefore we generalise the expansion (\ref{eq:S-alpha-exp}) up to the second order in the Reggeised gluon fields for the dressed adjoint Wilson line as follows:
\begin{eqnarray}
  &&  \Bbb{S}^{ab}(\T{x},Q) = \delta^{ab} + ig T^{c_1}_{ab} \int\limits_{\T{z}_1} R_Q^{(1)}(\T{x}-\T{z}_1) \alpha^{c_1}(\T{z}_1) + g d^{c_1ab} \int\limits_{\T{z}_1} R_Q^{(1,d)}(\T{x}-\T{z}_1) \alpha^{c_1}(\T{z}_1)  \nonumber \\ &&+ \frac{(ig)^2}{2} \sum\limits_{k=1}^{n_{\text{adj.}}} {\cal T}^{(k)c_1 c_2}_{ab} \int\limits_{\T{z}_1, \T{z}_2} R_Q^{(2,k)}(\T{x}-\T{z}_1,\T{x}-\T{z}_2) \alpha^{c_1}(\T{z}_1) \alpha^{c_2}(\T{z}_2) + O(g^3), \label{eq:Sbold-alpha-decomp}
\end{eqnarray}
where the basis of color structures ${\cal T}^{(j)c_1c_2}_{ab}$ in the second term spans the color-singlet subspace of the ${\bf 8}\otimes{\bf 8}\otimes {\bf 8}\otimes {\bf 8}$ and has $n_{\text{adj.}}=9$ elements for general $N_c$, see e.g. Appendix C in Ref.~\cite{Nikolaev:2005zj}. We are free to choose this basis in such a way that $\delta_{c_1c_2} {\cal T}^{(k)c_1c_2}_{ab}=0$ for $k>1$ and it is convenient to choose the first element to be ${\cal T}^{(1)c_1c_2}_{ab}=(T^{c_1} T^{c_2})_{ab}$. With this choice, the initial conditions for the resummation functions $R^{(1)}_Q$ and $R^{(2,1)}_Q$ can be read-off from Eq.~(\ref{eq:S-alpha-exp}):
\begin{eqnarray}
  &&  R_\Lambda^{(1)}(\T{z})=\delta^{(2)}(\T{z}), \label{eq:R1-init-cond} \\
  && R_\Lambda^{(2,1)}(\T{z}_1,\T{z}_2) = \delta^{(2)}(\T{z}_1)\delta^{(2)}(\T{z}_2),\; ~~~~ R_\Lambda^{(2,k)}(\T{z}_1,\T{z}_2) = 0~~\text{ for }k>1,\label{eq:R2-init-cond} \\
  && R_\Lambda^{(1,d)}(\T{z})=0. \label{eq:R1d-init-cond}
\end{eqnarray}
It turns out, that the term with $d^{abc}$ color structure in Eq.~(\ref{eq:Sbold-alpha-decomp}) does not contribute at this order. This is because the evolution equation for $R^{(1,d)}_Q(\T{z})$ is homogenous and  decoupled from the evolution of $R_Q^{(1)}(\T{z})$, thus $R^{(1,d)}_Q(\T{z})=0$ due to the initial condition (\ref{eq:R1d-init-cond}). We will drop this term in the discussion below.

 We note that the resummation functions $R^{(j)}_Q$ have a well defined physical meaning.  $R_Q^{(1)}(\T{x}-\T{z}_1)$ is the density of ``bare'' gluons at point $\T{z}_1$ in the dressed gluon state defined with resolution $Q$ at point $\T{x}$. Similarly, $R_Q^{(2)}(\T{x}-\T{z}_1,\T{x}-\T{z}_2)$ is the number density of bare gluon pairs  at transverse points $\T{z}_1$ and $\T{z}_2$ in the dressed gluon at $\T{x}$. {According to this probabilistic interpretation, the resummation functions should have the following properties:}
 \begin{eqnarray}
   && \int\limits_{\T{z}} R_Q^{(1)}(\T{z}) = 1, \label{eq:R1-norm-int} \\
   && \int\limits_{\T{z}_2} R_Q^{(2,1)}(\T{z}_1,\T{z}_2) = R_Q^{(1)}(\T{z}_1), \label{eq:R2-norm-int}
\end{eqnarray}
{which indeed hold for any $Q$ as  proven in Appendix~\ref{append:R-sum-rules} using the evolution equations for $R^{(j)}_Q$ derived in Sec.~\ref{sec:lin-res-coord}. }

 A parametrization similar to Eq.~(\ref{eq:Sbold-alpha-decomp}) can be written for the dressed fundamental Wilson line:
\begin{eqnarray}
  && \Bbb{V}_{ij}(\T{z}) = \delta_{ij} + ig t^{c_1}_{ij} \int\limits_{\T{z}_1} r_Q^{(1)}(\T{z}-\T{z}_1) \alpha^{c_1}(\T{z}_1) \nonumber \\
  && + \frac{(ig)^2}{2} \sum\limits_{k=1}^{n_{\text{fund.}}} {\cal T}^{(F,k)c_1 c_2}_{ij} \int\limits_{\T{z}_1 \T{z}_2} r_Q^{(2,k)}(\T{z}-\T{z}_1,\T{z}-\T{z}_2) \alpha^{c_1}(\T{z}_1) \alpha^{c_2}(\T{z}_2) + \ldots, \label{eq:Vbold-alpha-decomp}
\end{eqnarray}
where the basis of color structures ${\cal T}^{(F,k)c_1 c_2}_{ij}$  spanning the color-singlet subspace of ${\bf 3}\otimes \bar{\bf 3} \otimes {\bf 8} \otimes {\bf 8}$ has $n_{\text{fund.}}=3$ elements. By the same reasoning as above,  we take ${\cal T}^{(F,1)c_1 c_2}_{ij} = (t^{c_1} t^{c_2})_{ij}$ and assume  $\delta_{c_1 c_2} {\cal T}^{(F,1)c_1 c_2}_{ij} =0$ for $k>1$. The initial conditions for the resummation functions in the quark sector are:
\begin{eqnarray}
  &&  r_\Lambda^{(1)}(\T{z})=\delta^{(2)}(\T{z}), \label{eq:r1-init-cond} \\
  && r_\Lambda^{(2,1)}(\T{z}_1,\T{z}_2) = \delta^{(2)}(\T{z}_1)\delta^{(2)}(\T{z}_2),\\
  && r_\Lambda^{(2,k)}(\T{z}_1,\T{z}_2) = 0\text{ for }k>1,\label{eq:r2-init-cond}
\end{eqnarray}
following from
\begin{equation}
    V_{ij}(\T{x})=\exp\big[ i g t^c \alpha^c(\T{x}) \big]_{ij}.
\end{equation}
The resummation functions $r^{(1)}_Q$ and $r^{(2,1)}_Q$ satisfy the following sum-rules, identical to the sum-rules (\ref{eq:R1-norm-int}) and (\ref{eq:R2-norm-int}) for the gluon case:
 \begin{eqnarray}
         && \int\limits_{\T{z}} r^{(1)}_Q(\T{z}) = 1, \label{eq:r1-sum-rule} \\
         && \int\limits_{\T{z}_2} r^{(2,1)}_Q(\T{z}_1,\T{z}_2) = r^{(1)}_Q(\T{z}_1),\label{eq:r2-sum-rule}
 \end{eqnarray}
 {for any $Q$ as is proven in  Appendix~\ref{append:R-sum-rules}. }

 In Sec. \ref{sec:lin-res}, we reformulate Eqs. (\ref{eq:eqn-SQ-quarks}) and (\ref{eq:eqn-VQ}) as equations for $R_Q^{(j)}$ and $r_Q^{(j)}$ and then solve them. 

\subsubsection{Resummed BFKL kernel in coordinate space}\label{sec:res-BFKL-coord}
 Substituting the expansions (\ref{eq:JLR-exp}) and (\ref{eq:Sbold-alpha-decomp}) into the resummed Hamiltonian (\ref{eq:H-JIMWLK-res}) and expanding up to second order in $\alpha$-fields, one obtains:
\begin{eqnarray}
 \hat{H}_{\text{JIMWLK}}^{\text{(res. lin.)}}  &&  = 
\frac{\alpha_s}{2\pi^2} \int\limits_{\T{x},\T{y}}   \Bigg\{ f^{[c_1 a b_1} f^{c_2] a b_2} \bigg[ - \int\limits_{\T{z}}K(\T{x},\T{y},\T{z}) \alpha^{c_1}(\T{x}) \alpha^{c_2}(\T{y})  \nonumber  \\
    &&- \int\limits_{\T{z}_{1},\T{z}_2}K^{(2)}(\T{x},\T{y},\T{z}_1,\T{z}_2) \alpha^{c_1}(\T{z}_1) \alpha^{c_2}(\T{z}_2)  \label{eq:coll-res-K-Qt} \\
    && + \int\limits_{\T{z}}K^{(1)}(\T{x},\T{y},\T{z}) \bigg( \alpha^{c_1}(\T{z}) \alpha^{c_2}(\T{x}) + \alpha^{c_1}(\T{z})\alpha^{c_2}(\T{y})  \biggr)   \bigg] \frac{{\delta}^2}{ \delta \alpha^{b_1}(\T{x}) \delta \alpha^{b_2}(\T{y})} \nonumber \\
    && +C_A\delta^{(2)}(\T{y}-\T{x}) \int\limits_{\T{z}} [K(\T{x},\T{x},\T{z}) \alpha^{b}(\T{x})-K^{(1)}(\T{x},\T{x},\T{z}) \alpha^{b}(\T{z})] \frac{\delta}{ \delta \alpha^{b}(\T{x})} \Bigg\} , \nonumber
\end{eqnarray}
where $f^{[c_1 a b_1} f^{c_2] a b_2} = \big( f^{c_1 a b_1} f^{c_2 a b_2} + f^{c_2 a b_1} f^{c_1 a b_2} \big)/2$.  The kernels $K^{(1)}$ and $K^{(2)}$ are defined below:
\begin{eqnarray}
 && K^{(1)}(\T{x},\T{y},\T{z})= \int\limits_{\Tb{z}} K(\T{x},\T{y},\Tb{z}) R^{(1)}_{Q_\star (\T{x},\T{y},\Tb{z})}(\Tb{z}-\T{z}), \label{eq:KQt1-def} \\
 && K^{(2)}(\T{x},\T{y},\T{z}_1,\T{z}_2)= \int\limits_{\Tb{z}} K(\T{x},\T{y},\Tb{z}) R^{(2,1)}_{Q_\star(\T{x},\T{y},\Tb{z})}(\Tb{z}-\T{z}_1,\Tb{z}-\T{z}_2) \label{eq:KQt2-def}\,.
\end{eqnarray}
The kernels  incorporate the resummation functions $R^{(1)}_Q$ and $R^{(2,1)}_Q$ with either the scale choice (\ref{eq:scale choice}) { or (\ref{eq:scale choice-smooth}).} 

Acting with this linearised Hamiltonian on the color-singlet combination of two Reggeised gluons $\alpha^a(\T{x}) \alpha^a(\T{y})$ -- {\it the BFKL Pomeron}\footnote{The more common definition of the  Pomeron is $(\alpha^a(\T{x}) -\alpha^a(\T{y}))^2$. We will consider the action of the linearized Hamiltonian on this operator below as well.}, one obtains the resummed BFKL kernel in coordinate space:
\begin{eqnarray}
  &&\langle \hat{H}_{\text{JIMWLK}}^{\text{(res. lin.)}} \ \alpha^a(\T{x}) \alpha^a(\T{y}) \rangle_Y = \int\limits_{\T{z}_1,\T{z}_2}  K^{\text{(res.)}}(\T{x},\T{y},\T{z}_1,\T{z}_2)  \langle \alpha^a(\T{z}_1) \alpha^a(\T{z}_2) \rangle_Y, \label{eq:H-alpha-alpha}
\end{eqnarray}
where $\langle {\cal O} \rangle_Y\equiv\bra{T} {\cal O} \ket{T}_Y$ denotes the average of the operator $\cal{O}$  over the target state $\ket{T}$ evolved to rapidity $Y$.  The kernel appearing in Eq.~(\ref{eq:H-alpha-alpha}) is
\begin{eqnarray}
   && K^{\text{(res.)}}(\T{x},\T{y},\T{z}_1,\T{z}_2) = \frac{\alpha_sN_c}{2\pi^2} \bigg\{ \delta^{(2)}(\T{z}_1-\T{x})\delta^{(2)}(\T{z}_2-\T{y}) \bigg(\int\limits_{\T{z}} K_D(\T{x},\T{y},\T{z})\bigg)\nonumber \\
   &&-2K^{(2)}(\T{x},\T{y},\T{z}_1,\T{z}_2)  +\big( 2 K^{(1)}(\T{x},\T{y},\T{z}_1) - K^{(1)}(\T{y},\T{y},\T{z}_1) \big)\delta^{(2)}(\T{z}_2-\T{x}) \nonumber \\
   && + \big( 2 K^{(1)}(\T{x},\T{y},\T{z}_1) - K^{(1)}(\T{x},\T{x},\T{z}_1) \big)\delta^{(2)}(\T{z}_2-\T{y}) \bigg\}. \label{eq:K-res-def}
\end{eqnarray}

Another operator worth considering is obtained by expansion of the {\it color dipole} to second order in the $\alpha$-fields: % \comment{ML: I think $1/N_c$ is missing below}
\begin{equation}
   -\frac{4}{g^2}\big( \tr [ V(\T{x}) V^{\dagger}(\T{y})] - N_c \big) = (\alpha^a(\T{x})-\alpha^a(\T{y}))^2 + O(g).  
\end{equation}
With the help of  the sum-rules (\ref{eq:R1-norm-int}) and (\ref{eq:R2-norm-int}), the action of the resummed Hamiltonian (\ref{eq:coll-res-K-Qt}) on this dipole operator  can be put into  the following form:
    \begin{eqnarray}
  &&\hspace{-10mm}\langle \hat{H}_{\text{JIMWLK}}^{\text{(res. lin.)}} \ (\alpha^a(\T{x}) - \alpha^a(\T{y}))^2 \rangle_Y = \int\limits_{\T{z}_1,\T{z}_2}  K_D^{\text{(res.)}}(\T{x},\T{y},\T{z}_1,\T{z}_2)  \langle (\alpha^a(\T{z}_1)- \alpha^a(\T{z}_2))^2 \rangle_Y, \label{eq:H-(alpha-alpha)^2}
\end{eqnarray}
where:
\begin{eqnarray}
    && K_D^{\text{(res.)}}(\T{x},\T{y},\T{z}_1,\T{z}_2) = K^{\text{(res.)}}(\T{x},\T{y},\T{z}_1,\T{z}_2) + \frac{\alpha_sN_c}{2\pi^2} \bigg\{ K^{(2)}(\T{x},\T{x},\T{z}_1,\T{z}_2) + K^{(2)}(\T{y},\T{y},\T{z}_1,\T{z}_2) \nonumber \\
    && - K^{(1)}(\T{x},\T{x},\T{z}_1)\delta^{(2)}(\T{z}_2-\T{x}) - K^{(1)}(\T{y},\T{y},\T{z}_1) \delta^{(2)}(\T{z}_2-\T{y}) \bigg\}. \label{eq:K-Dip-res-coord}
\end{eqnarray}
Each term in the curly brackets in Eq.~(\ref{eq:K-Dip-res-coord}) depends  either on $\T{x}$ or  $\T{y}$. Thus they contribute only trivial terms proportional to $\delta^{(2)}(\T{k})$ in the Fourier transformed kernel, with $\T{k}$ being the momentum conjugate to $\T{x}-\T{y}$. Such terms are nullified by  convolution with the next iteration of the kernel or with  IR-safe impact factor. Therefore the kernel (\ref{eq:H-alpha-alpha}) is sufficient for the momentum space analysis in Sec.~\ref{sec:res-BFKL-mom}. Nevertheless, the representation (\ref{eq:H-(alpha-alpha)^2}) with the dipole kernel (\ref{eq:K-Dip-res-coord}) will come in handy in Sec.~\ref{sec:subl-coord-space} for  coordinate space analysis of the characteristic function. This is because the dipole kernel automatically leads to  coordinate space representation of the characteristic function in terms of explicitly IR-finite integrals.%, which can be computed numerically.

In the present paper, we will discuss  the forward scattering amplitude only. For this purpose it is sufficient to consider the average $\langle \alpha^a(\T{x}) \alpha^a(\T{y}) \rangle_Y$ which is independent of $(\T{x}+\T{y})/2$ and hence depends  only on $(\T{x}-\T{y})$. However we wish to emphasize that the anti-collinear resummation method described in Sec.~\ref{sec:resumm-general} is also applicable to a non-forward case, where the  momentum transfer conjugate to $(\T{x}+\T{y})/2$ is non zero. We will  consider the non-forward case in  future work.

  Traditionally, the BKFL evolution  is formulated  for  correlators of charge density operators in the target $\langle \rho_T^a(\T{x}) \rho_T^a(\T{y}) \rangle$ rather than for $\langle \alpha^a(\T{x}) \alpha^a(\T{y}) \rangle$. The target charge density and field operators are related as (see e.g.~\cite{Jalilian-Marian:1997qno}):
\begin{equation}
    \nabla^2 \alpha^a(\T{x}) = g\rho^a_T(\T{x}) + O(g^2\alpha^2), \label{eq:rhoT-alpha-rel}
\end{equation}
where by $O(g^2\alpha^2)$ we denote higher order terms in the fields $\alpha^a(\T{x})$, which do not contribute to the linear BFKL equation. The linear relation between $\alpha$ and $\rho_T$ is sufficient for studying the BFKL Green's function because in this approach  all the effects of higher order in $\rho_T$ on the evolution are neglected. Therefore, the rapidity evolution for the target charge density correlators takes the form:
\begin{eqnarray}
    \frac{\partial}{\partial Y} \langle \rho_T^a(\T{x}) \rho_T^a(\T{y}) \rangle_Y &=& -\int\limits_{\Tb{z}_{1,2},\T{z}_{1,2}} \nabla^2_{\T{x}} \nabla^2_{\T{y}}  K^{\text{(res.)}}(\T{x},\T{y},\Tb{z}_1,\Tb{z}_2) \nonumber \\
    && \times \nabla^{-2}(\Tb{z}_1-\T{z}_1) \nabla^{-2}(\Tb{z}_2-\T{z}_2)  \langle \rho_T^a(\T{z}_1) \rho_T^a(\T{z}_2) \rangle_Y, \label{eq:Evol-rho-rho}
\end{eqnarray}
where $\nabla^{-2}(\T{x})$ is the Green's function of $\nabla^2$: %\old{$\nabla_{\T{x-y}}^2 \nabla^{-2}(\T{y-z}) = \delta^{(2)}(\T{x-z})$ (with integration over $\T{y}$ implied)} 
{$\nabla_{\T{x}}^2 \nabla^{-2}(\T{x}) = \delta^{(2)}(\T{x})$}.

\subsubsection{Transforming the kernel to momentum space}\label{sec:res-BFKL-mom}

The momentum space version of  Eq.~(\ref{eq:Evol-rho-rho}) for the case of forward scattering is:
\begin{equation}
  \frac{\partial}{\partial Y} \langle \rho_T^a(\T{k}) \rho_T^a(-\T{k}) \rangle_Y = -\int\limits_{\T{q}} K_{\text{BFKL}}^{\text{(res.)}}(\T{k},\T{q})  \langle \rho_T^a(\T{q}) \rho_T^a(-\T{q}) \rangle_Y,  \label{eq:BFKL-mom}
\end{equation}
with the following definition of the Fourier transformed kernel:
\begin{equation}
     K^{\text{(res.)}}_{\text{BFKL}}(\T{k},\T{q}) = \frac{\T{k}^4}{\T{q}^4} \int\limits_{\T{x},\T{y},\T{z},\T{z}'}  \frac{e^{-i\T{k}(\T{x}-\T{y})}}{(2\pi)^{2}}   K^{\text{(res.)}}(\T{x},\T{y},\T{z},\T{z}') \frac{e^{i\T{q}(\T{z}-\T{z}')} + e^{-i\T{q}(\T{z}-\T{z}')} }{2 S_\perp}, \label{eq:Kres(k,q)-def}
\end{equation}
where $S_\perp=\int d^{2}\T{x}$ is the area of the transverse space. %While computing the Fourier transform we will be working in $(2-2\epsilon)$-dimensional transverse space to be able to track potential IR-divergences.  The final result for the momentum space kernel  is IR finite. 

{It is possible to compute the Fourier transform of the kernel (\ref{eq:K-res-def}) with the piecewise-smooth scale (\ref{eq:scale choice}) exactly. The details of this computation are given in Appendix~\ref{append:Fourier} and only the final result is presented here:}
\begin{eqnarray} 
 &&    K^{\text{(res.)}}_{\text{BFKL}}(\T{k},\T{q}) = \frac{\alpha_s N_c}{2\pi^2} \frac{\T{k}^4}{\T{q}^4} \bigg\{ -\frac{\T{q}^2}{\T{k}^2(\T{k}-\T{q})^2_+}  \label{eq:KD-res-drdQ-exact} \\
 &&     -2\frac{\T{k}\cdot (\T{k}-\T{q})}{\T{k}^2 (\T{k}-\T{q})^2}  \int\limits_{0}^\infty \frac{dQ^2}{Q^2} \frac{\partial R^{(1)}_Q(\T{q})}{\partial\ln Q^2} J_0\left(\frac{|\T{k}|}{Q} \right) J_0\left(\frac{|\T{k}-\T{q}|}{Q} \right) + \frac{1}{\T{k}^2} \int\limits_{0}^\infty \frac{dQ^2}{Q^2} \frac{\partial R_Q^{(2,1)}(\T{q})}{\partial \ln Q^2} J_0^2\left(\frac{|\T{k}|}{Q}\right) \nonumber \\ 
 &&    - \pi \delta^{(2)}(\T{k}-\T{q})\int\limits_{0}^\infty \frac{dQ^2}{Q^2} \frac{\partial R^{(1)}_Q(\T{q})}{\partial\ln Q^2} \int\limits_0^{Q^2} \frac{dq^2}{q^2} J_0\left( \frac{|\T{k}|}{q}\right)  + (\T{q}\to -\T{q}) \bigg\}, \nonumber
\end{eqnarray}
%\begin{eqnarray} 
% &&    K^{\text{(res.)}}_{\text{BFKL}}(\T{k},\T{q}) = \frac{\alpha_s N_c}{2\pi^2} \frac{\T{k}^4}{\T{q}^4} \bigg\{ -\frac{\T{q}^2}{\T{k}^2(\T{k}-\T{q})^2_+}  \label{eq:KD-res-drdQ-exact} \\
% &&     -2\frac{\T{k}\cdot (\T{k}-\T{q})}{\T{k}^2 (\T{k}-\T{q})^2}  \int\limits_{Q_0^2}^\infty \frac{dQ^2}{Q^2} \frac{\partial R^{(1)}_Q(\T{q})}{\partial\ln Q^2} J_0\left(\frac{|\T{k}|}{Q} \right) J_0\left(\frac{|\T{k}-\T{q}|}{Q} \right) + \frac{1}{\T{k}^2} \int\limits_{Q_0^2}^\infty \frac{dQ^2}{Q^2} \frac{\partial R_Q^{(2,1)}(\T{q})}{\partial \ln Q^2} J_0^2\left(\frac{|\T{k}|}{Q}\right) \nonumber \\ 
% &&    - \pi \delta^{(2)}(\T{k}-\T{q})\int\limits_{Q_0^2}^\infty \frac{dQ^2}{Q^2} \frac{\partial R^{(1)}_Q(\T{q})}{\partial\ln Q^2} \int\limits_0^{Q^2} \frac{dq^2}{q^2} J_0\left( \frac{|\T{k}|}{q}\right)  + (\T{q}\to -\T{q}) \bigg\}, \nonumber
%\end{eqnarray}
where the Fourier transforms of the resummation functions are defined as:
\begin{eqnarray}
     R_Q^{(1)}(\T{p}) &=& \int\limits_{\T{z}} R_Q^{(1)}(\T{z}) e^{-i\T{p}\T{z}}, \label{eq:R1-q-def} \\ 
     R_Q^{(2,1)}(\T{p}) &=& \int\limits_{\T{z}_1,\T{z}_2} R^{(2,1)}_Q(\T{z}_1,\T{z}_2) e^{i\T{p}(\T{z}_1-\T{z}_2)}. \label{eq:R2-q-def}
\end{eqnarray} 
and the $(+)$-distribution in transverse momentum space is defined as\footnote{In fact, the first term in Eq.~\eqref{eq:KD-res-drdQ-exact} that contains this distribution  is one of the possible forms of presenting the LO BFKL kernel. }:
\begin{equation}
     \int\limits_{\T{q}} \frac{f(\T{q})}{(\T{k}-\T{q})^2_+} = \int\limits_{\T{q}} \frac{f(\T{q})-f(\T{k})\theta(|\T{k}-\T{q}|<|\T{k}|)}{(\T{k}-\T{q})^2}\,.
\end{equation}
From Eqs.~(\ref{eq:R1-init-cond}) and (\ref{eq:R2-init-cond}) one finds the initial conditions for these quantities: 
\begin{eqnarray}
    R_\Lambda^{(1)}(\T{p}) = R_\Lambda^{(2,1)}(\T{p})=1. \label{eq:R1-R21-mom-init-conds}
\end{eqnarray}

 {In Sec.~\ref{sec:subl}}, we will present explicit calculations pertaining to the exact Fourier transform (\ref{eq:KD-res-drdQ-exact}) with the specific choices of $Q_{\star}(\T{x},\T{y},\T{z})$ introduced above. Meanwhile, it is also very informative  to extract  the leading-logarithmic (LL) terms $\propto \alpha_s^m \ln^m (\T{q}^2/\T{k}^2)$ in the anti-collinear limit $\T{q}^2\gg \T{k}^2$. Discarding all other spurious N${^{k\geq 1}}$LL contributions generated by the Fourier transform from position to momentum space turns out to regulate a potentially problematic behaviour of the resummed BFKL characteristic function at $\gamma=1/2$ related to the scale choice (\ref{eq:scale choice}). 
 
 The simplest way to {keep  the LL terms} only is  to approximate the Bessel functions appearing in  Eq.~(\ref{eq:KD-res-drdQ-exact})  according to the rule:
\begin{equation}
     J_0(x)\to \theta(1-x), \label{eq:J-theta-appr}
\end{equation}
{i.e. neglecting the decreasing tail of the Bessel functions at $x>1$. As we show in  Sec.~\ref{sec:subl},  the tails of the Bessel functions in Eq.~(\ref{eq:KD-res-drdQ-exact}) do not contribute to the characteristic function at the LLA in the anti-collinear limit.} Within the approximation (\ref{eq:J-theta-appr}), the result for the Fourier transformed kernel becomes more tractable:
\begin{eqnarray}
  && \hspace{-0.3cm} K^{\text{(res.,{$\theta$-approx})}}_{\text{BFKL}}(\T{k},\T{q}) = \frac{\alpha_s N_c}{2\pi^2} \frac{\T{k}^4}{\T{q}^4}  \bigg\{ -\frac{\T{q}^2}{\T{k}^2(\T{k}-\T{q})^2_+} + \frac{1}{\T{k}^2} \big(1- R_{|\T{k}|}^{(2,1)}(\T{q}) \big) \label{eq:KD-res-theta-appr-result}   \\
  &&  \hspace{-0.2cm}  -2\frac{\T{k}\cdot (\T{k}-\T{q})}{\T{k}^2 (\T{k}-\T{q})^2}   \big(1- R^{(1)}_{\max(|\T{k}|,|\T{k}-\T{q}|)}(\T{q}) \big)    - \pi \delta^{(2)}(\T{k}-\T{q}) \int\limits_{\T{q}^2}^\infty \frac{dQ^2}{Q^2} \frac{\partial R^{(1)}_Q(\T{q})}{\partial\ln Q^2} \ln\frac{Q^2}{\T{q}^2}  +(\T{q}\to -\T{q}) \bigg\}. \nonumber  
\end{eqnarray}
%\begin{eqnarray}
%  && K^{\text{(res.,$\theta$)}}_{\text{BFKL}}(\T{k},\T{q}) = \frac{\alpha_s N_c}{2\pi^2} \frac{\T{k}^4}{\T{q}^4}  \bigg\{ -\frac{\T{q}^2}{\T{k}^2(\T{k}-\T{q})^2_+} + \frac{1}{\T{k}^2} \big(1- R_{\max (|\T{k}|,Q_0)}^{(2,1)}(\T{q}) \big) \label{eq:KD-res-theta-appr-result}   \\
%  &&    -2\frac{\T{k}\cdot (\T{k}-\T{q})}{\T{k}^2 (\T{k}-\T{q})^2}   \big(1- R^{(1)}_{\max(|\T{k}|,|\T{k}-\T{q}|,Q_0)}(\T{q}) \big)   \nonumber \\ 
%  &&  - \pi \delta^{(2)}(\T{k}-\T{q}) \int\limits_{\max(\T{q}^2,Q_0^2)}^\infty \frac{dQ^2}{Q^2} \frac{\partial R^{(1)}_Q(\T{q})}{\partial\ln Q^2} \ln\frac{Q^2}{\T{q}^2}  +(\T{q}\to -\T{q}) \bigg\}. \nonumber  
%\end{eqnarray}
%\old{The scale $Q_0^2$ should be chosen of the order %$Q_P^2\sim(\T{x}-\T{y})^{-2}\sim \T{k}^2$. Let's analyse this %expression term-by-term, taking into account this choice.} 
The coefficient in front of $\delta^{(2)}(\T{k}-\T{q})$ is just an $O(\alpha_s)$ number, it has no  large logarithms $\sim\ln (\T{k}^2/\T{q}^2)$ and hence does not contribute to the LLA. To distil the contribution of the LLA in the limit $\T{k}^2\gg \T{q}^2$ from the remaining terms, we should anticipate another property  for the resummation functions $R_Q^{(1)}(\T{q})$ and $R_Q^{(2,1)}(\T{q})$.  The momentum $\T{q}$, conjugate to $(\T{z}-\T{z}')$ in  the function $\Bbb{D}^{ab}(\T{z},\T{z}')$, which probes the target in Eq.~(\ref{eq:Kres(k,q)-def}), is of the order of the correlation scale in the target ($Q_T^{-1}$). The anti-collinear resummation of Sec.~\ref{sec:resumm-general}, by construction, requires $Q_T\gg Q_P$. Consequently, both resummation functions $R_{Q_P}^{(j)}(\T{q})\to 1$ for $|\T{q}|<Q_P$. This means that the term proportional to $1-R^{(1)}_{\max(|\T{k}|,|\T{k}-\T{q}|)}(\T{q})$  drops-out both for $\T{k}^2\gg \T{q}^2$ and $\T{k}^2\ll \T{q}^2$. This property
 will become evident when we  discuss explicit solutions of the evolution equations in Sect.~\ref{sec:lin-res-mom}.

As a result, only the $R^{(2,1)}_{|\T{k}|}(\T{q})$-term in the first line of Eq.~(\ref{eq:KD-res-theta-appr-result}) represents the genuine LLA contribution and the final version of the LL-resummed BFKL kernel  in momentum space is
\begin{eqnarray}
    K^{\text{(res., LLA)}}_{\text{BFKL}}(\T{k},\T{q}) = \frac{\alpha_s N_c}{2\pi^2} \frac{\T{k}^4}{\T{q}^4}  \bigg[ -\frac{\T{q}^2}{\T{k}^2(\T{k}-\T{q})^2_+} + \frac{1}{\T{k}^2} \big(1- R_{|\T{k}|}^{(2,1)}(\T{q}) \big) + (\T{q}\to -\T{q})\bigg]. \label{eq:KD-res-LLA} 
\end{eqnarray}
%\old{Below we are going to compute the characteristic function %based on the 
%approximation (\ref{eq:KD-res-LLA}).}   
We are finally in a position to derive and solve the evolution equations for $R^{(j)}_Q(\T{q})$.

\subsection{Linearized resummation equations}\label{sec:lin-res}

\subsubsection{Coordinate space}\label{sec:lin-res-coord}

% and $\alpha_{c_1}(\T{z}_1)\alpha_{c_2}(\T{z}_2)$ and projecting the indices $c_{1,2}$ on $\delta_{c_1c_2}$

Substituting the expansions (\ref{eq:Sbold-alpha-decomp}) and (\ref{eq:Vbold-alpha-decomp}) into Eqs.~(\ref{eq:eqn-SQ-quarks}) and (\ref{eq:eqn-VQ}) and collecting the terms in front of $\alpha^{c_1}(\T{z}_1)$ one finds the following  system of coupled linear coordinate-space evolution equations for the first-order resummation functions $R^{(1)}_Q(\T{x})$ and $r^{(1)}_Q(\T{x})$:
\begin{eqnarray}
     \bigg[ \frac{\partial}{\partial\ln Q^2} + a_s\beta_0 \bigg] R_Q^{(1)}(\T{x}) &=& -a_s \int\limits_0^1 d\xi\int\limits_0^{2\pi} \frac{d\phi}{2\pi} \bigg[ 2N_c p_{gg}(\xi) R_Q^{(1)}\big( \T{x}-\xi Q^{-1} \T{n}_\phi \big) \nonumber \\
     &&+ 4T_F^2n_F p_{qg}(\xi) r_Q^{(1)}\big( \T{x}-\xi Q^{-1} \T{n}_\phi \big) \bigg],  \label{eq:R1-Qevol-quarks} 
     \end{eqnarray}
\begin{eqnarray}
      \bigg[ \frac{\partial}{\partial \ln Q^2} + 3C_F a_s \bigg] r_Q^{(1)}(\T{x}) &=& - a_s \int\limits_0^1 d\xi \int\limits_0^{2\pi} \frac{d\phi}{2\pi} p_{gq}(\xi) \bigg[ N_c R^{(1)}_Q\big( \T{x}+(1-\xi)Q^{-1}\T{n}_\phi \big) \nonumber \\
      &&-\frac{1}{N_c} r^{(1)}_Q\big( \T{x}-\xi Q^{-1}\T{n}_\phi \big) \bigg]\,.  \label{eq:r1-Qevol-quarks}
\end{eqnarray}
%\old{where $\T{n}_\phi$ is a unit vector in the transverse plane %with azimuthal angle $\phi$.} 
This homogeneous system of linear integro-differential equations has to be solved with the initial conditions (\ref{eq:R1-init-cond}) and (\ref{eq:r1-init-cond}). These equations are the general-$\xi$ versions of the equations for the function $\alpha_Q(\T{x})$, written in Sec. IV and V of Ref.~\cite{Kovner:2023vsy}. An approximation  adopted in Ref.~\cite{Kovner:2023vsy} for these equations can be obtained  by setting $\xi=1/2$ in the resummation functions in the r.h.s. and then integrating the rest over $\xi$.  

Since only the second order resummation function $R^{(2,1)}_Q$ survived in the LLA resummed result for the kernel (\ref{eq:KD-res-LLA}), we are first of all interested in deriving the evolution equations for $R^{(2,1)}_Q$  and for its fundamental-representation cousin $r^{(2,1)}_Q$. This can be achieved by collecting terms in front of $\alpha^{c_1}(\T{z}_1)\alpha^{c_2}(\T{z}_2)$ in the expansion of Eqs.~(\ref{eq:eqn-SQ-quarks}) and (\ref{eq:eqn-VQ}) and projecting the indices $c_{1,2}$ on $\delta_{c_1c_2}$. We obtain
\begin{eqnarray}
     &&\left[ \frac{\partial}{\partial \ln Q^2} + a_s\beta_0 \right] R_{Q}^{(2,1)}(\T{x}_1,\T{x}_2) = -2a_s \int\limits_0^1 d\xi\, \int\limits_0^{2\pi} \frac{d\phi}{2\pi}  \label{eq:R2-Qevol-quarks} \\
    && \times \bigg\{2N_c p_{gg}(\xi)   R_Q^{(2,1)} \left( \T{x}_1 + \xi Q^{-1} \T{n}_\phi ,\T{x}_2 + \xi Q^{-1} \T{n}_\phi \right)  \nonumber \\
    &&  + 2T_Fn_F \frac{C_F}{N_c} p_{qg}(\xi)   r_Q^{(2,1)} \left( \T{x}_1 + \xi Q^{-1} \T{n}_\phi ,\T{x}_2 + \xi Q^{-1} \T{n}_\phi \right) \nonumber \\
  &&   -N_c p_{gg}(\xi) R^{(1)}_Q\left( \T{x}_1 - (1-\xi) Q^{-1} \T{n}_\phi  \right) R_Q^{(1)}\left( \T{x}_2 + \xi Q^{-1} \T{n}_\phi  \right) \nonumber \\
  && + \frac{T_F n_F}{N_c^2} p_{qg}(\xi)  r^{(1)}_Q\left( \T{x}_1 - (1-\xi) Q^{-1} \T{n}_\phi  \right) r_Q^{(1)}\left( \T{x}_2 + \xi Q^{-1} \T{n}_\phi  \right) \bigg\}, \nonumber 
  \end{eqnarray}

  \begin{eqnarray}
     && \bigg[ \frac{\partial}{\partial \ln Q^2} + 3C_Fa_s \bigg] r_Q^{(2,1)}(\T{x}_1,\T{x}_2) = - 2a_s \int\limits_0^1 d\xi \, p_{gq}(\xi) \int\limits_0^{2\pi} \frac{d\phi}{2\pi}  \label{eq:r2-Qevol-quarks} \\
     &&\times \bigg\{ C_F  r_Q^{(2,1)} \left( \T{x}_1 +\xi Q^{-1} \T{n}_\phi ,\T{x}_2  +\xi Q^{-1} \T{n}_\phi \right)  \nonumber \\ 
    && +N_c  R_Q^{(2,1)} \left( \T{x}_1 - (1-\xi)Q^{-1} \T{n}_\phi ,\T{x}_2 - (1-\xi)Q^{-1} \T{n}_\phi \right) \nonumber \\
    && -{\frac{N_c}{2} \big( R^{(1)}_Q(\T{x}_1-(1-\xi)Q^{-1}\T{n}_\phi ) r^{(1)}_Q(\T{x}_2+\xi Q^{-1}\T{n}_\phi) } \nonumber \\
    && {+ r^{(1)}_Q(\T{x}_1+\xi Q^{-1}\T{n}_\phi) R^{(1)}_Q(\T{x}_2-(1-\xi)Q^{-1}\T{n}_\phi ) \big)  } \bigg\} . \nonumber 
\end{eqnarray}
These are also linear (and coupled) equations, but unlike Eqs.~(\ref{eq:R1-Qevol-quarks}) and (\ref{eq:r1-Qevol-quarks}) they contain inhomogeneous terms proportional to the square of the first order resummation functions. The initial conditions for the second order functions are given by Eqs.~(\ref{eq:R2-init-cond}) and (\ref{eq:r2-init-cond})\footnote{ The equations obtained above have similar structure to the equations for the functions $\Delta_Q(\T{z})$ and $B_Q^{ab}(\T{z})$ introduced in Ref.~\cite{Armesto:2025dwd} for $N_c=2$, however their meaning is somewhat different. In  \cite{Armesto:2025dwd}  the authors expressed the dressed gluon scattering matrix locally in terms of a complete basis of matrix functions  at the point $\T{x}$. In other words the equations in Ref.~\cite{Armesto:2025dwd} are written for the resummation functions multiplied by $\alpha^{c_1}(\T{z}_1)\alpha^{c_2}(\T{z}_2)$ as in Eqs.~(\ref{eq:Sbold-alpha-decomp}) and (\ref{eq:Vbold-alpha-decomp}), and integrated over $\T{z}_1$ and $\T{z}_2$. Our equations are significantly different since they determine directly the second order resummation functions $R^{(2,1)}_Q$ which depend on two coordinates $\T{z}_{1}$ and $\T{z}_{2}$}.

\subsubsection{Second order resummation equations in momentum space}\label{sec:lin-res-mom}

%\old{Since only one momentum space second order resummation %function $R^{(2,1)}_Q(\T{p})$ contributes to the LLA resummed %BFKL kernel (\ref{eq:KD-res-LLA}), }
Our main goal now is to determine the momentum space solution for the second order resummation function $R^{(2,1)}_Q(\T{p})$, which enters the LLA resummed BFKL kernel (\ref{eq:KD-res-LLA}). Below we will show  that the LLA solution for $R^{(2,1)}_Q(p)$ is essentially independent of the first order resummation functions $R^{(1)}_Q$ and $r^{(1)}_Q$. Hence, to avoid overloading the main text of the paper, we deposit  the discussion of the evolution of the first order functions  into  Appendix~\ref{append:lin-res-mom-1}. To obtain the evolution equations for the second order resummation functions, we multiply Eqs.~(\ref{eq:R2-Qevol-quarks}) and (\ref{eq:r2-Qevol-quarks}) by $e^{i\T{p}(\T{x}_1-\T{x}_2)}$ and integrate over $\T{x}_{1,2}$. The result can be expressed in a matrix form:
\begin{equation}
    \frac{\partial}{\partial \ln Q^2} {\cal R}_Q^{(2,1)}(\T{p}) = -a_s \bigg[ \Pi_2 {\cal R}_Q^{(2,1)}(\T{p}) + J_0\left(\frac{|\T{p}|}{Q} \right) \rho_Q^{(1)}(\T{p}) \bigg],\label{eq:R2-eqn-mom}
\end{equation}
where the caligraphic letters here and below denote two-dimensional column vectors, such as the vector ${\cal R}^{(2,1)}_Q$ of second order functions, ${\cal R}^{(2,1)}_Q(\T{p}) = \left( \begin{array}{c}
         R^{(2,1)}_Q(\T{p}) \\
         r^{(2,1)}_Q(\T{p}) 
    \end{array} \right)$. The constant matrix $\Pi_2$ is %(setting $T_F=1/2$):
\begin{equation}
    \Pi_2 = \left( \begin{array}{cc}
       \beta_0-2N_c\frac{11}{3}  & \frac{4C_F n_F}{3N_c}  \\
        -3N_c & 0
    \end{array} \right) , \label{eq:Pi2-matr}
\end{equation}
and the source $\rho_Q^{(1)}$ is
\begin{equation}
     \rho_Q^{(1)}(\T{p}) = \left( \begin{array}{c}
         \frac{11}{3}N_c [R_Q^{(1)}(\T{p})]^2 + \frac{2n_F}{3N_c^2} [r_Q^{(1)}(\T{p})]^2  \\
         3N_c R_Q^{(1)}(\T{p}) r_Q^{(1)}(\T{p}) 
    \end{array}\right). \label{eq:inhomog-part-quarks}
\end{equation}

Solutions of Eq.~(\ref{eq:R2-eqn-mom}) are most conveniently represented in the basis of eigenvectors ${\cal E}_{\pm}$ of the matrix $\Pi_2$:
\begin{equation}
    {\cal R}_Q^{(2,1)}(\T{p}) = \bar{R}_{Q,+}^{(2,1)}(\T{p}) {\cal E}_+ +  \bar{R}_{Q,-}^{(2,1)}(\T{p}) {\cal E}_- , \label{eq:R2-sol-EB-exp}
\end{equation}
where 
\begin{equation}
{\cal E}_{\pm}=\frac{1}{\sqrt{9N_c^2+\lambda_{\pm}^2}} \left( \begin{array}{c}
         -\lambda_\pm \\
          3N_c
    \end{array} \right) \label{eq:Pi2-eigen-basis} 
\end{equation}
are the normalised eigenvectors of the matrix (\ref{eq:Pi2-matr}) corresponding to eigenvalues
\begin{eqnarray}
     \lambda_{\pm} = \frac{\beta_0}{2}-\frac{11}{3}N_c \pm \sqrt{\left( \frac{\beta_0}{2}-\frac{11}{3}N_c \right)^2 - 4C_Fn_F} \; .
\end{eqnarray}
Then the coefficients $\bar{R}_{Q,+}^{(2,1)}(\T{p})$ and $ \bar{R}_{Q,-}^{(2,1)}(\T{p})$ of the expansion (\ref{eq:R2-sol-EB-exp}) are
\begin{eqnarray}
    \bar{R}_{Q,\pm}^{(2,1)}(\T{p}) = \left( \frac{Q^2}{\Lambda^2} \right)^{-a_s\lambda_{\pm}} \bigg[ \bar{R}_{\Lambda,\pm}^{(2,1)}(\T{p})  - a_s \int\limits_{\Lambda^2}^{Q^2} \frac{dq^2}{q^2} \left( \frac{q^2}{\Lambda^2} \right)^{a_s\lambda_{\pm}} J_0\left( \frac{|\T{p}|}{q} \right) \bar{\rho}_{q,\pm}^{(1)}(\T{p}) \bigg] . \label{eq:sol-eigen}
\end{eqnarray}
This is the exact solution of Eq. (\ref{eq:R2-eqn-mom}), with $\bar{R}_{\Lambda,\pm}^{(2,1)}(\T{p})$ and $\bar{\rho}_{q,\pm}^{(1)}(\T{p})$ being respectively the components of the initial conditions at the scale $\Lambda$ and the source vector (\ref{eq:inhomog-part-quarks}) in the eigenbasis (\ref{eq:Pi2-eigen-basis}).  Eq.~(\ref{eq:sol-eigen}) can be greatly simplified in the LLA in $\ln (\T{p}^2/Q^2)$. Within this accuracy, the functions $\bar{\rho}_{q,\pm}^{(1)}(\T{p})$ in the solution can be replaced with the corresponding initial conditions: $\bar{\rho}^{(1)}_{\Lambda,\pm}$ for $q>|\T{p}|$. On the other hand, the contribution of the region $q \ll|\T{p}|$ to the integral in Eq.~(\ref{eq:sol-eigen}), where the first order resummmation functions in $\bar{\rho}^{(1)}_{q,\pm}(\T{p})$ might have had some non-trivial behaviour, is suppressed by the oscillations of the Bessel function\footnote{Moreover, the resummation functions $R^{(1)}_Q$ and $r^{(1)}_Q$ exhibit strong Sudakov suppression in this region, as explained in Appendix~\ref{append:lin-res-mom-1}.}. Therefore the LLA in $\ln(\T{p}^2/Q^2)$ solution of  Eq.~(\ref{eq:R2-eqn-mom}) in the eigenbasis reads:
\begin{eqnarray}
  \hspace{-5mm}\bar{R}_{Q,\pm}^{(2,1),\text{LLA}}(\T{p}) = \left( \frac{Q^2}{\Lambda^2} \right)^{-a_s\lambda_{\pm}} \bigg[ \bar{R}_{\Lambda,\pm}^{(2,1)}(\T{p})  - \frac{1}{\lambda_{\pm}} \left( \left( \frac{\max (Q^2, \T{p}^2)}{\Lambda^2} \right)^{a_s\lambda_{\pm}} -1  \right) \bar{\rho}_{\Lambda,\pm}^{(1)}(\T{p}) \bigg]. 
\end{eqnarray}

Returning to the original basis and assuming $\Lambda^2>\max(Q^2,\T{p}^2)$, one finds that all explicitly $\Lambda$-dependent terms cancel, which reflects  $\Lambda$-independence of the solution anticipated in Sec.~\ref{sec:resumm-general}. Hence the LLA solution for $R^{(2,1)}_Q$ takes the form:
\begin{equation}
    R^{(2,1),\text{LLA}}_{Q}(\T{p}) = \frac{3\beta_0-11N_c^3-11N_c-3N_c^2\lambda_+}{3N_c^2(\lambda_- - \lambda_+)} \left( \frac{\max (Q^2, \T{p}^2)}{Q^2} \right)^{a_s \lambda_- } + (\lambda_+ \leftrightarrow \lambda_-). \label{eq:R2-LLA}
\end{equation}
Taking into account that $\lambda_+ + \lambda_- = \beta_0-22N_c/3$ one can check that $R^{(2,1)\text{,LLA}}_{Q}(\T{p})=1$ for $|\T{p}|<Q$ as  expected.

%%%END OF sec:AC-res %%%%%%%%%%%%%%%%%%%%%%%%%%%%%%%%%%%%%%%%%%%%%%%%%%%%%%%%%%%%%%%%%%%%%%%

\section{{LLA} DGLAP effects in the resummed characteristic function}\label{sec:DGLAP}

The BFKL-equation (\ref{eq:BFKL-mom}) is an integro-differential equation, and it is convenient to work in the basis of eigenfunctions of its kernel $K_{\text{BFKL}}(\T{k},\T{q})$. Due to scale invariance of the LO kernel,  eigenfunctions of the latter are just powers of $\T{q}^2$: $(\T{q}^2)^\gamma e^{in\phi_{\T{q}}}$. Beyond LO, the scale invariance is violated by effects of the running of QCD coupling. The following  generalized eigenvalue equation~\cite{Fadin:1998py} defines the  BFKL characteristic function $\chi(n,\gamma)$ beyond LO in $\alpha_s$:
\begin{eqnarray}
    \int\limits_{\T{q}} K_{\text{BFKL}}(\T{k},\T{q}) (\T{q}^2)^\gamma e^{in\phi_{\T{q}}} = -\frac{\alpha_s(\T{k}^2) N_c}{\pi} \chi(n,\gamma) (\T{k}^2)^\gamma e^{in\phi_{\T{k}}}, \label{eq:gen-eval-def} 
\end{eqnarray}
where:
\begin{eqnarray}
 &&   \chi(n,\gamma) = \chi_0(n,\gamma) + \sum\limits_{m=1}^\infty \big(a_s(\T{k}^2) N_c)^m \chi_m(n,\gamma), \label{eq:chiS-exp} \\
 &&   \chi_0(n,\gamma)= 2\psi(1)-\psi\left(\gamma+\frac{|n|}{2}\right) - \psi\left( 1-\gamma + \frac{|n|}{2} \right)\,.
\end{eqnarray}
 Here $\chi_0$ is the LO BFKL characteristic function
with  $\psi(z)=\frac{d}{dz}\ln\Gamma(z)$ being the Euler's $\psi$-function. 

This form of the eigenvalue problem contains explicit dependence of $\alpha_s$ on $\T{k}^2$ in the r.h.s., besides $(\T{k}^2)^\gamma$. This prevents one from constructing  the BFKL Green's function as straightforwardly as in the LO case. However, for the fixed-coupling approximation adopted in the present paper, the explicit argument of $\alpha_s$ in Eq.~(\ref{eq:gen-eval-def}) does not play any role.

 %In the present paper, 
 To construct proper eigenfunctions of the higher order kernel, we do  not employ  the procedure  proposed 
 in~\cite{Chirilli:2013kca,Chirilli:2014dcb}. The latter procedure applies  order-by-order in $\alpha_s$ only, whereas our work focuses on all-order resummation\footnote{Still, the formal solution of the BKFL equation with running coupling can be obtained from the well-known fixed-coupling solution (see e.g. Eq. (\ref{eq:G-BFKL-symm}) in  Appendix~\ref{append:rap-schemes}) by treating all the factors of $\alpha_s$ in it as operators in $\gamma$-space~\cite{Altarelli:2001ji,Marzani:2007gk}: $\hat{\alpha}_s\big(\ln (\T{q}^2_+/\T{q}^2_-) \to -\partial_\gamma)\chi(\gamma,n)$, with $\chi(\gamma,n)$ being the generalized characteristic function of Eq.~(\ref{eq:gen-eval-def}).  }.
 
 In the present paper we explore the all-order anti-collinear structure of the eigenvalue $\chi(n,\gamma)$ defined by Eq.~(\ref{eq:gen-eval-def}) at fixed coupling. It can be directly compared with the NLO BFKL results~\cite{Fadin:1998py,Kotikov:2000pm}.
 %\old{used for theoretical studies in fixed-coupling approximation (Sec.~\ref{sec:gamma-gamma}) and in the above-mentioned operator approach. }

\subsection{{LL} anti-collinear resummation in the characteristic function}\label{sec:DGLAP-chi}

We now substitute the LLA solution for $R^{(2,1)}_Q$, Eq.~(\ref{eq:R2-LLA}) into the resummed BFKL kernel (\ref{eq:KD-res-LLA}). Upon acting with the resulting kernel on the eigenfunction, one finds  the contribution due to the resummation:
\begin{eqnarray}
   \int\limits_{\T{q}} (\T{q}^2)^{\gamma-2} e^{in\phi_{\T{q}}}  \frac{\T{k}^2}{\pi}   \bigg[ R^{(2,1)\text{,LLA}}_{|\T{k}|}(\T{q}) -1 \bigg] = \delta_{n,0} (\T{k}^2)^{\gamma} {\Delta\chi_+^{\text{(anti-coll, LLA)}}(\gamma)}, 
\end{eqnarray}
where
\begin{eqnarray}
  &&  \Delta\chi^{\text{(anti-coll, LLA)}}_+(\gamma) =  \frac{1}{\gamma-1} - \bigg( \frac{3\beta_0-11N_c^3-11N_c-3N_c^2\lambda_+}{3N_c^2(\lambda_- - \lambda_+)(\gamma-1 +a_s\lambda_-)} + (\lambda_+ \leftrightarrow \lambda_-) \bigg) \nonumber \\
  && = \frac{1}{\gamma-1} - \frac{ N_c(\gamma-1)-\frac{4}{3}a_sC_Fn_F }{N_c  \big[ (\gamma-1)^2-a_s (\gamma-1) \big( \frac{22}{3}N_c-\beta_0 \big) + 4a_s^2C_Fn_F \big]} . \label{eq:DeltaChi-coll}
\end{eqnarray}
  In order to obtain the  eigenvalue with  anti-collinear resummation, Eq.~(\ref{eq:DeltaChi-coll}) should be added to the LO BFKL eigenvalue:
\begin{equation}
    \chi^{\text{(anti-coll,LLA)}}_{+}(n,\gamma)= \chi_0(n,\gamma) + \delta_{n,0} \Delta\chi_+^{\text{(anti-coll,LLA)}}(\gamma)\,. \label{eq:chi-anti-coll}
\end{equation}
 The subscript $(+)$ in Eq.~(\ref{eq:chi-anti-coll}) emphasizes that this result was obtained in the  $(+)$-scheme of rapidity factorisation, i.e. the resummation parameter $Y$ in the BFKL Eq.~(\ref{eq:BFKL-mom}) has the meaning of $\ln P^+$, the same as in the original JIMWLK equation (\ref{eq:JIMWLK}). 

Eq.(\ref{eq:DeltaChi-coll}) is 
the central result of the present paper. In the following subsections  this result is compared with the existing literature.
\subsection{{Characteristic function at $\gamma\to 1$}}

\begin{figure}
    \centering
    \includegraphics[width=0.65\linewidth]{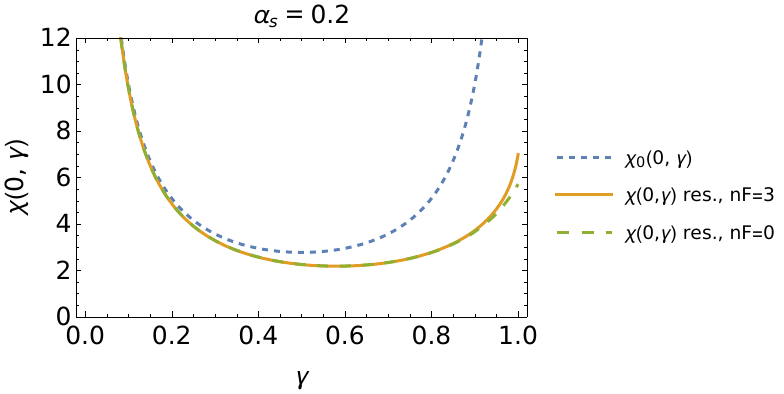}
    \caption{The plot of resummed characteristic function (\ref{eq:chi-anti-coll}) for $n=0$.} 
    \label{fig:chi-res}
\end{figure}
{The resummed characteristic function (\ref{eq:chi-anti-coll}) is plotted in Fig.~\ref{fig:chi-res} for $n=0$.} The anti-collinear resummation removes the pole at $\gamma=1$ (see Fig.~\ref{fig:chi-res}), which is present in $\chi_0(0,\gamma)$. The resummed characteristic function is finite at this point:
\begin{equation}
    \chi_+^{\text{(anti-coll, LLA)}}(0,\gamma\to 1) = \frac{\pi}{\alpha_s N_c} \left\{ \begin{array}{cc}  4/3 & \text{ for }n_F=3, \\ 12/11 & \text{ for }n_F=0. \end{array} \right. \label{eq:chi-AC_gamma-1}
\end{equation}
The ``collinear'' pole at $\gamma=0$ is not affected by the resummation, and in general the effect of the resummation on $\chi(0,\gamma)$ for $\gamma<1/2$ is insignificant.

It is interesting to compare the result (\ref{eq:chi-AC_gamma-1}) with the prediction of the DGLAP/BFKL duality %\old{proposal}
{approach} or Ref.~\cite{Altarelli:1999vw}. For $n_F=0$, it assumes validity of the following equation~\cite{Jaroszewicz:1982gr} to all orders in  the fixed-coupling:
\begin{equation}
    \frac{\alpha_s N_c}{\pi } \chi_{-}(0,\gamma_{gg}(\omega)) = \omega.\label{eq:BFKL-DGLAP-duality}
\end{equation}
Here $\gamma_{gg}(\omega)$ is the gluon DGLAP anomalous dimension while $\omega$ is the variable Mellin-conjugate to the light-cone momentum fraction. Due to the momentum sum rule for the gluon PDF, for $n_F=0$, the 
%\old{latter one}
{anomalous dimension}  satisfies $\gamma_{gg}(1)=0$ exactly. Therefore taking Eq.~(\ref{eq:BFKL-DGLAP-duality}) at $\omega=1$, Ref.~\cite{Altarelli:1999vw} concludes that for $n_F=0$:
\begin{equation}
    \chi_-(0,\gamma\to 0)  = \frac{\pi}{\alpha_s N_c}. \label{eq:chi-dual_gamma-1}
\end{equation}
The subscript $(-)$ of $\chi_- (n,\gamma)$ emphasizes that the characteristic function in Eq.~(\ref{eq:BFKL-DGLAP-duality}) is taken in the $(-)$ scheme of rapidity factorisation. That is the resummation parameter $Y$ in the BKFL equation is identified with the logarithm of the large light-cone momentum component of the target ($\ln P^-$). The latter scheme is suitable  for  description of the usual Bjorken limit of DIS and is used e.g. in the High-Energy Factorisation formalism of Refs.~\cite{Catani:1992rn,Catani:1993rn,Catani:1990xk,Catani:1990eg,Catani:1994sq,Collins:1991ty}. As is demonstrated in  Appendix~\ref{append:rap-schemes}, due  to the projectile-target symmetry, the following identity holds in the fixed-coupling approximation\footnote{Running coupling effects violate the symmetry $\gamma\leftrightarrow 1-\gamma$ in the generalized eigenvalue defined by Eq.~(\ref{eq:gen-eval-def}).}:
\begin{equation}
    \chi_+(0,\gamma) = \chi_-(0,1-\gamma).\label{eq:proj-target-sym}
\end{equation}
Consequently, we can compare the results (\ref{eq:chi-AC_gamma-1}) and (\ref{eq:chi-dual_gamma-1}): our result (\ref{eq:chi-AC_gamma-1}) agrees with (\ref{eq:chi-dual_gamma-1}) up to the coefficient $12/11$. Although numerically the discrepancy is small, the factor $12/11$ is clearly nontrivial and is not an artifact of any numerics. It indicates that the systematic anti-collinear resummation of Sec.~\ref{sec:resumm-general} contains physics beyond the simple duality assumption of Eq.~(\ref{eq:BFKL-DGLAP-duality}). 

We note also that there are results in the literature for the BFKL characteristic function which include resummation due to imposition of kinematical constraint, e.g. Ref.~\cite{Deak:2019wms}. This resummation contains very different physics and is completely unrelated to DGLAP resummation. In addition, the kinematical constraint approach of Ref.~\cite{Deak:2019wms} resums the leading anti-collinear poles in $(-)$-scheme, corresponding to the {\it collinear resummation in $(+)$-scheme}, which is the opposite of the regime considered in the present paper.

Indeed our resummation result (\ref{eq:chi-AC_gamma-1}) (and also (\ref{eq:chi-dual_gamma-1})) is very distinct from the one discussed in Ref.~\cite{Deak:2019wms}. Taking  $\gamma\to 0$ and $\gamma\to 1$ limits of Eqs. (67) and (68) in ~\cite{Deak:2019wms},  one finds: 
\begin{eqnarray}
&&\chi_-^{\text{(kin. constr.)}}(0,\gamma\to 0) = \frac{1}{\gamma} + O(\gamma) \Rightarrow \chi_+^{\text{(kin. constr.)}}(0,\gamma\to 1) = \frac{1}{1-\gamma} + O(1-\gamma) ,\nonumber \\
&&\chi_-^{\text{(kin. constr.)}}(0,\gamma\to 1) = \sqrt{\frac{\pi}{\alpha_s N_c}}=\chi^{\text{(kin. constr)}}_+(0,\gamma\to 0), \label{eq:KC-limit} 
\end{eqnarray}
 where  Eq.~(\ref{eq:proj-target-sym}) was used to relate the limits of $\chi_+(0,\gamma)$. 
 
  Note the distinct behaviour of the characteristic function $\sim 1/\sqrt{\alpha_s}$ in eqn.~(\ref{eq:KC-limit}), as opposed to  $\sim 1/\alpha_s$ expected from the {\it collinear} DGLAP resummation. The origin of this difference is clear: the kinematical constraint, as well as the classic ``$\gamma$-shift'' approach of Refs.~\cite{Salam98,SabioVera:05,SabioVera:07}, resum in the characteristic function only the leading anti-collinear poles of the type $\sim \alpha_s^m/(1-\gamma)^{2m+1}$ in $(-)$-scheme. These poles are completely determined by the transformations between  the rapidity factorisation schemes and have nothing to do with the DGLAP cascade. This is in sharp contrast to our DGLAP-based resummation. 
 
 In the next two subsections we will perform further tests of our result (\ref{eq:chi-anti-coll}).

\subsection{Comparison with the NLO BFKL eigenvalue}\label{sec:comp-NLO-BFKL}

We  have already mentioned the $(+)$ and $(-)$-schemes of rapidity factorization. The original result for the NLO BFKL kernel and eigenvalue was derived~\cite{Fadin:1998py,NLOCiafaloni1,NLOCiafaloni2,Kotikov:2000pm} in the so-called {\it symmetric scheme}, where the resummation parameter is the rapidity difference $Y$ between two produced final-state objects, e.g. Mueller-Navelet jets~\cite{MN:87}. The relations between the symmetric and $(\pm)$-schemes are explained in  Appendix \ref{append:rap-schemes}. %As it is shown there, \old{given the $\alpha_s$-expansion of }
The generalized characteristic function defined by  Eq.~(\ref{eq:gen-eval-def}), in the symmetric scheme at NLO reads:
\begin{equation}
    \chi_S(n,\gamma) = \chi_0(n,\gamma) + a_s(\T{k}^2)N_c \chi_{S1}(n,\gamma) + O(a_s^2).
\end{equation}
  The characteristic function in the $(\pm)$-scheme can be obtained as  solution of Eq.~(\ref{eq:rap-scheme-transf}). Up to NLO it reads:
 \begin{equation}
     \chi_{\pm}(n,\gamma) = \chi_0(n,\gamma) + a_s N_c \big( \chi_{S1}(n,\gamma) \pm 2\chi'_0(n,\gamma) \chi_0(n,\gamma) \big) + O(\alpha_s^2),   \label{eq:scheme-transf-NLO}
 \end{equation}
 where $\chi_0'(n,\gamma)=\partial\chi_0(n,\gamma)/\partial \gamma$. Our interest is in the behaviour of the last term in (\ref{eq:scheme-transf-NLO}) in the vicinity of $\gamma=-n/2$ (collinear pole) and $\gamma=1+n/2$ (anti-collinear pole):
 \begin{equation}
        2\chi'_0(n,\gamma) \chi_0(n,\gamma) = -\frac{2}{\left(\gamma+\frac{n}{2}\right)^3} + \frac{2}{\left( 1-\gamma + \frac{n}{2} \right)^3} + O\big( (\gamma+\frac{n}{2})^{0}, (1-\gamma+\frac{n}{2})^{0}   \big).
 \end{equation}

NLO corrections to the BFKL eigenvalue in the symmetric scheme, $\chi_{S1}(\gamma,n)$, 
%\old{for general conformal spin $n$ }
was computed in Ref.~\cite{Kotikov:2000pm} following  computations of the NLO correction to the kernel in Refs.~\cite{Fadin:1998py,NLOCiafaloni1,NLOCiafaloni2}. It has the following collinear and anti-collinear structure:
\begin{eqnarray}
    \chi_{S1}(n,\gamma)&=& -\frac{2}{\left( \gamma+n/2 \right)^3} - \frac{\delta_{n,0}}{\gamma^2} \left( \frac{11}{3}+\frac{2n_F}{3N_c^3} \right) + 2\frac{\psi(1)-\psi(n+1)}{\big( \gamma+\frac{n}{2} \big)^2}   \label{eq:NLO-chi-exp-poles} \\
    && -\frac{2}{\left(1-\gamma + \frac{n}{2} \right)^3} -\underbracket{\frac{\delta_{n,0}}{(1-\gamma)^2} \left( \frac{11}{3}+\frac{2n_F}{3N_c^3} \right)} + \frac{-\beta_0/N_c + 2\big( \psi(1)-\psi(n+1) \big) }{\big( 1-\gamma+\frac{n}{2} \big)^2} \nonumber \\
    && + O\big( (\gamma+\frac{n}{2})^{-1}, (1-\gamma+\frac{n}{2})^{-1}   \big).\nonumber 
\end{eqnarray}
The leading anti-collinear pole of this expression $\propto -2/(1-\gamma+\frac{n}{2})^3$, cancels in the $(+)$-scheme due to Eq.~(\ref{eq:scheme-transf-NLO}). This agrees with the $\alpha_s$-expansion of our resummed result (\ref{eq:chi-anti-coll}):
\begin{eqnarray}
     \chi^{\text{(anti-coll., LLA)}}_{+}(n,\gamma\rightarrow 1) &&= \chi_0(n,\gamma) - \underbracket{a_s N_c \delta_{n,0} \frac{11+2n_F/N_c^3}{3(1-\gamma)^2}} \label{eq:chi-anti-coll-NLO} \\
    && +\frac{(a_s N_c)^2 \delta_{n,0}}{(1-\gamma)^3} \bigg[ \frac{121}{9} +\frac{4n_F}{9N_c} +\frac{40n_F}{9N_c^3} + \frac{4n_F^2}{9N_c^4} \bigg] + O(\alpha_s^3), \nonumber
\end{eqnarray}
which has only the subleading pole $1/(1-\gamma)^2$ at NLO (the underlined terms in both expressions): both the $n_F$ and $N_c$-dependencies of the subleading pole proportional to $\delta_{n,0}$  agree between Eqs.~(\ref{eq:NLO-chi-exp-poles}) and (\ref{eq:chi-anti-coll-NLO}). %Note that the subleading collinear pole $\sim 1/\gamma^2$ in (\ref{eq:NLO-chi-exp-poles}) has the same coefficient as  the anti-collinear pole.

In Eq. (\ref{eq:chi-anti-coll-NLO}) we have expanded our result up to $\alpha_s^2$ order. This is our prediction for the NNLO coefficient in QCD. This prediction could be confronted with partial information about the BFKL at NNLO available in the literature for ${\cal N}=4$ supersymmetric Yang-Mills theory (${\cal N}=4$ SYM). In ${\cal N}=4$ SYM, the coefficient of the  $\sim a_s/(1-\gamma)^2$ pole vanishes~\cite{Kotikov:2002ab}.  At NNLO~\cite{Gromov:2015vua,Velizhanin:2015xsa,Caron-Huot:2016tzz}
the coefficient of the  anti-collinear pole 
$a_s^2/(1-\gamma)^3$ is non-zero and equals
$-(4N_c)^2\zeta(2)$ ~\cite{Deak:2019wms}. %\old{$\zeta(2)$ reflects the maximal %transcendentality principle put forward %in~\cite{Kotikov:2002ab}.} 
The coefficient of  $a_s^2/(1-\gamma)^3$ in our result 
(\ref{eq:chi-anti-coll-NLO}) does not have $\zeta(2)$, which  violates the maximal transcendentality principle {put forward in~\cite{Kotikov:2002ab}}. Although the relative numerical contribution of such maximally-transcendental terms in QCD is always sub-dominant in comparison to the terms of lower transcendental weight, it is still interesting to investigate their physical origin, which is a subject of our ongoing work. 

In Eq.~(\ref{eq:NLO-chi-exp-poles}),  the term $\propto -\beta_0/(1-\gamma)^2$  contribute to the $\gamma\to 1$ asymptotics at $n=0$. Such a term is absent in our result. We believe the reason is that this term is related to the running-coupling corrections which are not included in the present study. 
We substantiate this statement by presenting an observation inspired by  Ref.~\cite{Fadin:1998py}.
 The following NLO term in the characteristic function 
\begin{equation}
    -a_s \beta_0 \chi_0'(n,\gamma\rightarrow 1+n/2) = \frac{-a_s \beta_0}{(1-\gamma+n/2)^2} + O((1-\gamma+n/2)^{-1}) 
\end{equation}
is generated via  modification of the LO kernel:
\begin{equation}
 K_{\text{BFKL}}^{\text{(LO)}}(\T{k},\T{q})\to     \frac{\alpha_s(\T{q})}{\alpha_s(\T{k})} K_{\text{BFKL}}^{\text{(LO)}}(\T{k},\T{q}).
\end{equation}
This can be understood as a particular scale choice for the strong coupling factor in Eq.~(\ref{eq:rhoT-alpha-rel}). 
 
 Finally, the terms $\propto \psi(1)-\psi(n+1)$ in Eq.~(\ref{eq:NLO-chi-exp-poles}) contribute only  for $n\neq 0$, i.e. they probe the angular dependence of the collinear splittings, and they are not 
 %\old{captured} 
 addressed by the resummation of Sec.~\ref{sec:resumm-general}.
% \old{\new{in its present form. }}
 We note however, that the familiar DGLAP splitting functions in Eqns.~(\ref{eq:eqn-SQ-quarks}) and (\ref{eq:eqn-VQ}) appear only upon angular averaging, which in principle can be undone. This must introduce  azimuthal angle dependence for splittings that produce gluons close to the position of the source. This dependence will not change the $n=0$ results, but will affect the characteristic function at $n\neq 0$. We plan to study this question in future.

\subsection{Target-Bjorken limit of the $\gamma^*\gamma^*$-scattering}\label{sec:gamma-gamma}

\begin{figure}
    \centering
   \begin{tabular}{cc}
    \includegraphics[width=0.5\linewidth]{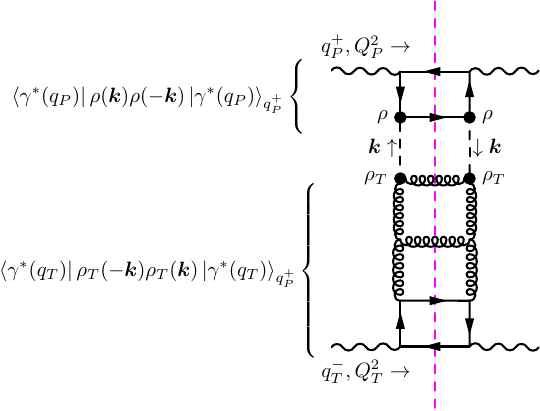} & \includegraphics[width=0.4\linewidth]{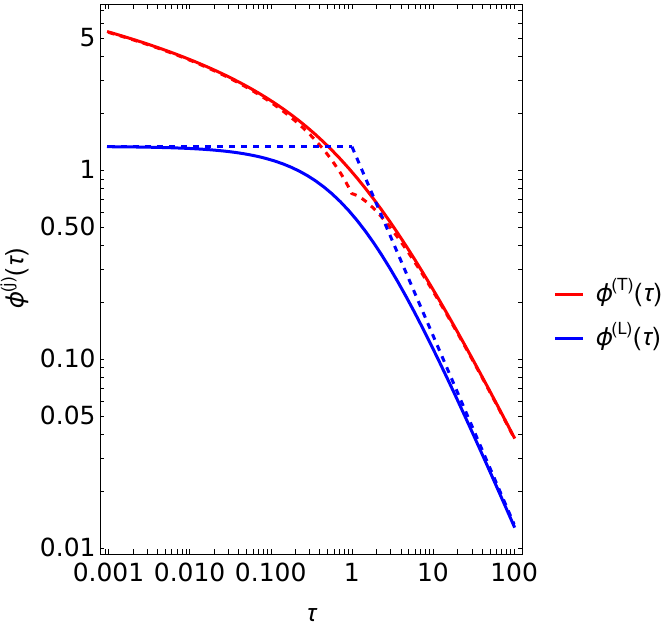} \\
    (a) & (b) 
    \end{tabular}
    \caption{ Panel (a) -- factorisation for the imaginary part of the $\gamma^*\gamma^*$-scattering amplitude in the Regge limit. Dashed lines denote Reggeised gluon ($\alpha^a(\T{x})$) exchanges; Panel (b) -- plots of the photon impact factor ``wave functions'' of Eq. (\ref{eq:gamma-WF-def}) as a function of $\tau=\T{k}^2/Q^2$, solid lines -- exact expressions (\ref{eq:phiL-exact}) and (\ref{eq:phiT-exact}), dashed lines -- approximate expressions (\ref{eq:appr-phi}).}
    \label{fig:gamma-gamma}
\end{figure}

The high-energy behaviour of the scattering amplitude of two off-shell photons is a classic problem in High-Energy QCD~\cite{Balitsky:1979ns,BFKL3,Bialas:2001ks,Balitsky:2010ze,Balitsky:2012bs,Chirilli:2014dcb,Ivanov:2014hpa,Colferai:2023dcf}. For the present study it is useful as the controllable playground to perform all-order in $\alpha_s$ tests of our resummation.
%\old{proposal}. 
In the leading power in the $\gamma^*\gamma^*$ scattering energy\footnote{We use $k^{\pm}=k^0 \pm k^3$.} $s=q_P^+q_T^-$, the imaginary part of the forward-scattering amplitude of the process:
\begin{equation}
    \gamma^*(q_P) + \gamma^*(q_T) \to \gamma^*(q_P) + \gamma^*(q_T),
\end{equation}
with $q_{T,P}^2=-Q_{T,P}^2$, $q_P^-=-Q_P^2/q_P^+$, $q_T^+=-Q_T^2/q_T^-$, can be represented as (Fig.~\ref{fig:gamma-gamma}(a)):
\begin{eqnarray}
      {\rm Im}\,{\cal A} &=& g^4 \int \frac{d^2 \T{k}}{2 (2\pi)^4} \langle \gamma^*(q_P) |\rho^a(\T{k}) \rho^b(-\T{k})| \gamma^*(q_P) \rangle_{0} \nonumber \\
      && \times\left(\frac{1}{2\T{k}^2} \right)^2 \langle \gamma^*(q_T) |\rho^a_T(-\T{k}) \rho^b_T(\T{k})| \gamma^*(q_T)\rangle_{\ln q_P^+/q_T^+}. \label{eq:A-gamma-gamma-0}
\end{eqnarray}
In  two-Reggeon exchange (BFKL) approximation,   the projectile charge density operators in LCPT are 
\begin{equation}
    \rho^a(\T{x}) = \int\limits_{k^+>\Lambda^+} a_i^{\dagger\,b}(k^+,\T{x}) T^a_{bc} a_i^c(k^+,\T{x}),\label{eq:rho-P-def}
\end{equation}
 with $a^a_i(k^+,\T{x})$ and $a_i^{\dagger\,a}(k^+,\T{x})$ being the annihilation and creation operators of  gluons with light-cone momentum component $k^+$ and transverse position $\T{x}$. The factor $(1/(2\T{k}^2))^2$ can be understood either as a product of propagators\footnote{The factors of 2 are due to the $n^+n^-=2$ normalisation of Sudakov basis vectors, used in the present study.} of the Reggeized gluons in  Fig.~\ref{fig:gamma-gamma}(a) or as  Fourier transform of the $\nabla^{-2}$-factors in Eq.~(\ref{eq:rhoT-alpha-rel}) relating $\rho_T$ and $\alpha$.  The details of the computation of the $\langle \rho \rho \rangle$-correlator over an off-shell photon state are discussed in  Appendix~\ref{append:gamma-IF}. The correlators can be represented as follows:
\begin{equation}
     {\cal P}_j^{\mu\nu} g^2 \langle \gamma_\mu^*(q) |\rho^a(\T{k}) \rho^b(-\T{k})| \gamma_\nu^*(q) \rangle_Y = 8 e_q^2 \delta^{ab} \alpha \alpha_s \frac{\sqrt{s}\T{k}^2}{Q^2} 
     h_j(x_B) \phi_Y^{(j)} (\T{k}^2/Q^2), \label{eq:gamma-WF-def}
\end{equation}
where $e_q$ is the quark charge in units of positron charge and the index $j=T,L$ denotes the type of structure functions on which the photon polarisation indices $\mu$ and $\nu$ can be projected,  $h_j(x_B)=\delta_{jT}+x_B\delta_{jL}$ with $x_B=Q^2/s$ (see Appendix~\ref{append:gamma-IF} for details). The amplitude (\ref{eq:A-gamma-gamma-0}) can be expressed in terms of ``wave functions'' $\phi^{(j)}_Y(\tau)$ as follows:
\begin{eqnarray}
  && \hspace{-8mm}  {\rm Im}\,{\cal A}_{i,j}(s,Q_P,Q_T) = \frac{N_c^2-1}{2} \left( \frac{4e_q^2 \alpha \alpha_s}{(2\pi)^2} \right)^2 \frac{\pi s}{Q_T^2} h_i(x_B^{(P)}) h_j(x_B^{(T)}) F_{i,j}(x_B^{(T)},Q_P,Q_T), \\
  &&  F_{i,j}(x_B^{(T)},Q_P,Q_T)=\int \frac{d^2 \T{k}}{\pi Q_P^2} \phi_0^{(i)}(\T{k}^2/Q_P^2) \,  \phi^{(j)}_{\ln q_P^+/q_T^+}(\T{k}^2/Q_T^2), \label{eq:SFs-WFs}
\end{eqnarray}
where we have introduced the Bjorken variable of the target (projectile): $x_B^{(T)}=Q_T^2/s$ ($x_B^{(P)}=Q_P^2/s$).  The target wave function  is to be evolved from the scale $q_T^+$ up to the scale $q_P^+$ with   the resummed  BFKL equation (\ref{eq:BFKL-mom}),  to resum large energy logarithms. The {\it target-Bjorken limit} we proceed to consider, consists of first taking $x_B^{(T)}\ll 1$, which justifies the approximation (\ref{eq:A-gamma-gamma-0}) for the amplitude, and then $Q_T^2\gg Q_P^2$. 

The exact LO in $\alpha_s$ expression for the wave-functon $\phi_0^{(j)}(\tau)$ is given in Eqs. (\ref{eq:phiL-exact}) and (\ref{eq:phiT-exact}) of Appendix~\ref{append:gamma-IF}. We will however use a convenient approximation:  
\begin{equation}
    \tilde{\phi}_0^{(j)}(\tau)= \theta(1-\tau) \big( f^{(j)}_1 + f^{(j)}_{2}\ln \frac{1}{\tau} \big) +  \frac{\theta(\tau-1)}{\tau}\big( f^{(j)}_{1} + f^{(j)}_{2} \ln\tau \big),\label{eq:appr-phi}
\end{equation}
with $f_1^{(T)}=3/4, f_1^{(L)}=4/3$, $f_2^{(T)}=2/3$ and $f_2^{(L)}=0$, which incorporates its asymptotic behaviour, see Fig.~\ref{fig:gamma-gamma}(b). This approximate formula admits the following simple Mellin-space representation:
\begin{eqnarray}
    && \tilde{\phi}^{(j)}_0(\tau) = \int\limits_{1/2-i\infty}^{1/2+i\infty} \frac{d\gamma}{2\pi i} \tau^{\gamma-1} \tilde{\phi}^{(j)}_0(\gamma), \\
    && \tilde{\phi}_0^{(j)}(\gamma) = f^{(j)}_1 \left(\frac{1}{\gamma} + \frac{1}{1-\gamma}  \right) + f^{(j)}_{2} \left( \frac{1}{\gamma^2} + \frac{1}{(\gamma-1)^2} \right). \label{eq:WFs-gamma}
\end{eqnarray}
Using the fact that the $\langle \rho \rho \rangle_{\ln \Lambda_+}$-correlator satisfies the BFKL equation (\ref{eq:BFKL-mom}), it is possible to write-down an expression for the evolved wave function of the target 
in the fixed-coupling approximation:
\begin{eqnarray}
     \phi^{(j)}_{\ln q^+_T\to \ln q^+_P} (\T{k}^2/Q_T^2) &=& \int\frac{d\gamma}{2\pi i} \left( \frac{\T{k}^2}{Q_T^2}\right)^{\gamma-1} \phi^{(j)}_0(\gamma) \exp \bigg[\hat{\alpha}_s \chi_+(0,\gamma) \ln\frac{q_P^+}{|q_T^+|} \bigg] \nonumber \\
    &=& \int \frac{d\omega}{2\pi i} \int \frac{d\gamma}{2\pi i} \left( {x_B^{(T)}} \right)^{-\omega} \left( \frac{\T{k}^2}{Q_T^2}\right)^{\gamma-1} \frac{\phi^{(j)}_0(\gamma)}{\omega - \hat{\alpha}_s \chi_+(0,\gamma)}, \label{eq:phi-target-evolved}
\end{eqnarray}
where $q_T^+=-Q^2/q_T^-\ll q_P^+$. The contour in the $\omega$-plane goes parallel and in the same direction as the  imaginary axis, to the right of all singularities.  $\hat{\alpha}_s = \alpha_s N_c/\pi$.

Substituting Eq.~(\ref{eq:phi-target-evolved}) for the target wavefunction and Eq.~(\ref{eq:appr-phi}) for the projectile wavefunction into Eq.~(\ref{eq:SFs-WFs}), one obtains the following expression for the structure functions:
\begin{equation}
    F_{i,j}(x_B^{(T)},Q_P,Q_T) =  \int \frac{d\omega}{2\pi i} \int\limits_{1/2-i\infty}^{1/2+i\infty} \frac{d\gamma}{2\pi i} \left( {x_B^{(T)}} \right)^{-\omega} \left( \frac{Q_P^2}{Q_T^2}\right)^{\gamma-1} \frac{\tilde{\phi}^{(i)}_0(1-\gamma) \tilde{\phi}_0^{(j)}(\gamma) }{ \omega - \hat{\alpha}_s \chi_+(0,\gamma) }. \label{eq:F2-target}
\end{equation}

The behaviour of the structure functions at $Q_T\gg Q_P$ is determined by the singularities in the $\gamma$-plane located to the right of the contour. The product of wave functions (\ref{eq:WFs-gamma}) in the numerator of Eq.~(\ref{eq:F2-target}) has the pole at $\gamma=1$. The simple pole {from the impact factors} corresponds to the Bjorken-scaling behaviour of the structure function while the second order pole introduces the scaling violating correction $\sim \ln Q_T^2/Q_P^2$. 

In principle, our formalism can be generalized to include the anti-collinear resummation in the target impact factor, in the spirit of Ref.~\cite{Colferai:2023dcf}. It is likely that the target impact factor should contain a dipole made of two dressed Wilson lines, $\tr\big[\Bbb{V}_Q(\T{x}) \Bbb{V}_Q^\dagger(\T{y})\big]$. We do not pursue this direction any further in this paper and instead stick to the discussion of resummation in the evolution.

The denominator in Eq.~(\ref{eq:F2-target}) has a pole at $\gamma=1-\gamma_T(\omega)$ with the anomalous dimension $\gamma_T(\omega)=O(\alpha_s)$. The position of this pole is determined by the equation similar to the  duality equation (\ref{eq:BFKL-DGLAP-duality}):
\begin{equation}
    \hat{\alpha}_s \chi_+(0,1-\gamma_T(\omega)) = \omega. \label{eqn:duality-target}
\end{equation}
This pole corresponds to the {energy-dependent} all-order-in-$\alpha_s$ violation of the Bjorken scaling. Substituting $\chi_+^{\text{(anti-coll.,LLA)}}$ of Eq.~(\ref{eq:chi-anti-coll}) as $\chi_+$ and solving  Eq.~(\ref{eqn:duality-target}) order-by-order in $\alpha_s$ one obtains:
\begin{eqnarray}
    \gamma_T(\omega)&=& 4N_c a_s \left( \frac{1}{\omega} - \frac{11}{12} - \frac{n_F}{6N_c^3} \right) + a_s^2\cdot 0 + a_s^3 \cdot 0  \nonumber \\
    &&+ (4N_c a_s)^4 \: 2\zeta(3) \left(\frac{1}{\omega^4} - \frac{11}{6\omega^3} - \frac{n_F}{3N_c^3 \omega^3} +O(\omega^{-2})  \right) + O(a_s^5). \label{eq:gamma-star-exp}
\end{eqnarray}
The LO in $\alpha_s$ term of this expansion agrees with the corresponding term of the expansion of the anomalous dimension in  Eq.~(24) of Ref.~\cite{Fadin:1998py}. It can also be compared with the low-$\omega$ limit of the largest eiganvalue of the anomalous-dimension matrix:
 \begin{equation}
     \Gamma= \left( \begin{array}{cc}
        \gamma_{gg}(\omega)  & 2n_F \gamma_{gq}(\omega)  \\
         \gamma_{qg}(\omega) &  \gamma_{qq}(\omega)
     \end{array} \right) = \left( \begin{array}{cc}
         \frac{2N_c}{\omega} + \left( \frac{\beta_0}{2}-\frac{11}{3}N_c \right) & \frac{4 n_F C_F}{\omega} - 3n_FC_F \\
         \frac{1}{3} & 0 
     \end{array} \right) + O(\omega), \label{eq:anom-dim-matr}
 \end{equation}
 which controls the coupled DGLAP evolution  of gluon and $n_F$ massless quarks PDFs, in  fixed-coupling approximation. Assuming that all $f_q=f_{\bar{q}}$ are the same:
 \begin{equation}
     \frac{\partial}{\partial \ln\mu^2} \left( \begin{array}{c}
          f_g(\omega,\mu^2) \\
          f_q(\omega,\mu^2) 
     \end{array} \right) = \frac{\alpha_s}{2\pi} \Gamma \left( \begin{array}{c}
          f_g(\omega,\mu^2) \\
          f_q(\omega,\mu^2) 
     \end{array} \right).
 \end{equation}
 The largest eigenvalue of the matrix (\ref{eq:anom-dim-matr}) is equal to
 \begin{equation}
     \Gamma_+ = \frac{2N_c}{\omega} + \frac{\beta_0}{2} - \frac{11}{3} N_c + \frac{2n_F}{3N_c} C_F + O(\omega).
 \end{equation}
 Both $\propto 1/\omega$ and $O(\omega^0)$ terms agree with the corresponding terms in the LO in $\alpha_s$ part of  Eq.~(\ref{eq:gamma-star-exp}). 

 The familiar pattern~\cite{Jaroszewicz:1982gr} of ``accidental'' zeroes at NLO and NNLO in the LLA small-$x$ resummed DGLAP anomalous dimension has also emerged in  Eq. (\ref{eq:gamma-star-exp}). Yet,  the full NLO DGLAP anomalous dimension has a term $\propto a_s^2 n_F/\omega$ (NLL low-$x$ term, see again  Eq.~(24) in Ref.~\cite{Fadin:1998py}),  which is absent in our Eq.~(\ref{eq:gamma-star-exp}).  This suggests that the higher order NLL terms $\propto \alpha_s^n/\omega^{n-1}$ in our result are probably incomplete as well. 
 
  To conclude, the anti-collinear resummation of  Sec.~\ref{sec:resumm-general} indeed correctly implements the LO DGLAP information into the JIMWLK and BFKL kernels, leading to the correct scaling-violation behaviour of the amplitude in the target-Bjorken limit, at least as long as the fixed-coupling approximation is considered.

%%%END OF sec:DGLAP %%%%%%%%%%%%%%%%%%%%%%%%%%%%%%%%%%%%%%%%%%%%%%%%%%%%%%%%%%%%%%%%%%%%%%%%

\subsection{Comparison with the ``all-poles'' resummation prescription}

 As has been first understood in Ref.~\cite{Salam98}, the relation between $(+/-)$ and symmetric rapidity-factorisation schemes can be used to resum the leading (anti-)collinear poles $\propto \alpha_s^m/\gamma^{2m+1}$ ($\alpha_s^m/(1-\gamma)^{2m+1}$) (see also Appendix~\ref{append:rap-schemes}).  Later this resummation procedure has been generalized to subleading poles in Refs.~\cite{SabioVera:05,SabioVera:07}, which  includes the $\alpha_s^m/(1-\gamma)^{m+1}$-poles resummed in the present paper, Eq.~(\ref{eq:chi-anti-coll-NLO}). This ``all-poles'' resummation procedure, albeit not proven to provide controlled resummation, has become quite popular in phenomenology~\cite{Szymanowski:10,Szymanowski:13,Szymanowski:14,Hentschinski:2013id,Hentschinski:2012kr,Chernyshev:2025kfe}. So it is useful to also compare our results with those of the all-poles resummation. 

Following Refs.~\cite{SabioVera:05,SabioVera:07}, the resummed  characteristic function in the fixed-coupling approximation, in the symmetric scheme can be written as follows:
\begin{eqnarray}
 \chi_S^{\text{(all-poles)}}(n, \gamma) & = &
\frac{1}{\hat{\alpha}_s}\sum_{m = 0}^\infty \bigg[
-\frac{\hat{\alpha}_s(\kappa_{0,1}+\hat{\alpha}_s\kappa_{1,1})}{m+1}+  \kappa_{1, 2} \, \hat{\alpha}_s  - d_m(n,\gamma)
\label{eq:chi-AP} \\ & + &
\sqrt{2 \hat{\alpha}_s \left( \kappa_{0, 1} + \kappa_{1, 1} \, \hat{\alpha}_s \right)
+ \left( \kappa_{1, 2} \, \hat{\alpha}_s - d_m(n,\gamma) \right)^2}  \bigg]
  +(\gamma \to 1 - \gamma),
\nonumber
\end{eqnarray}
where $d_m(n,\gamma) = m+\gamma+n/2$. Anti-collinear poles are contained in the $m=0$ term.  In the equation above  $\kappa_{0,1}=1$. The coefficients $\kappa_{1,2}$, $\kappa_{1,1}$ can be read off  the expansion of the NLO correction to the characteristic function (\ref{eq:NLO-chi-exp-poles})
at the collinear pole:
\begin{equation}
    \chi_{S1}(n=0,\gamma\rightarrow 0) = -\frac{2}{\gamma^3} + \frac{4\kappa_{1,2}}{\gamma^2} + \frac{4\kappa_{1,1}}{\gamma}+O(\gamma^0),
\end{equation}
where for simplicity we consider the case of $n=0$.

When determining the coefficients $\kappa_{1,1}$ and $\kappa_{1,2}$ one must be careful not to include the running-coupling effects, so one considers only the part of the NLO BFKL eigenvalue~\cite{Kotikov:2000pm} which is not proportional to $\beta_0$ and contributes to collinear and anti-collinear poles. 
{The coefficients in front of the poles of both kinds are the same in this $\beta_0$-independent part of the eigenvalue. We quote only the collinear expansion}:
\begin{eqnarray}
  &&  \Delta\chi_{S1}^{\text{(non-$\beta_0$)}}(0,\gamma) = \biggl(\frac{67}{9}-2\zeta (2)-\frac{10}{9}\frac{n_{f}}{N_{c}}\biggr) %
\chi_0 (0,\gamma) \nonumber \\% -2\Phi (0,\gamma )-2\Phi(0,1-\gamma )  \nonumber \\
  && -\frac{\pi ^{2}\cos (\pi \gamma )}{\sin ^{2}(\pi \gamma
)(1-2\gamma )}\biggl(3+\biggl(1+\frac{n_{f}}{N_{c}^{3}}\biggr)
\frac{2+3\gamma (1-\gamma )}{(3-2\gamma )(1+2\gamma )}\biggr) + \psi ^{\prime \prime }(\gamma)+\psi ^{\prime
\prime }(1-\gamma)  \nonumber \\
&& \mathop{=}\limits^{\gamma\rightarrow 0} -\frac{2}{\gamma^3} -\frac{1}{\gamma^2}\bigg(\frac{11}{3} + \frac{2n_F}{3N_c^3} \bigg) -\frac{n_F}{\gamma}\bigg(\frac{13}{9N_c^3}+\frac{10}{9N_c}\bigg) + O(\gamma^0) .
\end{eqnarray}
From where we determine $\kappa_{1,2}=-\frac{11}{12}-\frac{n_F}{6N_c^3}$, $\kappa_{1,1}=-n_F\bigg(\frac{13}{36N_c^3}+\frac{10}{36N_c}\bigg)$. 
%The function $\Phi(n,\gamma)$ is introduced in Eq.~(28) of %Ref.~\cite{Kotikov:2000pm} and \new{contributes to the expansion only for non-%zero values of $n$.} %has the following collinear expansion:
%\begin{eqnarray}
%    &&\Phi(n,\gamma) = -\frac{\psi(n+1)-\psi(1)}{(\gamma + n/2)^2}  + \frac{1}{\gamma+n/2}\bigg[\psi'(n+1)-\psi'(1) \nonumber \\
%    &&+\frac{1}{4}\bigg( \psi'\big(\frac{n}{2}+1 \big) - \psi'\big(\frac{n+1}{2} \big) + \psi'(1) - \psi'(1/2) \bigg) \bigg]+O((\gamma+n/2)^0).
%\end{eqnarray}

For the purpose of comparison with Eq. (\ref{eq:chi-AP}) it turns out to be the easiest to convert our resummed result (\ref{eq:chi-anti-coll}) form the $(+)$-scheme to the symmetric scheme by solving  Eq.~(\ref{eq:rap-scheme-transf-2}) order-by-order in $\alpha_s$. For instance, the $O(\alpha_s^3)$-term of the expansion (\ref{eq:chiS-exp}) of our result, transformed in this way to the symmetric scheme has the form:
\begin{eqnarray}
    && \chi_{S}^{\text{(anti-coll,LLA)}}(0,\gamma\rightarrow 1)|_{\alpha_s^3} = -\frac{40}{(1-\gamma)^7} -\frac{40}{(1-\gamma)^6}\bigg(\frac{11}{3}+\frac{2n_F}{3N_c^3} \bigg) \label{eq:chi3S-anti-coll-exp} \\
    && -\frac{1}{(1-\gamma)^5} \bigg( \frac{484}{3} + \frac{496 n_F}{9N_c^3} + \frac{32 n_F}{9N_c} + \frac{16 n_F^2}{9N_c^6} + \frac{32 n_F^2}{9N_c^4} \bigg) \nonumber \\
    && - \frac{1}{(1-\gamma)^4} \bigg( \frac{1331}{27} - \frac{638 n_F}{27 N_c^3} - \frac{88n_F}{27 N_c} - \frac{4n_F^2}{3N_c^6} - \frac{88n_F^2}{27N_c^4} - \frac{8n_F^2}{27N_c^2} - \frac{8n_F^3}{27N_c^5} -16\zeta(3) \bigg) \nonumber \\
    && + O((1-\gamma)^{-3}).\nonumber
\end{eqnarray}
On the other hand, the corresponding term of the expansion of Eq.~(\ref{eq:chi-AP}) is:
\begin{eqnarray}
     && \chi_{S}^{\text{(all-poles)}}(0,\gamma\rightarrow 1)|_{\alpha_s^3} = \boldsymbol{-\frac{40}{(1-\gamma)^7}} \boldsymbol{ -\frac{40}{(1-\gamma)^6}\bigg(\frac{11}{3}+\frac{2n_F}{3N_c^3} \bigg) }\\
    && -\frac{1}{(1-\gamma)^5} \bigg( \boldsymbol{\frac{484}{3}} + \frac{280 n_F}{3N_c^3} + \frac{80 n_F}{3N_c} + \frac{16 n_F^2}{3N_c^6}  \bigg) \nonumber \\
    && - \frac{1}{(1-\gamma)^4} \bigg( \boldsymbol{\frac{1331}{27}} - \frac{814 n_F}{9 N_c^3} - \frac{440n_F}{9 N_c} - \frac{148n_F^2}{9N_c^6} - \frac{80n_F^2}{9N_c^4}  - \frac{8n_F^3}{27N_c^9} \bigg) + O((1-\gamma)^{-3}),\nonumber
\end{eqnarray}
where we have highlighted in bold the terms which agree with  Eq.~(\ref{eq:chi3S-anti-coll-exp}).  At each order in $\alpha_s$, the leading-$N_c$ terms agree between our result (\ref{eq:chi-anti-coll}) and the ``all-poles'' resummation (\ref{eq:chi-AP}), modulo the $\zeta(m)$-terms (e.g. the very last term in (\ref{eq:chi3S-anti-coll-exp})). The $n_F$-dependence is different everywhere except the first subleading $\alpha_s^m/(1-\gamma)^{2m}$-term.

%%%END OF sec:ALL-POLES %%%%%%%%%%%%%%%%%%%%%%%%%%%%%%%%%%%%%%%%%%%%%%%%%%%%%%%%%%%%%%%%%%%%

\section{Subleading effects and the role of a smooth scale choice}\label{sec:subl}

In the previous section we have computed the BFKL characteristic function using the LLA form of the resummed kernel. This form was obtained by approximating the kernel originally formulated in the coordinate space in \cite{Kovner:2023vsy}. The original formulation therefore contains subleading corrections and it is interesting to see what could be their effect on the resummed characteristic function. In this section we re-compute the characteristic function, starting from the exact Fourier transform of the kernel (\ref{eq:KD-res-drdQ-exact}) and confirm that it differs from the LLA one (\ref{eq:chi-anti-coll}) by subleading logarithmic terms only. These subleading logarithmic terms, albeit having no effect for values of $\gamma$ close to one, turn out to be highly-sensitive to the form of the scale choice function $Q_{\star}(\T{x},\T{y},\T{z})$ for small values of $\gamma$. For example, the scale choice (\ref{eq:scale choice}) leads to appearance of an unphysical pole in the characteristic function at $\gamma=1/2$, as we show in Sec.~\ref{sec:subl-mom-space}. This pole is removed by the smooth scale choice (\ref{eq:scale choice-smooth}) as we show by the coordinate space computation of the characteristic function in Sec.~\ref{sec:subl-coord-space}. Yet the sensitivity of the subleading logarithmic corrections to the smearing parameter $\lambda$ remains. 

\subsection{Momentum space analysis with the default scale choice}\label{sec:subl-mom-space}

The resummed characteristic function, corresponding to the scale choice (\ref{eq:scale choice}) can be computed by substituting the momentum space kernel (\ref{eq:KD-res-drdQ-exact}) and taking into account that in the fixed coupling approximation the logarithmic $Q$ - derivatives of resummation functions are  functions of the ratio $t=\T{p}^2/Q^2$ :
\begin{equation}
    \frac{\partial R^{(j)}_Q(\T{p})}{\partial\ln Q^2} = - \frac{\partial R^{(j)}(t)}{\partial\ln t}.
\end{equation}
This allows one to represent the resummation contribution to the characteristic function in the form:
\begin{eqnarray}
    \Delta \chi^{\text{(anti-coll)}}_{+}(\gamma)= \int\limits_0^\infty dt\, t^{\gamma-2}\bigg[ 2h_1(\gamma,t)\frac{\partial R^{(1)}(t)}{\partial\ln t}-h_2(\gamma)\frac{\partial R^{(2,1)}(t)}{\partial\ln t} \bigg], \label{eq:DelChi-exact}
\end{eqnarray}
where
\begin{eqnarray}
    h_1(\gamma,t)&=&\int\frac{d^2\T{q}}{\pi \T{k}^2} \bigg(\frac{\T{q}^2}{\T{k}^2} \bigg)^{\gamma-2} \frac{(\T{k}-\T{q}\sqrt{t})\cdot\T{k}}{(\T{k}-\T{q}\sqrt{t})^2 } J_0\bigg(\frac{|\T{k}-\T{q}\sqrt{t}|}{|\T{q}|} \bigg)  J_0\bigg(\frac{|\T{k}|}{|\T{q}|} \bigg), \\
    h_2(\gamma)&=&h_1(\gamma,0) = \frac{\Gamma(1-\gamma)\Gamma(\gamma-1/2)}{\sqrt{\pi}\Gamma^2(\gamma)} \label{eq:h2(gamma)}. 
\end{eqnarray}
We immediately note that both $h_1(\gamma,t)$ and $h_2(\gamma)$ have a pole at $\gamma=1/2$:
\begin{equation}
    h_2(\gamma\rightarrow 1/2)=\frac{1}{\pi (\gamma-1/2)} + O((\gamma-1/2)^0) = h_1(\gamma\rightarrow 1/2,t),
\end{equation}
which potentially leads to the appearance of an unphysical pole at $\gamma=1/2$ in the resummed characteristic function. In principle, one could hope that  the contributions of $R^{(1)}_Q$ and $R^{(2,1)}_Q$ may conspire to cancel this pole. However the condition for such cancellation has the form:
\begin{equation}
    \int\limits_0^\infty \frac{dQ}{|\T{p}|} \bigg[ \frac{\partial R^{(2)}_Q(\T{p})}{\partial\ln Q^2} - 2 \frac{\partial R^{(1)}_Q(\T{p})}{\partial\ln Q^2} \bigg]\mathop{=}\limits^{?}0,
\end{equation}
and is not fulfilled by the evolution equations. Therefore, the unphysical pole at $\gamma=1/2$ is indeed a feature of the scale choice (\ref{eq:scale choice}). In the next subsection we show that the smooth scale choice (\ref{eq:scale choice-smooth}) is free from this unphysical pole.

The appearance of the pole at $\gamma=1/2$ does not affect the LLA discussed in  Sec.~\ref{sec:DGLAP}, because the LLA terms in the momentum space kernel $\sim \alpha_s^n \ln^n(\T{q}^2/\T{k}^2)$ at leading power for $\T{q}^2\gg \T{k}^2$ correspond in $\gamma$-space to terms $\sim \alpha_s^n/(1-\gamma)^{n+1}$. For $\gamma\to 1$ the functions $h_1$ and $h_2$ behave as:
\begin{eqnarray}
    h_1(\gamma\rightarrow 1,t) &=& O((1-\gamma)^0), \\
    h_2(\gamma\rightarrow 1) &=& \frac{1}{1-\gamma} + O((1-\gamma)^0).
\end{eqnarray}
Therefore only the $1/(1-\gamma)$-term in $h_2(\gamma)$ and, correspondingly, only the $R^{(2,1)}_Q$-term in Eq.~(\ref{eq:DelChi-exact}) contribute to the  LLA. Discarding the $R_Q^{(1)}$-contribution, replacing $h_2(\gamma)\to 1/(1-\gamma)$ in Eq.~(\ref{eq:DelChi-exact}) and substituting the LLA solution for $R^{(2,1)}_Q$, Eq.~(\ref{eq:R2-LLA}), one reproduces the LLA characteristic function (\ref{eq:DeltaChi-coll}).

\subsection{The coordinate space analysis, a smooth scale choice}\label{sec:subl-coord-space}

It is challenging to obtain a closed-form expression for the Fourier transform of the kernel with generic scale choice function such as (\ref{eq:scale choice-smooth}).  Therefore in this subsection we will study  effects of this scale choice on the characteristic function in coordinate space. Notice that upon Fourier transform, the  eigenfunction $(\T{q}^2)^\gamma$  becomes:
\begin{equation}
    \nabla_x^{-2} \nabla_y^{-2}\int\limits_{\T{q}} \frac{e^{i(\T{x}-\T{y})\T{q}}}{\pi} (\T{q}^2)^{\gamma}= \frac{\Gamma(\gamma-1)4^{\gamma-1}}{\Gamma(2-\gamma)} ((\T{x}-\T{y})^2)^{1-\gamma}.
\end{equation}
Thus the characteristic function can be obtained by  acting with the coordinate space kernel (\ref{eq:K-Dip-res-coord}) directly on the eigenfunction $((\T{z}_1-\T{z}_2)^2)^{1-\gamma}$. We  use the fact that the resummation functions in the fixed-coupling approximation are functions of $t=\T{p}^2/Q^2$. Hence the resummation correction to the characteristic function takes identical form as in Eq.~(\ref{eq:DelChi-exact}), up to replacement of  the functions $h_{1,2}$ with the functions  $h_1[Q_\star]$ and $h_2[Q_\star]$. It is convenient to represent each of these functions as a sum of two terms:
$h_1[Q_\star](\gamma,t)= h_1^{(1)}[Q_\star](\gamma,t) + h_1^{(2)}[Q_\star](\gamma,t)$ and $h_2[Q_\star](\gamma)= h_2^{(1)}[Q_\star](\gamma) + h_2^{(2)}[Q_\star](\gamma)$, where
\begin{eqnarray}
  &&  h_2^{(1)}[Q_\star](\gamma) = f(\gamma) \int\limits_{\T{z}} \bigg\{  \frac{ [Q^2_\star(\T{x},\T{x},\T{z})]^{\gamma-1} - [Q^2_\star(\T{x},\T{y},\T{z})]^{\gamma-1}}{ (\T{x}-\T{z})^2 ((\T{x}-\T{y})^2)^{1-\gamma} } + (\T{x}\leftrightarrow \T{y}) \bigg\}, \label{eq:h2-1-coord} \\
  && h_2^{(2)}[Q_\star](\gamma) = f(\gamma) \int\limits_{\T{z}}  \frac{ ((\T{x}-\T{y})^2)^{\gamma} }{(\T{x}-\T{z})^2 (\T{y}-\T{z})^2} \big[Q_\star^2(\T{x},\T{y},\T{z}) \big]^{\gamma-1}, \label{eq:h2-2-coord}
\end{eqnarray}
with $f(\gamma)=\Gamma(2-\gamma) 2^{1-2\gamma}/(\pi\Gamma(\gamma))$, and
\begin{eqnarray}
   && h_1^{(1)}[Q_\star](\gamma,t) = \frac{f(\gamma)}{2} \int\limits_0^{2\pi} \frac{d\phi}{2\pi} \\
   && \times \int\limits_{\T{z}} \bigg\{  \frac{([(\T{x}-\T{z})\sqrt{t}-\T{n}_\phi Q^{-1}_\star(\T{y},\T{y},\T{z}) ]^2)^{1-\gamma} - ([(\T{x}-\T{z})\sqrt{t}-\T{n}_\phi Q^{-1}_\star(\T{x},\T{y},\T{z}) ]^2)^{1-\gamma}}{(\T{y}-\T{z})^2 ((\T{x}-\T{y})^2)^{1-\gamma} }     \nonumber \\
    && + \frac{([(\T{x}-\T{z})\sqrt{t}-\T{n}_\phi Q^{-1}_\star(\T{x},\T{x},\T{z}) ]^2)^{1-\gamma} - ([(\T{x}-\T{z})\sqrt{t}-\T{n}_\phi Q^{-1}_\star(\T{x},\T{y},\T{z}) ]^2)^{1-\gamma}}{(\T{x}-\T{z})^2 ((\T{x}-\T{y})^2)^{1-\gamma}  }   + (\T{x}\leftrightarrow \T{y})\bigg\}, \nonumber
\end{eqnarray}

\begin{eqnarray}
   &&  h_1^{(2)}[Q_\star](\gamma,t)=f(\gamma)\int\limits_0^{2\pi}\frac{d\phi}{2\pi} \int\limits_{\Tb{z}} \frac{ ((\T{x}-\T{y})^2)^\gamma}{(\T{y}-\T{z})^2 (\T{x}-\T{z})^2 } \nonumber \\
   && \times \bigg[  ([(\T{x}-\T{z})\sqrt{t}-\T{n}_\phi Q_\star^{-1}(\T{x},\T{y},\T{z}) ]^2)^{1-\gamma} - (t (\T{x}-\T{z})^2)^{1-\gamma}  \bigg] ,
\end{eqnarray}
Although the expressions for $h_1^{(1)}$ and $h_1^{(2)}$ are somewhat complicated, they reduce to $h_2^{(1)}[Q_\star](\gamma)$ and $h_2^{(2)}[Q_\star](\gamma)$ in the limit $t\to 0$.

We use this representation for the following reason. In the coordinate space kernel (\ref{eq:K-Dip-res-coord}), the leading order dipole kernel $K^{\text{(res.)}}_D(\T{x},\T{y},\T{z})$ is split into two parts. The WW kernel $K(\T{x},\T{y},\T{z})$ contains the resolution scale  $Q_\star(\T{x},\T{y},\T{z})$. On the other hand, the parts that correspond to emission and absorption from the same source, $\propto 1/(\T{x}-\T{z})^2$ and 
$\propto 1/(\T{y}-\T{z})^2$, contain the resolution scales  $Q_\star(\T{x},\T{x},\T{z})$ and  $Q_\star(\T{y},\T{y},\T{z})$ respectively. Yet, the IR behaviour of $K_D$ and  $\frac{1}{(\T{x}-\T{z})^2}$ (or $\frac{1}{(\T{y}-\T{z})^2}$ ) is different:  the former decreases at large $\T{z}$ as $1/\T{z}^4$ while the latter  as $1/\T{z}^2$. Thus it makes sense  to split the resummed kernel into terms that contain $K_D$, and those that contain  the factors   $1/(\T{x}-\T{z})^2$ or $1/(\T{y}-\T{z})^2$ only.   In the above representation the terms $h_1^{(2)}$ and $h_2^{(2)}$ contain the dipole kernel and correspond to the contribution to the characteristic function in which the scale is chosen as $Q_\star(\T{x},\T{y},\T{z})$, while the terms $h_1^{(1)}$ and $h_2^{(1)}$ originate form the contributions of the type $K^{(1)}(\T{x},\T{x},\T{z})$ and $K^{(2)}(\T{x},\T{x},\T{z}_1,\T{z}_2)$ in the coordinate space kernel (\ref{eq:K-Dip-res-coord}).
 
 The unphysical pole at $\gamma=1/2$ originates from  $h_1^{(1)}$ and $h_2^{(1)}$.  It's relation with the piecewise-smooth behaviour of the scale (\ref{eq:scale choice}) becomes clear if one considers e.g. Eq.~(\ref{eq:h2-1-coord}) with the scale choice (\ref{eq:scale choice}):
\begin{eqnarray}
   && h_2^{(1)}[\max(X^{-2},Y^{-2})](\gamma) = 2 f(\gamma) \nonumber \\ \nonumber \\
    &&\times \int\limits_{\T{z}} \frac{ [(\T{x}-\T{z})^2]^{1-\gamma} - [(\T{y}-\T{z})^2]^{1-\gamma}}{ (\T{x}-\T{z})^2 ((\T{x}-\T{y})^2)^{1-\gamma} } \theta((\T{y}-\T{z})^2< (\T{x}-\T{z})^2). \label{eq:1/2-pole-expl}
\end{eqnarray}
At large $\T{z}^2$ the numerator can be expanded as:
\begin{equation}
   [(\T{x}-\T{z})^2]^{1-\gamma} - [(\T{y}-\T{z})^2]^{1-\gamma} = 2(\gamma -1) \underbrace{((\T{x}-\T{z})^2)^{-\gamma}  (\T{z}-\T{x})\cdot (\T{x}-\T{y})}_{\sim (\T{z}^2)^{1/2-\gamma}}  + O(((\T{x}-\T{z})^2)^{-\gamma}) . \label{eq:h2-num-exp}
\end{equation}
Were the leading term $\sim(\T{z}^2)^{1/2-\gamma}$ to vanish upon integration over azimuthal angle of $\T{z}$, the result would be finite for any $\gamma<1$.  However due to the $\theta$-function in Eq.~(\ref{eq:1/2-pole-expl}) this does not happen and the $\T{z}$ - integral produces the $1/(\gamma-1/2)$-pole. 

It is clear  that for any smooth scale choice there would not be any rigid $\theta$-function constraint. The leading term of the expansion of the numerator in Eq.~(\ref{eq:h2-1-coord}), similar to the one in Eq.~(\ref{eq:h2-num-exp}), would vanish upon the angular integration, and no extraneous pole for $\gamma<1$  would arise. 

\begin{figure}
    \centering
\includegraphics[width=0.7\linewidth]{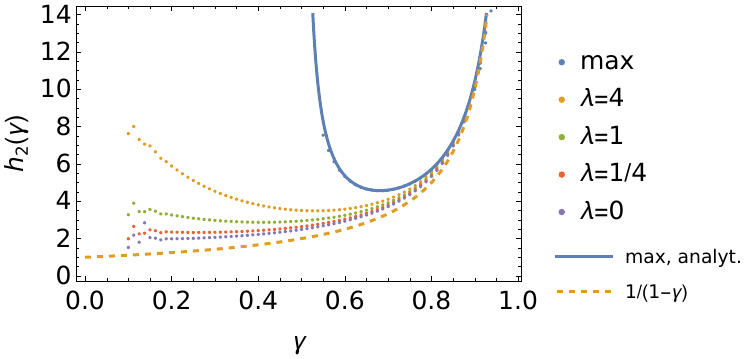}
    \caption{Dots -- numerical results for the function $h_2[Q_{\star}](\gamma)$ (eqns. (\ref{eq:h2-1-coord})+(\ref{eq:h2-2-coord})) with the default scale choice (\ref{eq:scale choice}) and smooth scale choice (\ref{eq:scale choice-smooth}) for several values of $\lambda$. Solid line -- function $h_2(\gamma)$ from Eq.~(\ref{eq:h2(gamma)}), dashed line -- the LLA function $h_2^{\text{(LLA)}}(\gamma)=1/(1-\gamma)$.}
    \label{fig:h2-smooth-scale}
\end{figure}

While the unphysical pole disappears, there remains a significant sensitivity of the characteristic function at $\gamma< 1/2$ on the scale choice as is demonstrated in Fig.~\ref{fig:h2-smooth-scale}. In Fig.~\ref{fig:h2-smooth-scale} the function $h_2[Q_{\star \lambda}](\gamma)$ evaluated numerically, 
is plotted for the smooth scale choice (\ref{eq:scale choice-smooth}). 
%evaluating numerically the integrals (\ref{eq:h2-1-%coord}) and (\ref{eq:h2-2-coord})
This sensitivity may be used to estimate the influence of subleading logarithmic corrections absent in our resummation. From Fig.~\ref{fig:h2-smooth-scale} one can see that with increasing $\lambda\gg 1$, the function $h_2[Q_{\star\lambda}]$ with the smooth scale choice (\ref{eq:scale choice-smooth}) approaches the function $h_2$ defined with the default scale choice (\ref{eq:scale choice}), as expected\footnote{The numerical results for the default scale-choice (\ref{eq:scale choice}) are also shown in the Fig.~\ref{fig:h2-smooth-scale} by the points labelled as ``max''. These numerical results are in good agreement with the analytic result (\ref{eq:h2(gamma)}) away from the singularities at $\gamma=1/2$ and $\gamma=1$, demonstrating the soundness of the numerical technique for those values of $\gamma$. Accordingly, we do not show the numerical results with the smooth scale-choice (\ref{eq:scale choice-smooth}) for $\gamma<0.1$, where numerical integration becomes unstable.}. For $\lambda\to 0$ the smooth scale choice $Q_{\star \lambda\to 0}(X,Y) \to (X^{-2}+Y^{-2})/2$ and the corresponding $h_2$-function behaves similarly to it's LLA version $1/(1-\gamma)$.

The contributions of $R^{(1)}_Q$ and, as a result, of $h_1(\gamma,t)$, are  logarithmically subleading. This was discussed above. Taking these contributions into account would go beyond the accuracy, which is under control at the moment. Therefore we do  not discuss them here.

\section{Conclusions and outlook}\label{sec:conclusions}

 In this work we have studied the effects of DGLAP resummation in JIMWLK evolution proposed in \cite{Kovner:2023vsy}, in linear regime. 
We have linearized the resummed JIMWLK Hamiltonian thereby obtaining the modified BFKL equation which includes the DGLAP resummation of anti-collinear logarithms.

We calculated the resummed characteristic function in the LLA in the fixed coupling approximation. Our main result is given in Eq. (\ref{eq:DeltaChi-coll}). We find that, compared to LO BFKL, the characteristic function is strongly modified close to $\gamma=1$, as expected. Particularly, the pole  at $\gamma=1$ present in LO BFKL, disappears.  The value of the characteristic function for 
pure glue theory is $\chi(n=0,\gamma=1)=\frac{12}{11}\frac{\pi}{4\alpha_sN_c}$, while $\chi(n=0,\gamma=1)=\frac{4}{3}\frac{\pi}{4\alpha_sN_c}$ for QCD with 3 massless quarks.

We have compared the results of our calculation with the existing literature. 
We find agreement wherever agreement is due. In particular, upon expansion in $\alpha_s$, 
the structure and the residues (both $N_c$ and $n_f$ dependencies) of the leading and subleading poles in $1-\gamma$ in our calculation are the same as in the  NLO BFKL for $n=0$
(the angular independent part), 
up to running coupling effects.

We also considered
$\gamma^*(Q_P)+\gamma^*(Q_T)$ elastic scattering with $Q_T^2\gg Q_P^2$ and recovered the LO DGLAP anomalous dimension corresponding to the PDF of the photon with smaller virtuality, $Q_P^2$. Similarly, our resummation is also valid for
DIS of a projectile hadron on a target photon (the target Bjorken limit).

Further comparison involves \cite{Altarelli:1999vw}. An interesting point here is that the resummed value of $\chi(n=0,\gamma=1)$ we obtained differs from the value conjectured in \cite{Altarelli:1999vw} by a factor $12/11$ (the pure glue case). This suggests that the simple duality assumption of \cite{Altarelli:1999vw} does not account for all the physics of anti-collinear resummation.

% \comment{ML: this summary needs a revisions} 
 Comparison with the ``all poles'' resummation of \cite{SabioVera:05} reveals that, at every order in the $\alpha_s$ expansion, the leading in $N_c$ terms nearly agree between the two expressions (see  the discussion in the main text), but $n_f$ dependence is different with the exception of the residue of the first subleading pole. The disagreement is not particularly surprising, given that the ``all poles'' resummation is an ad hoc prescription for the resummation of single logarithms.

Beyond the LLA we studied the resummation with an explicit choice of the scale that determines the transverse size of the dressed gluons, $Q_\star(\T{x},\T{y},\T{z})$ suggested in \cite{Kovner:2023vsy}. We find that close to $\gamma=1$ it gives the same result as the LLA discussed above. However somewhat surprisingly, it leads to an unphysical pole in the characteristic function at $\gamma=1/2$. 

In itself this is not an issue, as going too far away from $\gamma=1$ we are certainly moving out of the regime where our resummation is supposed to be applicable. Nevertheless, it would be pleasing to have an explicit framework which, while resumming anti-collinear logs, does not produce unphysical singularities. Fortunately we find that it is possible to modify slightly the choice of scale so that the singularity disappears. We identify as the origin of the singularity the fact that the scale choice  of \cite{Kovner:2023vsy}, that is $Q^{2}_\star(\T{x},\T{y},\T{z})=\max\{{(\T{x}-\T{z})^{-2}};{(\T{y}-\T{z})^{-2}}\}$, is only piece wise smooth. Albeit continuous as function of $(\T{x}-\T{y})$, it has a discontinuous derivative at $(\T{x}-\T{z})^2=(\T{y}-\T{z})^2$.  As discussed in Sec.~\ref{sec:subl}, it is sufficient to make the function everywhere smooth to eliminate the unphysical singularity at $\gamma=1/2$. In fact for a sufficiently slowly varying scale (as a function of $\T{z}$) the result for the characteristic function is very similar to LLA down to small $\gamma$\footnote{We did not make a special effort to study $\chi(\gamma)$ very close to $\gamma=0$, as this point is clearly outside of the range of our original approximation.}.

We stress again that pushing our expressions to $\gamma$ much smaller than one (say, $\gamma\le 1/2$) is to some extent an exercise in cosmetics. Small values of $\gamma$ are affected mostly by DGLAP logarithms in the collinear, rather than anti-collinear regime. We thus believe, that once one correctly resums collinear DGLAP logarithms in addition to the 
anti-collinear ones, the issue of unphysical pole will automatically disappear. Work along these lines is currently in progress.

Our current approach can be extended to include the running coupling effects. This should be done along the lines discussed in 
Section 3. It is interesting to see whether anti-collinear resummation and running coupling effects are simply ``additive'' or if some interesting interplay between the two can arise. The anti-collinear resummation can be also straightforwardly applied to non-forward scattering amplitudes, in addition to the forward case  discussed in the present paper. These results will be reported in our forthcoming publications. 

Finally we note that the resummation of \cite{Kovner:2023vsy} is not limited to the linear regime discussed here, but is formulated for the full nonlinear theory. The most interesting outstanding problem is to understand quantitatively the behaviour of the resummed Hamiltonian away from the linear regime and the effects of the resummation on the approach to saturation.   
%%%%%%%%%%%%%%%%%%%%%%%%%%%%%%%%%%%%%%%%%%%%%%%%%%%%%%%%%%%%%%%%%%%%%%%%%%%%%%%%%%%%%%%%%%%%
\appendix

\section{Coordinate-space sum-rules for $R^{(1)}_Q$ and $R^{(2,1)}_Q$}\label{append:R-sum-rules}

The sum-rules (\ref{eq:R1-norm-int}) and (\ref{eq:r1-sum-rule}) are easy to prove.  For the Fourier transformed resummation functions $R_Q^{(1)}(\T{p})$ and $r_Q^{(1)}(\T{p})$, defined by Eq.~(\ref{eq:R1-q-def}), the sum-rules  are equivalent to
\begin{equation}
    R_Q^{(1)}(\T{p}=0) = r_Q^{(1)}(\T{p}=0) = 1.
\end{equation}
This property is satisfied by the initial condition at $Q=\Lambda$, Eq.~(\ref{eq:R1-init-cond}) and (\ref{eq:r1-init-cond}). Therefore, all we need to show is that
\begin{eqnarray}
    \frac{\partial R_Q^{(1)}(\T{p}=0)}{\partial\ln Q^2} =0, \ \ ~~~~~~~~
    \frac{\partial r_Q^{(1)}(\T{p}=0)}{\partial\ln Q^2} =0.\label{A2}
\end{eqnarray}
Eq. (\ref{A2}) is indeed true due to the momentum space evolution equations (\ref{eq:R1-eqn})  combined with the fact that the functions $\pi_{ij}(p)$ introduced there satisfy
$\pi_{ij}(p\to 0)\to 0$. 

To prove the sum-rules (\ref{eq:R2-norm-int}) and (\ref{eq:r2-sum-rule}) for the second order functions, we introduce the functions:
\begin{eqnarray}
    \Sigma^{(2)}_Q(\T{x}_1) = \int\limits_{\T{x}_2} R_Q^{(2,1)}(\T{x}_1,\T{x}_2), ~~~~~~~~~~~~~
    \sigma^{(2)}_Q(\T{x}_1) = \int\limits_{\T{x}_2} r_Q^{(2,1)}(\T{x}_1,\T{x}_2),
\end{eqnarray}
so that the sum-rules are equivalent to
\begin{eqnarray}
\Sigma^{(2)}_Q(\T{x}_1)=R_Q^{(1)}(\T{x}_1), \label{eq:R2-SR-Sigma} ~~~~~~~~~~~~
\sigma^{(2)}_Q(\T{x}_1)=r_Q^{(1)}(\T{x}_1).\label{eq:r2-SR-sigma}
\end{eqnarray}
The evolution equation for $\Sigma^{(2)}_Q$ can be derived by integrating  Eq.~(\ref{eq:R2-Qevol-quarks}) over $\T{x}_2$:
\begin{eqnarray}
         &&\left[ \frac{\partial}{\partial \ln Q^2} + a_s\beta_0 \right] \Sigma_{Q}^{(2)}(\T{x}_1) = -2a_s \int\limits_0^1 d\xi\, \int\limits_0^{2\pi} \frac{d\phi}{2\pi}  \label{eq:Sigma2-evol} \\
    && \times \bigg\{2N_c p_{gg}(\xi)   \Sigma_Q^{(2)} \left( \T{x}_1 + \xi Q^{-1} \T{n}_\phi  \right)  + 2T_Fn_F \frac{C_F}{N_c} p_{qg}(\xi)  \sigma_Q^{(2)} \left( \T{x}_1 + \xi Q^{-1} \T{n}_\phi  \right) \nonumber \\
  &&   -N_c p_{gg}(\xi) R^{(1)}_Q\left( \T{x}_1 - (1-\xi) Q^{-1} \T{n}_\phi  \right)  + \frac{T_F n_F}{N_c^2} p_{qg}(\xi)  r^{(1)}_Q\left( \T{x}_1 - (1-\xi) Q^{-1} \T{n}_\phi  \right)  \bigg\}. \nonumber
\end{eqnarray}
Integrating-out $\T{x}_2$ in Eq.~(\ref{eq:r2-Qevol-quarks}) we obtain the equation for $\sigma^{(2)}_Q$:
 \begin{eqnarray}
     &&  \bigg[ \frac{\partial}{\partial \ln Q^2} + 3C_Fa_s \bigg] \sigma_Q^{(2)}(\T{x}_1) = - 2a_s \int\limits_0^1 d\xi \, p_{gq}(\xi) \int\limits_0^{2\pi} \frac{d\phi}{2\pi}  \label{eq:sigma2-quarks} \\
     &&\times \bigg\{ C_F  \sigma_Q^{(2)} \left( \T{x}_1 +\xi Q^{-1} \T{n}_\phi \right) +N_c  \Sigma_Q^{(2)} \left( \T{x}_1 - (1-\xi)Q^{-1} \T{n}_\phi  \right) \nonumber \\
    && -{ \frac{N_c}{2} \big( R^{(1)}_Q(\T{x}_1-(1-\xi)Q^{-1}\T{n}_\phi \big) + r^{(1)}_Q(\T{x}_1-(1-\xi)Q^{-1}\T{n}_\phi) \big) }   \bigg\} . \nonumber 
\end{eqnarray}
Next, substitute $\Sigma^{(2)}_Q(\T{x}_1)\to R^{(1)}_Q(\T{x}_1)$, $\sigma^{(2)}_Q(\T{x}_1)\to r^{(1)}_Q(\T{x}_1)$ into Eq.~(\ref{eq:Sigma2-evol}) and use Eq.~(\ref{eq:R1-Qevol-quarks}) to express the integral of $R^{(1)}_Q$:
\begin{eqnarray}
 &&\int\limits_0^1 d\xi\int\limits_0^{2\pi} \frac{d\phi}{2\pi} p_{gg}(\xi) R_Q^{(1)}\big( \T{x}-\xi Q^{-1} \T{n}_\phi \big)  \\
 && = -\frac{1}{2N_c a_s}\bigg[ \frac{\partial}{\partial\ln Q^2} + a_s\beta_0 \bigg] R_Q^{(1)}(\T{x}) - \frac{2T_F^2n_F}{N_c} \int\limits_0^1 d\xi\int\limits_0^{2\pi} \frac{d\phi}{2\pi}  p_{qg}(\xi) r_Q^{(1)}\big( \T{x}-\xi Q^{-1} \T{n}_\phi \big). \nonumber
\end{eqnarray}
We obtain
\begin{eqnarray}
    && \text{(r.h.s. of Eq. (\ref{eq:Sigma2-evol}))} = \bigg[ \frac{\partial}{\partial\ln Q^2} + a_s\beta_0 \bigg] R_Q^{(1)}(\T{x}_1) \\
    &&+ a_s \int\limits_0^1 d\xi \, p_{qg}(\xi) \int\limits_0^{2\pi} \frac{d\phi}{2\pi} r_Q^{(1)}(\T{x}_1-\xi Q^{-1}\T{n}_\phi) \underbrace{\bigg( 4T_F^2 n_F - 2T_F n_F \bigg( \frac{2C_F}{N_c} + \frac{1}{N_c^2} \bigg) \bigg)}_{=0}. \nonumber 
\end{eqnarray}
The end result is identical to the left-hand side, taking into account that $T_F=1/2$.

Similarly, substituting   $\Sigma^{(2)}_Q(\T{x}_1)\to R^{(1)}_Q(\T{x}_1)$, $\sigma^{(2)}_Q(\T{x}_1)\to r^{(1)}_Q(\T{x}_1)$  into  Eq.~(\ref{eq:sigma2-quarks}) and excluding the integral of $r^{(1)}_Q$ with the help of Eq.~(\ref{eq:r1-Qevol-quarks}):
\begin{eqnarray}
&& \int\limits_0^1 d\xi \int\limits_0^{2\pi} \frac{d\phi}{2\pi} p_{gq}(\xi) r^{(1)}_Q\big( \T{x}-\xi Q^{-1}\T{n}_\phi \big) \\
  &&  = \frac{N_c}{a_s}\bigg[ \frac{\partial}{\partial \ln Q^2} + 3C_F a_s \bigg] r_Q^{(1)}(\T{x}) +N_c^2 \int\limits_0^1 d\xi \int\limits_0^{2\pi} \frac{d\phi}{2\pi} p_{gq}(\xi)   R^{(1)}_Q\big( \T{x}+(1-\xi)Q^{-1}\T{n}_\phi \big)  \nonumber
\end{eqnarray}
the right-hand side of this equation reads:
\begin{eqnarray}
    &&\text{(r.h.s. of Eq.~(\ref{eq:sigma2-quarks}))} = \frac{a_s}{N_c}\bigg\{ \frac{N_c}{a_s} \bigg[ \frac{\partial}{\partial \ln Q^2} + 3C_F a_s \bigg] r_Q^{(1)}(\T{x}_1) \nonumber \\
    && + N_c^2  \int\limits_0^1 d\xi \, p_{qg}(\xi) \int\limits_0^{2\pi} \frac{d\phi}{2\pi} R_Q^{(1)}(\T{x}_1-(1-\xi)Q^{-1}\T{n}_\phi) \bigg\} \nonumber \\
    && - a_s N_c \int\limits_0^1 d\xi \, p_{qg}(\xi) \int\limits_0^{2\pi} \frac{d\phi}{2\pi} R_Q^{(1)}(\T{x}_1-(1-\xi)Q^{-1}\T{n}_\phi).
\end{eqnarray}
The result is  identical to the left-hand side of Eq.~(\ref{eq:sigma2-quarks}). Thus we have proven that the ansatz (\ref{eq:R2-SR-Sigma})  is the solution of Eqns. (\ref{eq:Sigma2-evol}) and (\ref{eq:sigma2-quarks}). Therefore the sum-rules (\ref{eq:R2-norm-int}) and (\ref{eq:r2-sum-rule}) are consistent with the evolution equations and  trivially satisfied by the initial conditions (\ref{eq:R2-init-cond}) and (\ref{eq:r2-init-cond}).

%%%%%%%%%%%%%%%%%%%%%%%%%%%%%%%%%%%%%%%%%%%%%%%%%%%%%%%%%%%%%%%%%%%%%%%%%%%%%%%%%%%%%%%%%%%%

\section{Fourier transform of the kernel in transverse momentum space}\label{append:Fourier}

 In this Appendix, we detail computations of the Fourier transform (\ref{eq:Kres(k,q)-def}) of the resummed BFKL kernel (\ref{eq:K-res-def}) defined in coordinate space. We isolate the LO BFKL kernel from the contribution  due to the resummation denoted as $\Delta K^{\text{(res.)}}$:
 \begin{eqnarray}
   && K^{\text{(res.)}}(\T{x},\T{y},\T{z}_1,\T{z}_2) = K^{\text{(LO)}}_{\text{BFKL}} + \Delta K^{\text{(res.)}}.
 \end{eqnarray}
The LO BFKL kernel is
 \begin{eqnarray}
   &&  K^{\text{(LO)}}_{\text{BFKL}} = \frac{\alpha_sN_c}{2\pi^2} \bigg\{ \delta^{(2)}(\T{z}_1-\T{x})\delta^{(2)}(\T{z}_2-\T{y}) \bigg(\int\limits_{\T{z}} K_D(\T{x},\T{y},\T{z})\bigg) \nonumber \\
   && +\big( 2 K(\T{x},\T{y},\T{z}_1) - K(\T{y},\T{y},\T{z}_1) \big)\delta^{(2)}(\T{z}_2-\T{x}) \nonumber \\
   && + \big( 2 K(\T{x},\T{y},\T{z}_1) - K(\T{x},\T{x},\T{z}_1) \big)\delta^{(2)}(\T{z}_2-\T{y})  -2K(\T{x},\T{y},\T{z}_1) \delta^{(2)}(\T{z}_1-\T{z}_2)\bigg\},\label{eq:K-LO-coord}
\end{eqnarray}
 To facilitate  computations, we introduce, in the part of Eq.~(\ref{eq:K-res-def}) containing the resummation functions, an integral over an auxiliary scale variable $Q^2$ fixed by $\delta(Q^2-Q_\star^2(\T{x},\T{y},\T{z}))$. Then $\Delta K^{\text{(res.)}}$ can be separated into contributions of two types:
  \begin{eqnarray}
 \Delta K^{\text{(res.)}}(\T{x},\T{y},\T{z}_1,\T{z}_2) &&= \Delta K^{(I)}_{\text{res.}}(\T{x},\T{y},\T{z}_1,\T{z}_2)\nonumber \\
  &&-\Delta K^{(II)}_{\text{res.}}(\T{x},\T{y},\T{z}_1,\T{z}_2)-\Delta K^{(II)}_{\text{res.}}(\T{y},\T{x},\T{z}_1,\T{z}_2).  
  \end{eqnarray}
The terms of the first type are those proportional to $K(\T{x},\T{y},\T{z})$:
\begin{eqnarray}
   \Delta K^{(I)}_{\text{res.}} &&= 2 \frac{\alpha_s N_c}{2\pi^2} \int\limits_{0}^\infty dQ^2 \int\limits_{\Tb{z}}  K(\T{x},\T{y},\Tb{z}) \delta(Q^2-Q^2_\star(\T{x},\T{y},\Tb{z})) \nonumber \\
  && \times \bigg\{ -(R^{(2,1)}_Q(\Tb{z}-\T{z}_1,\Tb{z}-\T{z}_2)-\delta^{(2)}(\Tb{z}-\T{z}_1)\delta^{(2)}(\Tb{z}-\T{z}_2)) \nonumber \\
  && + (R^{(1)}_Q(\Tb{z}-\T{z}_1)-\delta^{(2)}(\Tb{z}-\T{z}_1))\delta^{(2)}(\T{z}_2-\T{x}) \nonumber \\
  && + (R^{(1)}_Q(\Tb{z}-\T{z}_1)-\delta^{(2)}(\Tb{z}-\T{z}_1))\delta^{(2)}(\T{z}_2-\T{y}) \bigg\}.
\end{eqnarray}
 The terms of the second type are proportional to $K(\T{x},\T{x},\T{z})$:
\begin{eqnarray}
    \Delta K^{(II)}_{\text{res.}} =  \frac{\alpha_s N_c}{2\pi^2} \int\limits_0^\infty dQ^2 \int\limits_{\Tb{z}}  K(\T{x},\T{x},\Tb{z}) \delta(Q^2-Q^2_\star(\T{x},\T{x},\Tb{z})) \nonumber \\
    \times (R^{(1)}_Q(\Tb{z}-\T{z}_1)-\delta^{(2)}(\Tb{z}-\T{z}_1))\delta^{(2)}(\T{z}_2-\T{y}).
\end{eqnarray}

The Fourier transforms of the kernels are defined in (\ref{eq:Kres(k,q)-def}). Applied to
 $\Delta K^{(I)}_{\text{res.}}$ and $\Delta K^{(II)}_{\text{res.}}$ they take the forms:
\begin{eqnarray}
   && \frac{\T{q}^4}{\T{k}^4} \Delta K^{(I)}_{\text{res.}}(\T{k},\T{q}) =    \frac{\alpha_s N_c}{\pi^2} \int\limits_{0}^\infty dQ^2 \bigg\{ -k^{(I)}_Q(-\T{k},\T{k}) (R_{\max(Q,Q_0)}^{(2,1)}(\T{q})-1) \nonumber \\
  && +\big( k_Q^{(I)}(-\T{k}-\T{q},\T{k})+ k_Q^{(I)}(-\T{k},\T{k}+\T{q})  \big) (R_{\max(Q,Q_0)}^{(1)}(\T{q})-1)\bigg\} + (\T{q}\to -\T{q}), \label{eq:DelK-I_kI}\\
  &&\hspace{-1cm} \frac{\T{q}^4}{\T{k}^4}  \Delta K_{\text{res.}}^{(II)}(\T{k},\T{q}) = \frac{\alpha_s N_c}{2\pi^2} \int\limits_{0}^\infty dQ^2\, k_Q^{(II)}(-\T{k},\T{k}+\T{q}) (R_{\max(Q,Q_0)}^{(1)}(\T{q})-1) + (\T{q}\to -\T{q}),\label{eq:DelK-II_kII}
\end{eqnarray}
where the kernels $k_Q^{(I)}$ and $k_Q^{(II)}$  defined as
\begin{eqnarray}
    &&  k^{(I)}_Q(\T{K},\T{P})=\int\limits_{\T{x},\T{y},\T{z}} \frac{K(\T{x},\T{y},\T{z})}{(2\pi)^{2-2\epsilon}2S_\perp}  e^{i\T{K}(\T{x}-\T{z})} e^{i\T{P}(\T{y}-\T{z})} \times \delta(Q^2-\max[(\T{x}-\T{z})^{-2},(\T{y}-\T{z})^{-2}])\nonumber \\
    && k^{(II)}_Q(\T{K},\T{P})=\int\limits_{\T{x},\T{y},\T{z}} \frac{K(\T{x},\T{x},\T{z})}{(2\pi)^{2-2\epsilon}2S_\perp}  \delta(Q^2-(\T{x}-\T{z})^{-2}) e^{i\T{K}(\T{x}-\T{z})} e^{i\T{P}(\T{y}-\T{z})} .\label{eq:kI-def},
\end{eqnarray}
The integral over $\T{z}$ cancels against $S_\perp$ due to translational invariance of the integrands. The remaining integrals over $\T{x}$ and $\T{y}$ do not have any divergences and can be computed setting $\epsilon=0$. The rwesult for  $k^{(I)}_Q$:
\begin{eqnarray}
  &&  k^{(I)}_Q(\T{K},\T{P})= \kappa_Q(\T{K},\T{P}) + \kappa_Q(\T{P},\T{K}), \label{eq:kI-sym} \\ \nonumber \\ 
  &&  \kappa_Q(\T{K},\T{P}) = \int\limits_{\T{x},\T{y}} \frac{e^{i\T{K}\T{x}} e^{i\T{P}\T{y}}}{2(2\pi)^{2}} \frac{(\T{x}\cdot \T{y}) \T{x}^2 }{\T{y}^2}  \delta(\T{x}^2-Q^{-2}) \theta(\T{y}^2>\T{x}^2).\label{eq:Kappa-I-def}
\end{eqnarray}
The result for  $k^{(II)}_Q$:
%straightforwardly evaluates to:
\begin{equation}
    k_Q^{(II)}(\T{K},\T{P})=  \frac{\pi}{Q^2}   J_{0}\bigg(\frac{|\T{K}|}{Q} \bigg) \delta^{(2)}(\T{P}) .
\end{equation}
It can be also written in the following form:
\begin{equation}
    k_Q^{(II)}(\T{K},\T{P})= \frac{\pi}{Q^2}  \delta^{(2)}(\T{P}) \frac{\partial}{\partial \ln Q^2} \int\limits_0^{Q^2}\frac{dq^2}{q^2} J_0\bigg(\frac{|\T{K}|}{q} \bigg). \label{eq:kII-res}
\end{equation}

Going back to $k^{(I)}_Q$ and $\kappa_Q$ we note that the integrals in Eq.~(\ref{eq:Kappa-I-def}) can  be evaluated:
\begin{eqnarray}
 &&  \kappa_Q(\T{K},\T{P}) =  \frac{(-\nabla_{\T{K}}\cdot \nabla_{\T{P}})}{8} J_{0}\bigg(\frac{|\T{K}|}{Q}\bigg) \int\limits_{Q^{-2}}^\infty \frac{d\T{y}^2}{Q^2 \T{y}^2} J_{0}\big(|\T{P}| |\T{y}| \big)\nonumber  \\
 && = \frac{-\T{K}\cdot \T{P}}{4  |\T{K}| |\T{P}|} J_1\bigg(\frac{|\T{K}|}{Q}\bigg) \int\limits_{Q^{-1}}^\infty\frac{d|\T{y}|}{Q^3} J_{1}\big(|\T{P}| |\T{y}| \big) =  \frac{-\T{K}\cdot \T{P}}{4  |\T{K}| \T{P}^2 Q^3 } J_1\bigg(\frac{|\T{K}|}{Q}\bigg) J_0\bigg(\frac{|\T{P}|}{Q}\bigg) . 
\end{eqnarray}
Taking into account Eq.~(\ref{eq:kI-sym}),  the result for $k^{(I)}_Q$ is
\begin{equation}
   k^{(I)}_Q(\T{K},\T{P})= - \frac{\T{K}\cdot \T{P}}{2Q^2\T{K}^2 \T{P}^2}  \frac{\partial}{\partial \ln Q^2} \bigg[  J_{0}\bigg(\frac{|\T{K}|}{Q}\bigg)  J_{0}\bigg( \frac{|\T{P}|}{Q}\bigg)  \bigg]. \label{eq:kI-res} 
\end{equation}

The results (\ref{eq:kII-res}) and (\ref{eq:kI-res}) are suggestive of  integrating by parts in the $Q^2$-integrals in Eqns.~(\ref{eq:DelK-I_kI}) and (\ref{eq:DelK-II_kII}). The boundary terms have the form:
\begin{equation}
    (R^{(j)}_{Q}(\T{q})-1) J_0\bigg(\frac{a}{Q} \bigg)J_0\bigg(\frac{b}{Q} \bigg)
\end{equation}
and vanish both for $Q\to 0$ (because of the Bessel functions) and for $Q\to \infty$ (since $R^{(j)}_{Q}(\T{q})\to 1$). Thus the derivatives $\partial/\partial\ln Q^2$ can be integrated by parts to act on the factors $R^{(j)}_{Q}$:
\begin{eqnarray}
     && \frac{\T{q}^4}{\T{k}^4} \Delta K^{\text{(res.)}}(\T{k},\T{q}) =   \frac{\alpha_s N_c}{2\pi^2}\bigg\{ \int\limits_{Q_0^2}^\infty \frac{dQ^2}{Q^2} \bigg[ \frac{\partial R^{(2,1)}_Q(\T{q})}{\partial\ln Q^2} \frac{1}{\T{k}^2} J_{0}^2\bigg(\frac{|\T{k}|}{Q} \bigg) \nonumber \\
   && -\frac{\partial R_Q^{(1)}(\T{q})}{\partial\ln Q^2} \frac{2(\T{k}+\T{q})\cdot \T{k}}{(\T{k}+\T{q})^2 \T{k}^2}  J_{0}\bigg(\frac{|\T{k}+\T{q}|}{Q} \bigg) J_{0}\bigg(\frac{|\T{k}|}{Q} \bigg) \nonumber \\
   && -\pi \delta^{(2)}(\T{k}+\T{q}) \frac{\partial R^{(1)}_Q(\T{q})}{\partial\ln Q^2}  \int\limits_0^{Q^2}\frac{dq^2}{q^2} J_0\bigg(\frac{|\T{k}|}{q} \bigg) \bigg]\bigg\} +(\T{q}\to -\T{q}),
\end{eqnarray}
which together with the Fourier transform of Eq.~(\ref{eq:K-LO-coord}) leads to Eq.~(\ref{eq:KD-res-drdQ-exact}).

%%%%%%%%%%%%%%%%%%%%%%%%%%%%%%%%%%%%%%%%%%%%%%%%%%%%%%%%%%%%%%%%%%%%%%%%%%%%%%%%%%%%%%%%%%%%
\section{First order resummation equations in momentum space}\label{append:lin-res-mom-1}

\begin{figure}
    \centering
    \includegraphics[width=0.6\linewidth]{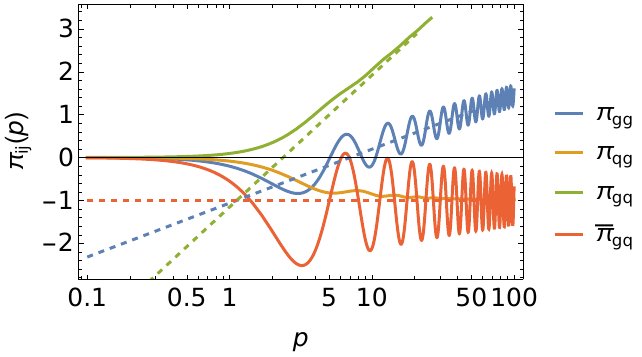}
    \caption{Solid lines -- plots of functions $\pi_{ij}(p)$ defined in Eq.~(\ref{eq:pi_ij-def}), dashed lines -- asymptotic expressions (\ref{eq:pi_gg-asy}) -- (\ref{eq:pib_gq-pi_qg-asy}).}
    \label{fig:pi_ij-plots}
\end{figure}

\begin{figure}
    \centering
    \includegraphics[width=0.7\linewidth]{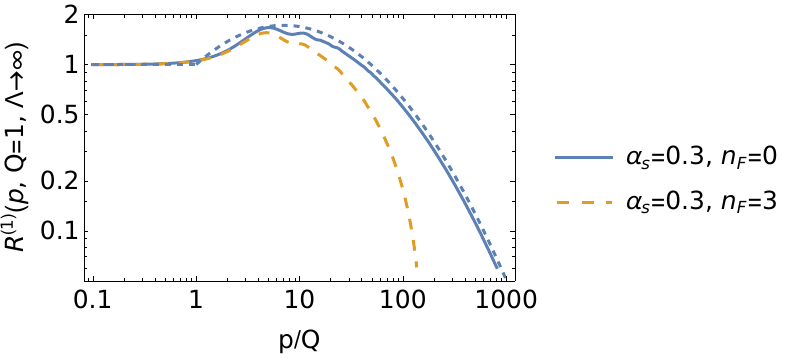}
    \caption{Solid line -- $R_Q^{(1)}(\T{p})$ obtained from the numerical solution of the Eq.~(\ref{eq:R1-eqn}) at $n_F=0$, dashed line -- the same, but for $n_F=3$, short-dashed line -- asymptotic solution (\ref{eq:R1-asy-sol}) for $n_F=0$.}
    \label{fig:R1-numerics}
\end{figure}

The  resummation equations for the first order
functions $R^{(1)}_Q(\T{x})$ and $r^{(1)}_Q(\T{x})$ are most easily solved in momentum space. Multiplying  Eqs.~(\ref{eq:R1-Qevol-quarks}) and (\ref{eq:r1-Qevol-quarks}) by $e^{i\T{p}\T{x}}$ and integrating over $\T{x}$:
\begin{eqnarray}
      \frac{\partial}{\partial \ln Q^2} {\cal R}_Q^{(1)}(\T{p}) = - a_s \Pi_1(|\T{p}|/Q) {\cal R}_Q^{(1)}(\T{p}). \label{eq:R1-eqn}
\end{eqnarray}
Here we have combined the first order functions into the vector ${\cal R}^{(1)}_Q(\T{p}) \equiv \left( \begin{array}{c}
         R^{(1)}_Q(\T{p}) \\
         r^{(1)}_Q(\T{p}) 
    \end{array} \right)$.  The matrix $\Pi_1$ has the form:
\begin{equation}
     \Pi_1(p) = \left( \begin{array}{cc}
       -\frac{11 N_c}{3} \pi_{gg}(p)  & \frac{2n_F}{3}\pi_{qg}(p)  \\
       -\frac{3}{2}N_c \bar{\pi}_{gq}(p)  & \frac{3}{2} \frac{1}{N_c} \pi_{gq} (p) 
    \end{array} \right).
\end{equation}
The functions $\pi_{ij}(p)$ and $\bar{\pi}_{ij}(p)$ are defined as follows:
\begin{equation}
     \pi_{ij}(p) = \frac{\int\limits_0^1d \xi \, p_{ij}(\xi) \, J_0(p\xi)}{\int\limits_0^1 d\xi \, p_{ij}(\xi)}-1, \, ~~~~~ \bar{\pi}_{ij}(p) = \frac{\int\limits_0^1d \xi \, p_{ij}(1-\xi) \, J_0(p\xi)}{\int\limits_0^1 d\xi \, p_{ij}(\xi)}-1.\label{eq:pi_ij-def}
\end{equation}
Plots of the functions $\pi_{ij}(p)$ are shown in  Fig.~\ref{fig:pi_ij-plots}.
For $p\gg 1$ these functions have the following asymptotic behaviour, on top of the oscillations induced by the Bessel functions:
\begin{eqnarray}
    \pi_{gg}(p) &\simeq& \frac{6}{11}\left( \ln p -\frac{11}{6} - \ln 2 + \gamma_E \right), \label{eq:pi_gg-asy} \\
    \pi_{gq}(p) &\simeq& \frac{4}{3}\left( \ln p - \frac{3}{4} - \ln 2 + \gamma_E \right), \label{eq:pi_gq-asy} \\
    \bar{\pi}_{gq}(p) &\simeq& \pi_{qg}(p) \simeq -1. \label{eq:pib_gq-pi_qg-asy}
\end{eqnarray} 
In particular, for $n_F=0$, Eq.~(\ref{eq:pi_gg-asy}) makes it possible to construct the following asymptotic solution for the $R_Q^{(1)}(\T{p})$:
\begin{eqnarray}
    R_Q^{(1)}(\T{p}) \simeq \exp \bigg[ -a_sN_c\bigg( \frac{1}{2} \ln^2 \frac{Q^2}{\T{p}^2} + \big( \frac{11}{3}+2(\ln 2-\gamma_E) \big)\ln\frac{Q^2}{\T{p}^2} \bigg)  \bigg].\label{eq:R1-asy-sol}
\end{eqnarray}
 This asymptotic solution agrees reasonably well with the numerical solution of  Eq.~(\ref{eq:R1-eqn}) for $n_F=0$. For $n_F>0$ the  suppression of the region $Q^2\gg \T{p}^2$ is even stronger, see Fig.~\ref{fig:R1-numerics}.

 We note that (\ref{eq:R1-asy-sol}) has a familiar form of a Sudakov form factor of a gluon TMD-distribution, although the coefficient in front of the double logarithm is smaller by a factor of two. 
 
 The origin of this Sudakov-like form factor is somewhat mysterious to us for two reasons. First, the double logarithmic part of the evolution  is already included in the evolution in $Y$. Second, as discussed in the main text, the physical meaning of    $ R_Q^{(1)}(\T{z}) $ is that of the density of gluons at point $\T{z}$. Since $\T{z}$ is the ``impact parameter'' inside the dressed gluon, the momentum $\T{p}$, which is conjugate to it, is not a momentum of some gluon. We therefore are not sure whether the similarity of the exponential factor in (\ref{eq:R1-asy-sol}) with a Sudakov form factor has any physical meaning, or  it is purely superficial.

%%%%%%%%%%%%%%%%%%%%%%%%%%%%%%%%%%%%%%%%%%%%%%%%%%%%%%%%%%%%%%%%%%%%%%%%%%%%%%%%%%%%%%%%%%%

\section{Rapidity factorisation schemes}\label{append:rap-schemes}

\begin{figure}
    \centering
    \includegraphics[width=\linewidth]{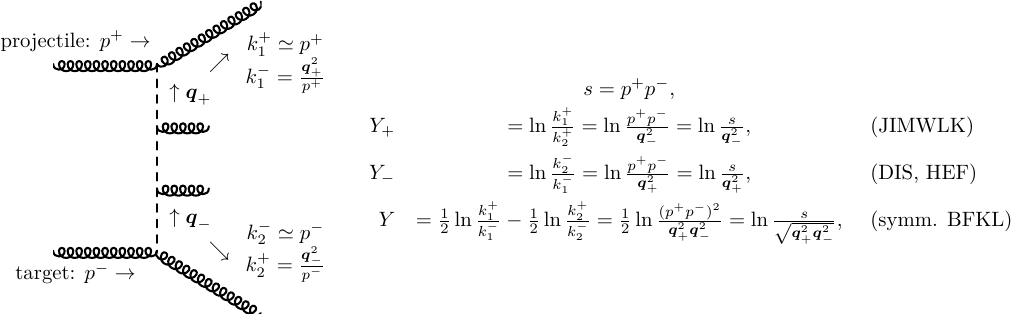}
    \caption{Kinematics of production of a pair of highly separated in rapidity jets in a collision of two on-shell gluons with the center of mass energy squared $s=p^+p^-$. The expressions for the resummation parameters of the $(\pm)$-schemes ($Y_{\pm}$) and of the symmetric scheme $Y$, which is the rapidity difference between jets, are derived in terms of the jet transverse momenta $\T{q}_{\T{\pm}}$ and collision energy $\sqrt{s}$. The dashed line denotes a Reggeized gluon exchange.}
    \label{fig:rapidity-schemes}
\end{figure}

 The material of this Appendix can be found in many sources, see e.g. Refs.~\cite{Fadin:1998py,NLOCiafaloni2,Salam98,RGIBFKL,Deak:2019wms}. For convenience we collect here the relevant facts concerning the transformation of the BFKL eigenvalue between different rapidity-factorisation schemes. We limit ourselves to the fixed-coupling approximation while adopting notations and conventions of the present paper. The kinematic relationships between different schemes are most easy to explain on a specific example of production of two gluon jets in a collision of two on-shell initial-state gluons at the centre of mass energy $s$, Fig.~\ref{fig:rapidity-schemes}. 

In the BFKL approach, the energy-dependence of the cross section is controlled by the BKFL Green's function, which is a solution of Eq.~(\ref{eq:BFKL-mom}) with initial condition $\delta^{(2)}(\T{q}_{+}-\T{q}_{-})$. In the fixed-coupling approximation the solution has the form:
\begin{eqnarray}
    G(s,\T{q}_+,\T{q}_-) = \frac{1}{\pi \T{q}_+^2} \sum\limits_{n=-\infty}^{+\infty} \int\frac{d\omega}{2\pi i} \int\frac{d\gamma}{2\pi i} \bigg( \frac{s}{\sqrt{\T{q}_{+}^2 \T{q}_{-}^2}} \bigg)^{\omega} \left( \frac{\T{q}_{+}^2}{\T{q}_{-}^2} \right)^\gamma \frac{e^{in\Delta\phi_{+-}}}{\omega - \hat{\alpha}_s \chi_S(n,\gamma)}, \label{eq:G-BFKL-symm}
\end{eqnarray}
where $\Delta \phi_{+-}$ is the azimuthal angle between vectors $\T{q}_+$ and $\T{q}_-$, $\hat{\alpha}_s=\alpha_s N_c/\pi$ and $\chi_S(n,\gamma)$ is the characteristic function in the symmetric scheme\footnote{The contours in $\gamma$ and $\omega$ planes go parallel to the imaginary axis. The former goes through the point $\gamma=1/2$ and the latter one is located to the right of all the singularities in the $\omega$-plane. }. Due to the projectile-target symmetry, $G(s,\T{q}_+,\T{q}_-) = G(s,\T{q}_-,\T{q}_+)$, which implies 
\begin{equation}
    \chi_S(n,1-\gamma) = \chi_S(n,\gamma).\label{eq:chi_S-symmetry}
\end{equation}
The representation (\ref{eq:G-BFKL-symm}) can be rewritten in a form corresponding to the $(+)$ or $(-)$ schemes: 
\begin{eqnarray}
    G(s,\T{q}_+,\T{q}_-) = \frac{1}{\pi \T{q}_+^2} \sum\limits_{n=-\infty}^{+\infty} \int\frac{d\omega}{2\pi i} \int\frac{d\gamma}{2\pi i} \left( \frac{s}{\T{q}_{\pm}^2} \right)^{\omega} \left( \frac{\T{q}_{+}^2}{\T{q}_{-}^2} \right)^\gamma \frac{e^{in\Delta\phi_{+-}}}{\omega - \hat{\alpha}_s \chi_S(n,\gamma\mp \frac{\omega}{2})}, \label{eq:G-BFKL-asymm}
\end{eqnarray}
where the upper signs correspond to the $(-)$-scheme and lower ones to the $(+)$-scheme. The denominator now defines a new pole in the $\omega$-plane:
\begin{equation}
    \frac{1}{\omega - \hat{\alpha}_s \chi_S(n,\gamma\mp \frac{\omega}{2})} = \frac{R_{\mp}(n,\gamma)}{\omega - \chi_{\mp}(n,\gamma)} + \ldots,
\end{equation}
where the non-pole part of the $\omega$-dependence is denoted by the ellipsis. The residues $R_{\pm}(n,\gamma)$ renormalize the impact factors\footnote{Note that the multiplication in $\gamma$-space is equivalent to convolutions over $\T{q}_{\pm}$.}, ensuring that the cross section is scheme independent, while the characteristic functions in the $(\pm)$-schemes are related with the characteristic function in the symmetric scheme by the equation:
\begin{equation}
    \chi_{\pm}(n,\gamma) = \chi_S\big(n,\gamma \pm \frac{\hat{\alpha}_s}{2} \chi_{\pm}(n,\gamma) \big). \label{eq:rap-scheme-transf}
\end{equation}

From this equation, and using the property (\ref{eq:chi_S-symmetry}), one derives:
\begin{eqnarray}
    &&\chi_+(n,\gamma)-\chi_-(n,1-\gamma) = \chi_S\big(n,\gamma + \frac{\hat{\alpha}_s}{2} \chi_{+}(n,\gamma) \big) - \chi_S\big(n,\gamma + \frac{\hat{\alpha}_s}{2} \chi_{-}(n,1-\gamma) \big) \nonumber \\
    && = \sum\limits_{k=1}^\infty \frac{\partial^k\chi_S(n,\gamma)}{\partial\gamma^k} \frac{\hat{\alpha}_s^k}{2^k k!} \bigg( \chi^k_+(n,\gamma) - \chi^k_-(n,1-\gamma) \bigg). 
\end{eqnarray}
The latter equation can be satisfied independently of $\alpha_s$ if and only if:
\begin{equation}
    \chi_+(n,\gamma) = \chi_-(n,1-\gamma),\label{eq:chi-PM-relation}
\end{equation}
which gives Eq.~(\ref{eq:proj-target-sym}). Indeed, let's assume instead that the quantity $\chi_+(n,\gamma) - \chi_-(n,1-\gamma) = O(\alpha_s^{n_0})$ with $n_0\geq 1$. Then 
\begin{eqnarray}
  \underbrace{\chi_+(n,\gamma)-\chi_-(n,1-\gamma)}_{O(\alpha_s^{n_0})} \mathop{=}\limits^{\text{?}} \underbrace{\chi_S'(n,\gamma)}_{O(1)} \underbrace{\frac{\hat{\alpha}_s}{2}}_{O(\alpha_s)}  \underbrace{\big(\chi_+(n,\gamma)-\chi_-(n,1-\gamma) \big)}_{O(\alpha_s^{n_0})} + O(\alpha_s^{n_0+2}), 
\end{eqnarray}
which is clearly a contradiction.

By the same reasoning, one can derive an inverse relation between characteristic functions of the symmetric and asymmetric schemes:
\begin{equation}
    \chi_{S}(n,\gamma) = \chi_{\pm}\big(n,\gamma \mp \frac{\hat{\alpha}_s}{2} \chi_{\pm}(n,\gamma) \big). \label{eq:rap-scheme-transf-2}
\end{equation}

{ Starting from the seminal work of Ref.~\cite{Salam98} it is known that the scheme-transformation discussed in this Appendix determines the structure of the leading collinear or anti-collinear poles of $\chi_S$. If one assumes that the leading collinear poles $\sim \alpha_s^n/\gamma^{2n+1}$ are absent in the $(-)$-scheme, then using Eq.~(\ref{eq:rap-scheme-transf-2}) it is possible to constrain their coefficients in the symmetric scheme. The same relation holds between the coefficients of the leading anti-collinear poles $\sim \alpha_s^n/(1-\gamma)^{2n+1}$ in the $(+)$ and symmetric schemes. For example one can assume that for $\gamma\ll 1$, $\chi_-(0,\gamma\rightarrow 0)\simeq 1/\gamma$, i.e. the leading collinear poles at higher orders in $\alpha_s$ are absent. Plugging this ansatz into Eq.~(\ref{eq:rap-scheme-transf}) and solving for $\chi_S$ one obtains:  }
\begin{eqnarray}
    \chi_S(0,\gamma\rightarrow 0) &=& \frac{2}{\gamma + \sqrt{\gamma^2 + 2\hat{\alpha}_s}} = \frac{1}{\hat{\alpha}_s} \big[\sqrt{\gamma^2+2\hat{\alpha}_s} - \gamma \big]  \nonumber \\
 &=&\frac{1}{\gamma} - \frac{2a_sN_c}{\gamma^3} + \frac{8(a_sN_c)^2}{\gamma^5}+O(a_s^3), 
\end{eqnarray}
{which agrees with the NLO BFKL result (\ref{eq:NLO-chi-exp-poles}) in QCD and  NNLO BFKL eigenvalue known in ${\cal N}=4$ SYM~\cite{Gromov:2015vua,Velizhanin:2015xsa,Caron-Huot:2016tzz}. }

%%%%%%%%%%%%%%%%%%%%%%%%%%%%%%%%%%%%%%%%%%%%%%%%%%%%%%%%%%%%%%%%%%%%%%%%%%%%%%%%%%%%%%%%%%%

%%%END OF append:Fourier %%%%%%%%%%%%%%%%%%%%%%%%%%%%%%%%%%%%%%%%%%%%%%%%%%%%%%%%%%%%%%%%%%%

\section{LO impact factor of a virtual photon}\label{append:gamma-IF}

The impact factor of a virtual photon is a well-studied object~\cite{Balitsky:1979ns,BFKL3,Bialas:2001ks,Chirilli:2014dcb,Ivanov:2014hpa,Colferai:2023dcf}, including the NLO in $\alpha_s$-corrections ~\cite{Balitsky:2010ze,Balitsky:2012bs}. Still, the technical details of the calculation of the LO impact factor presented below,  to our knowledge are not available in the literature. 

\begin{figure}
    \centering
    \includegraphics[width=0.5\linewidth]{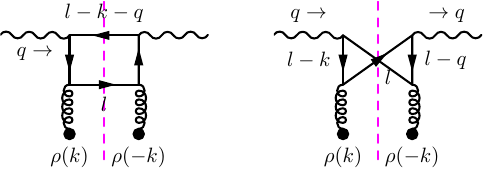}
    \caption{Feynman diagrams contributing to the photon impact factor. The diagrams with the opposite directions of the fermion lines should be added.}
    \label{fig:diags-gamma-IF}
\end{figure}

The expectation value of $\hat{\rho}(\T{k}) \hat{\rho}(-\T{k})$ operator over the virtual photon state, used in Sec.~\ref{sec:gamma-gamma} can be computed as a sum of two Feynman diagrams:
\begin{eqnarray}
    g^2 \bra{\gamma^*_\mu(q)} \hat{\rho}^a(\T{k}) \hat{\rho}^b(-\T{k}) \ket{\gamma^*_\nu(q)} = 2\big({\cal D}_1^{\mu\nu} + {\cal D}_2^{\mu\nu} \big) \tr(T^a T^b) = W^{\mu\nu} \tr(t^a t^b) ,
\end{eqnarray}
where the overall factor of 2 stands for the contribution of the diagrams with  opposite directions of the fermion lines\footnote{To check the Ward identity w.r.t. both $\mu$ and $\nu$, those diagrams should be added explicitly}. The contributions of the diagrams shown in Fig.~\ref{fig:diags-gamma-IF} are:
\begin{eqnarray}
    {\cal D}_1^{\mu\nu} &=& e_q^2 g^2 \int dk_- \int\frac{d^4l \, \delta_+(l^2) \delta_+((l+k-q)^2) }{(2\pi)^{3} [(l-k)^2]^2 } \nonumber \\
    && \times \tr[\slashed{l} \slashed{n}^+ (\slashed{l}-\slashed{k}) \gamma^\mu (\slashed{l}-\slashed{k}-\slashed{q}) \gamma^\nu (\slashed{l}-\slashed{k}) \slashed{n}^+], \\
    {\cal D}_2^{\mu\nu} &=&  e_q^2 g^2 \int dk_- \int\frac{d^4l \, \delta_+(l^2) \delta_+((l-k-q)^2) }{(2\pi)^{3} (l-k)^2 (l-q)^2 }  \nonumber \\
    && \times\tr [\slashed{l} \slashed{n}^+ (\slashed{l}-\slashed{k}) \gamma^\mu (\slashed{l}-\slashed{k}-\slashed{q})\slashed{n}^+ (\slashed{l}-\slashed{q}) \gamma^\nu],
\end{eqnarray}
where we have used the Feynman rule $(-\T{k}^2)n_+^\mu$ (``non-sense polarisation'') for the LO gluon coupling to the Reggeized gluon current (see e.g. the discussion of Feynman rules in Lipatov's EFT in Ref.~\cite{Nefedov:2019mrg}). 

We use the standard decomposition of the tensor $W^{\mu\nu}$ in terms of structure functions:
\begin{eqnarray}
    && W^{\mu\nu} = q^+ \bigg[ \bigg( -g^{\mu\nu} + \frac{q^\mu q^\nu}{q^2} \bigg)F_T \nonumber \\
    &&+ \frac{1}{Pq}\bigg( P^\mu - \frac{q^\mu(Pq)}{q^2} \bigg)\bigg( P^\nu - \frac{q^\nu(Pq)}{q^2} \bigg) \big(F_L+2x_B F_T \big) \bigg].
\end{eqnarray}
Here $P^\mu=P^- n_+^\mu/2$ is the four-momentum of the target photon, while the projectile photon has momentum $q^\mu = \big( q^+n_-^\mu - Q^2 n^+_\mu/q_+ \big)/2$ with $q^2=-Q^2$ and $x_B=Q^2/(2Pq)$. The projectors on the structure functions $F_L$ and $F_T$ are:
\begin{eqnarray}
    {\cal P}_L^{\mu \nu} &=& \frac{2Q^4}{P_-q_+^4} n_+^\mu n_+^\nu, \\
    {\cal P}_T^{\mu \nu} &=& \frac{-q_+^2 g^{\mu\nu} + Q^2 n_+^\mu n_+^\nu}{2q_+^3}.
\end{eqnarray}
With these projectors and the definition of the photon ``wave function'' (\ref{eq:gamma-WF-def}), one obtains:
\begin{eqnarray}
   &&\phi^{(L)}(\T{k}^2/Q^2) = \int\limits_0^1 \frac{dz Q^2}{\pi \T{k}^2} \int \frac{ d^2\T{l}\, (-4) Q^2  (1-z)^2 z^2 \left(\T{k}^{2}-2\left(\T{k}\cdot\T{l}\right)\right)^2}{ \left(\T{l}^{2}+Q^2 (1-z) z \right)^2 \left((\T{l}-\T{k})^2+Q^2 z(1-z)\right)^2}, \label{eq:phiL-00} \\
   &&\phi^{(T)}(\T{k}^2/Q^2) = \int\limits_0^1\frac{dz Q^2 }{\pi \T{k}^2} \int \frac{ d^2\T{l}\, (-1/2) (z^2+(1-z)^2)}{\left(\T{l}^{2}+Q^2
   z(1-z) \right)^2 \left((\T{l}-\T{k})^2+Q^2 z(1-z)\right)^2} \big[ 4 Q^2 (z-1) z (\T{k}\cdot\T{l})^{2} \nonumber \\
   && +\T{k}^{2} \left(\left(\T{l}^{2}-Q^2 (z-1) z\right)^2-2\left(\T{k}\cdot \T{l}\right)\left(\T{l}^{2}+Q^2 (z-1)z\right)\right)+\T{k}^{2}
   \T{l}^{2}\big], \label{eq:phiT-00}
\end{eqnarray}
where we have introduced $z=l^+/q^+$.

One can integrate-out the transverse momentum via  reduction of  Eqs. (\ref{eq:phiL-00}) and (\ref{eq:phiT-00}) in terms of the following ``master'' integrals without numerators:
\begin{eqnarray}
    && j_1(\T{k}^2,z) = \int\frac{d^2 \T{l}}{\pi \big(\T{l}^2+Q^2z(1-z)\big) \big((\T{k}-\T{l})^2 + Q^2z(1-z) \big)} \nonumber \\
    && = \frac{1}{4Q^2 z(1-z)} \frac{1}{\sqrt{t(1+t)}} \ln\bigg[ 1+4(2t+1)\sqrt{t(1+t)}+8t(1+t) \bigg] , \\
    && j_2(\T{k}^2,z) = \int\frac{d^2\T{l}}{\pi(\T{l}^2+Q^2z(1-z))^2} = \frac{1}{Q^2 z (1-z)},
\end{eqnarray}
where $t=\T{k}^2/(4Q^2 z(1-z))$. The Mellin transforms of these integrals can also be computed with the help of Feynman parametrisation:
\begin{eqnarray}
  &&  \tilde{j}_1(\gamma,z) = \int_{\T{k}} \bigg(\frac{\T{k}^2}{Q^2} \bigg)^{-\gamma} \frac{j_1(\T{k}^2,z)}{\pi \T{k}^2} =  z^{-1-\gamma}(1-z)^{-1-\gamma} \frac{\pi \csc(\pi\gamma) \Gamma^2(1+\gamma)}{\Gamma(2\gamma+2)}, \\
  && \tilde{j}_2(\gamma,z) = \int_{\T{k}}  \bigg(\frac{\T{k}^2}{Q^2} \bigg)^{-\gamma} \frac{j_2(\T{k}^2,z)}{\pi \T{k}^2}=0.
\end{eqnarray}
Using these results one can derive  exact expressions for the wavefunctions in transverse momentum space:
\begin{eqnarray}
    &&\phi^{(L)}(\tau) = \frac{1}{\tau}\int\limits_0^1 dz\, 8z(1-z) \label{eq:phiL-exact}  \\
    &&\times\bigg\{ \frac{1}{4\sqrt{t(z)(1+t(z))}}\ln\bigg[ 1+4(2t(z)+1)\sqrt{t(z)(1+t(z))}+8t(z)(1+t(z)) \bigg] -1 \bigg\}, \nonumber \\
    && \phi^{(T)}(\tau) = -\frac{1}{\tau}\int\limits_0^1 dz\, \big((1-z)^2+z^2\big) \label{eq:phiT-exact} \\
    && \times \bigg\{ \frac{2t(z)+1}{4\sqrt{t(z)(1+t(z))}}\ln\bigg[ 1+4(2t(z)+1)\sqrt{t(z)(1+t(z))}+8t(z)(1+t(z)) \bigg] -1 \bigg\}, \nonumber
\end{eqnarray}
where $t(z)=\tau/(4z(1-z))$. In Mellin space,
\begin{eqnarray}
    \phi^{(L)}(\gamma) &=& \frac{4\pi^2\gamma (\gamma-1) \cos(\pi\gamma)}{(2\gamma-3)(2\gamma-1)(2\gamma+1) \sin^2(\pi\gamma)}, \\
    \phi^{(T)}(\gamma) &=& \frac{\pi^2(\gamma-2)(\gamma+1)\cos(\pi\gamma)}{(2\gamma-3)(2\gamma-1)(2\gamma+1)\sin^2(\pi\gamma)}.
\end{eqnarray}
These results agree e.g. with Eqs. (3.19) in Ref.~\cite{Colferai:2023dcf}. The approximations (\ref{eq:appr-phi}) correspond to retaining only the poles at $\gamma=0$ and $\gamma=1$.

%%%END OF append:gamma-IF %%%%%%%%%%%%%%%%%%%%%%%%%%%%%%%%%%%%%%%%%%%%%%%%%%%%%%%%%%%%%%%%%%

\acknowledgments

%This is the most common positions for acknowledgments. A macro is available to maintain the same layout and spelling of the heading.

%\paragraph{Note added.} This is also a good position for notes added after the paper has been written.

The research was supported by  the  Binational Science Foundation grants \#2022132, 2021789 and by VATAT (Israel planning and budgeting committee) grant for supporting
theoretical high energy physics. This work is also supported by the U.S. Department of Energy, Office of Science, Office of Nuclear Physics, within the framework of the Saturated Glue (SURGE) Topical Theory Collaboration.
 The work of  A.K. is supported by the NSF Nuclear Theory grant \#2514546.
The work of ML and MN is supported by the ISF grant \#910/23. The work of V.S. is supported by the U.S. Department of Energy, Office of
Nuclear Physics through contract DE-SC0020081. \\
We  thank Physics Departments of the Ben Gurion University of the Negev and University of Connecticut for hospitality during mutual visits. \\

%%%%%%%%%%%%%%%%%%%%%%%%%%%%%%%%%%%%%%%%%%%%%%%%%%%%%%%%%%%%%%%

\bibliographystyle{JHEP}
\bibliography{mybibfile}

\end{document}